\documentclass[%
 reprint,
superscriptaddress,
 amsmath,amssymb,
 aps,
]{revtex4-2}

\usepackage{graphicx}
\usepackage{dcolumn}
\usepackage{bm}

\usepackage{isotope}
\usepackage{subcaption}

\begin{document}

\preprint{APS/123-QED}

\title{Mean-field derived IBM-1 Hamiltonian with intrinsic triaxial deformation}

\author{Polytimos Vasileiou}
\email{polvasil@phys.uoa.gr}
\affiliation{Department of Physics, National \& Kapodistrian University of Athens, Zografou Campus, GR--15784, Greece}
\author{Dennis Bonatsos}
\affiliation{Institute of Nuclear and Particle Physics, National Center for Scientific Research ``Demokritos'', GR--15310, Aghia Paraskevi, Greece}
\author{Theo J. Mertzimekis}
\affiliation{Department of Physics, National \& Kapodistrian University of Athens, Zografou Campus, GR--15784, Greece}
%
%

%
\date{\today}

\begin{abstract}
An interacting-boson-model-1 (IBM-1) Hamiltonian, derived from self-consistent mean-field calculations using a Skyrme energy density functional is employed for the study of energy spectra and $B(E2)$ transition strengths in the even-even \isotope[162-184]{Hf} and \isotope[168-186]{W}. An intrinsic triaxial deformation, derived from fermionic proxy-SU(3) irreps, is incorporated into the IBM-1 potential energy curve, which is subsequently mapped to the fermionic one, in order to derive the parameters of the IBM-1 Hamiltonian. It is shown that the inclusion of the intrinsic triaxial deformation derived from the proxy-SU(3) irreps leads to a significantly improved agreement between the theoretical predictions and experimental data for the low-lying quadrupole bands in the examined isotopes, without the need of higher-order terms in the IBM-1 Hamiltonian. The calculated $B(E2)$ transition strengths are also improved, compared to the axially symmetric case. The recently suggested preponderance of triaxial deformation over extended regions of the nuclear chart is obtained as a by-product. Future potential improvements and extensions to this mapping approach are also discussed.
\end{abstract}

\keywords{Suggested keywords}
\maketitle


\section{\label{sec:level1}Introduction}

The nuclear shell model~\cite{mayer55,heyde90,talmi93} is widely considered to provide a detailed microscopic description of atomic nuclei in terms of their constituent protons and neutrons. The spherical shell model~\cite{mayer55} is based on the isotropic 3-dimensional harmonic oscillator, possessing the SU(3) symmetry~\cite{wybourne74,moshinsky96,iachello06}, to which the spin-orbit force is added, which destroys the SU(3) symmetry beyond the $sd$ shell. In most cases only the valence protons and valence neutrons outside spherical closed shells are taken into account, in order to reduce the computational demands, although recently no-core shell model approaches~\cite{navratil00,dytrych08,launey20} are being developed for light nuclei. 

An alternative description of the collective properties of atomic nuclei is provided by the macroscopic collective model of Bohr and Mottelson~\cite{bohr52,bohr_mot98a,bohr_mot98b}, in which nuclear shapes are described in terms of the collective variables $\beta$ and $\gamma$, corresponding to the departure from sphericity and to the departure from axial symmetry respectively.  

A severe truncation of the shell model space in even-even nuclei is provided by the phenomenological Interacting Boson Model (IBM)~\cite{iachello87}, having an overall U(6) symmetry, in which the collective properties of medium-mass and heavy nuclei are described in terms of $s$ and $d$ bosons, possessing angular momentum 0 and 2 respectively. These bosons correspond to correlated fermion pairs, equal in number to the valence proton pairs and valence neutron pairs, measured from the nearest closed shell. Only one- and two-body interactions among the bosons are taken into account. Three dynamical symmetries exist in IBM: U(5)~\cite{arima76}, SU(3)~\cite{arima78}, and O(6)~\cite{arima79}, corresponding to vibrational (near-spherical), axially deformed, and soft to triaxial deformation (called $\gamma$-unstable) nuclei respectively. In the original version of the model, called IBM-1, no distinction is made between bosons corresponding to valence protons and neutrons. When this distinction is made, IBM-2 occurs. In the classical limit of IBM \cite{ginocchio80,ginocchio80a,dieperink80}, built through the use of coherent states (also called intrinsic states~\cite{iachello87}), potential energy surfaces (PESs) corresponding to IBM Hamiltonians are obtained in terms of the collective variables $\beta$ and $\gamma$. 

Along a different path, self-consistent mean-field many-body methods have been developed for atomic nuclei, using non-relativistic interactions~\cite{bender03,delaroche10,erler11} or relativistic energy density functionals~\cite{lalazissis97,vretenar05,niksic14}. PESs are readily constructed within this framework for nuclei of any mass, but the calculation of spectra and electromagnetic transition rates becomes demanding.  

A bridge between the mean field and IBM approaches has been built by Nomura \textit{et al.}~\cite{nomura08,nomura10,nomura11}. Along this path, the PES derived from the mean-field calculations is used to determine the free parameters appearing in IBM, by fitting the IBM PES to the mean-field PES. In this way one can subsequently use the IBM codes~\cite{casperson12} to derive predictions for spectra and  electromagnetic transition rates for specific nuclei without having to fit any free parameters to the data of each individual nucleus, as was done in the early days of IBM. In other words, one is able to obtain detailed spectroscopic predictions for a specific nucleus from an IBM Hamiltonian with microscopically derived parameters. 

On the other hand, triaxiality in even-even nuclei~\cite{davydov58,davydov59,meyertervehn75} has been attracting recently considerable attention. While the number of nuclei in which triaxiality is dominant is rather small, as indicated by the experimental odd-even  staggering within their $\gamma$-bands~\cite{zamfir91,mccutchan07}, it has been recently pointed out that some degree of triaxiality is present almost everywhere on the nuclear chart~\cite{tsunoda21,otsuka23,rouoof24}.    

It is known that within the standard IBM-1, in which only one- and two-body interactions are included, no triaxial shapes are obtained~\cite{pvisacker81}. Triaxial shapes can be obtained by taking into account three-body terms~\cite{pvisacker81,heyde84,thiamova10,fortunato11}. An alternative path is to use the standard IBM-2 with only one- and two-body interactions, in which distinction between protons and neutrons is made. Triaxial shapes then occur when valence protons are particles and valence neutrons are holes, or vice versa~\cite{dieperink82}. In the SU(3) framework of IBM, particles are described by prolate irreduclible representations (irreps) of the type $(N_1,0)$, while holes are described by oblate irreps of the form $(0,N_2)$. Combining these irreps one gets for the whole nucleus the irrep $(N_1,N_2)$, which is triaxial. This approach has been called the SU(3)$^*$ limit of IBM-2~\cite{dieperink82,walet87,sevrin87}.      

In the present work we propose an alternative path for the description of triaxiality in IBM-1, insisting on the use of one- and two-body interactions only but introducing an intrinsic triaxial deformation of microscopic origin. The value of the intinsic triaxial deformation is obtained through the proxy-SU(3) approximation~\cite{bonatsos17a,bonatsos17b,bonatsos23} to the shell model, described below. The parameters of the IBM-1 Hamiltonian are obtained by fitting its PES to the PES obtained from mean-field calculations employing the Skyrme energy density functional~\cite{skyax}. The present approach resembles the method of Nomura \textit{et al.}~\cite{nomura08,nomura10}, with two main differences: a) the use of an intrinsic triaxial deformation in the present case, and b) the use of different energy density functionals. While the latter difference is not expected to inflict major changes, since all energy density functionals in use are known to provide good results, the first one should reveal substantial differences, to be estimated through comparison of the present results to earlier approaches~\cite{nomura11a,nomura11b,rudigier15}. 

Discussion on the microscopic origin of the intrinsic triaxial deformation to be used within IBM-1 is now in place.  

A bridge between the spherical shell model and nuclear deformation has been built by Elliott~\cite{elliott58a,elliott58b,elliott63}, who pointed out that the wave functions in a degenerate oscillator level, classified according to the irreps of SU(3), for which the symbol $(\lambda,\mu)$ is used~\cite{elliott58a}, can be expressed as integrals of the Hill-Wheeler type over intrinsic states, and furthermore that all states belonging in a band involve the same intrinsic state in the integral~\cite{elliott58b}, with states in a band having all other quantum numbers equal being distinguished by an additional quantum number $K$ (the ``missing quantum number'' in the decomposition from SU(3) to SO(3)). As a consequence, simple expressions are obtained for the quadrupole moments, which resemble those of a rotational model with permanent deformation. In other words, Elliott revealed deformation within the spherical shell model. 

Elliott's SU(3) symmetry is destroyed in shells beyond the $sd$ shell, because the spin-orbit interaction pushes down in energy the orbitals of each shell bearing the  highest total angular momentum $j$~\cite{mayer55}. As a consequence a shell beyond the $sd$ one consists of its initial orbitals, minus the deserting orbitals (those with the highest $j$, which went to the shell below), plus the intruder orbitals coming from the shell above (in which they were having the highest $j$). The proxy-SU(3) approximation for even-even nuclei~\cite{bonatsos17a,bonatsos17b,bonatsos23} proved that a unitary transformation~\cite{martinou20} allows the replacement of the intruder orbitals (except the single level with the highest projection of $j$) by the deserting orbitals, thus reestablishing the SU(3) symmetry of the shell, with the exception of the left-over single level, which however is found to lie highest in energy, thus not influencing most nuclei in the shell.  The validity of the proxy-SU(3) symmetry has been first proved~\cite{bonatsos17a,bonatsos20} within the Nilsson's deformed shell model~\cite{nilsson55,nilsson95}, while later its connection to the spherical shell model has also been clarified~\cite{martinou20}. 

In proxy-SU(3) symmetry the SU(3) irreps to which the valence protons and the valence neutrons correspond are used, labeled by $(\lambda_p,\mu_p)$ and $(\lambda_n,\mu_n)$ respectively, coupled to the most stretched irrep $(\lambda,\mu)= (\lambda_p+\lambda_n, \mu_p+\mu_n)$~\cite{bonatsos17b}. What is of utmost importance is that for each kind of nucleons the highest weight irrep is used, which is the most symmetric one~\cite{martinou21b}, as required by the Pauli principle and the short-range nature of the nucleon-nucleon interaction, as discussed in detail in Refs.~\cite{bonatsos17b,martinou21b}. It should be noticed that the highest weight irrep is identical to the irrep possessing the highest eigenvalue of the second order Casimir operator of SU(3), related to the quadrupole-quadrupole interaction, up to the middle of the shell, but this is not the case any more beyond midshell, as it can be seen in Table I of Ref.~\cite{bonatsos17b}. 

The Elliott labels $\lambda$ and $\mu$ are known to be connected to the shape variables $\beta$ and $\gamma$ of the collective model, this connection being achieved  by mapping the eigenvalues of the invariant operators of the two theories~\cite{castanos88,draayer89}. Using this mapping from the $(\lambda,\mu)$ irrep characterizing a nucleus, parameter-independent predictions for its deformation variable $\beta$ and its axial symmetry variable $\gamma$ are obtained~\cite{bonatsos17b}, providing among other results an explanation for the dominance of prolate over oblate shapes~\cite{hamamoto09,hamamoto12} in the ground states of even-even nuclei, as well as an argument in favor of the recently suggested preponderance of triaxiality in heavy deformed nuclei~\cite{tsunoda21,otsuka23,rouoof24} supported by empirical evidence, provided by the experimental ratio of the band-head of the $\gamma$-band over the first excited state of the ground state band, $E(2_\gamma^+)/E(2_g^+)$~\cite{ensdf}, through use of the triaxial rotor model~\cite{davydov58,casten00,esser97}, as seen in Sec. VI of~\cite{bonatsos17b}.      

Medium-mass and heavy deformed nuclei are described in IBM-1 within its SU(3) dynamical symmetry~\cite{arima78,iachello87}, in which the ground state band belongs to the SU(3) irrep $(2N_B,0)$, where $N_B$ is the number of bosons corresponding to the specific nucleus. Since in this irrep one has $\mu=0$, the value of $\gamma$ is also close to zero. This is not the case in the proxy-SU(3) scheme, in which almost all nuclei have $\mu \neq 0$, as seen for example in Tables II and III of Ref.~\cite{bonatsos17b}. This means that $\gamma$ can obtain values away from zero, in agreement with the experimental expectations, as seen in Figs. 5 and 6 of Ref.~\cite{bonatsos17b}. It is therefore reasonable to consider an IBM description of medium-mass and heavy nuclei assuming an intrinsic non-zero value of $\gamma$. The Hf and W series of isotopes provide an appropriate test-ground for this assumption, since data exist for several isotopes extending from moderate to strong deformations~\cite{ensdf}.   

It should be noticed that in IBM-1 the ground state band (gsb) sits alone in the $(2N_B,0)$ SU(3) irrep, while the $\gamma$ ($K=2$) and $\beta$ ($K=0$) bands lie in the next irrep, $(2N_B-4,2)$~\cite{arima78,iachello87}. As a consequence, no $B(E2)$ transitions are allowed to connect the $\gamma$ band to the gsb, since they belong to different irreps, which cannot be connected by the quadrupole operator, which is a generator of SU(3), thus the SU(3) symmetry has to be broken~\cite{casten88}, since the experimental $B(E2)$s connecting the $\gamma$ band to the gsb are known to be substantial~\cite{ensdf}.  This problem is avoided in the proxy-SU(3) scheme, since the gsb and the $\gamma$ band belong to the same irrep with $\mu \neq 0$, thus no \textit{a priori} need for breaking the SU(3) symmetry arises. The problem is known to be avoided also within the pseudo-SU(3) scheme~\cite{ratnaraju73,draayer82,draayer84}, an alternative way of re-establishing the SU(3) symmetry of the 3-dimensional harmonic oscillator in medium-mass and heavy nuclei.  

In short, Elliott proved microscopically that deformation occurs within the spherical shell model based on an intrinsic state. In the present approach, we exploit the fact that the proxy-SU(3) irreps, dictated in a parameter-free way by the Pauli principle and the short range nature of the nucleon-nucleon interaction, exhibit non-vanishing  values of the Elliott label $\mu$, i.e., non-vanishing values of the collective variable $\gamma$, while in IBM-1 the SU(3) irreps containing the ground state band have 
$\mu=0$, because of the bosonic nature of the constituent particles. By introducing an intrinsic non-vanishing value of $\gamma$ in IBM-1, we therefore incorporate into the classical limit of IBM-1 the triaxiality feature required by the Pauli principle and the short-range nature of the nucleon-nucleon interaction.   

In different words, Elliott, working within the microscopic spherical shell model, discovered an intrinsic state corresponding to departure from sphericity in the form of axial deformation, prolate or oblate. Within the microscopic spherical shell model the Pauli principle is explicitly taken into account, since protons and neutrons are used, which are fermions. In addition, the short-range nature of the nucleon-nucleon interaction is taken into account through the form of the harmonic oscillator potential. The question is how to handle departure from axial deformation, i.e., triaxial shapes, within the phenomenological IBM-1, in which the constituent particles are bosons, ignoring the Pauli principle, without adding higher order (cubic) terms, which would lead us beyond the original definition of IBM-1, and without introducing the distinction between protons and neutrons, which would lead us to IBM-2. The way proposed in the present approach is to start from the classical limit of IBM-1, formed with the use of coherent states, and add to it the triaxial deformation in the form of a non-vanishing value of the deformation variable $\gamma$. The specific value of $\gamma$ is not treated as a free parameter, but is obtained in a parameter-free way from the proxy-SU(3) approximation to the shell model, in which both the Pauli principle and the short-range nature of the nucleon-nucleon interaction are explicitly taken into account, leading to the choice of the highest weight SU(3) irrep as the one describing the nucleus.

\section{\label{sec:models}Theoretical Procedure}
We begin our analysis by performing constrained Skyrme Hartree-Fock + BCS (denoted HF + BCS, for the sake of brevity) calculations for each of the isotopes under study, with the constraint being placed on the quadrupole deformation parameter, $\beta$. We employ the SkyAx~\cite{skyax} code to carry out the EDF calculations on a two-dimensional mesh on the $r-z$ plane, where the mesh spacing of $dr=dz=0.7$~fm is kept constant throughout. The SV-bas~\cite{klupfel08} EDF is used for the calculations. This functional is the starting point for the systematic variation (SV) set of parameterizations --- a set of relatively modern parameterizations to the Skyrme interaction, which has reached a lot of success, being employed in various theoretical studies in recent years --- and as such, it was chosen for our calculations (see also Ref.~\cite{abolghasem24}, and references therein, for a more detailed discussion on the SV-family of EDFs). For the pairing, a density-dependent delta-force was employed (see Ref.~\cite{skyax} for more details). However, it should be noted that, similar to the case of Refs.~\cite{nomura08,nomura10,nomura11,nomura11a}, the choice of Skyrme interaction does not essentially impact the results of the study, as long as the usual ones are considered. For each isotope, a set of constrained HF + BCS calculations is carried out for each value of $\beta$, leading to the construction of the corresponding potential energy curve (PEC). Since the total energy is used for the mapping procedure, all of the ingredients, including those related to kinetic terms, are supposed to be taken sufficiently into consideration~\cite{nomura11a}.

Having obtained the HF + BCS PEC for each isotope, we move on to the IBM description of the PEC. The Extended Consistent Q Formalism (ECQF)~\cite{lipas85}, first introduced in~\cite{warner82,warner83}, is adopted to write the IBM-1 Hamiltonian in the form~\cite{zamfir02,werner05}: 
\begin{equation}
    \label{eq:Hecqf}
    H(\zeta,\chi) = c\Bigl[(1-\zeta)\hat{n}_d-\frac{\zeta}{4N_B}\hat{Q}^{\chi}\hat{Q}^{\chi}\Bigr],
\end{equation}
where $N_B$ is the number of valence bosons, $\hat{n}_d=d^{\dagger}\cdot \tilde{d}$, $\hat{Q}^{\chi}=(s^{\dagger}\tilde{d}+d^{\dagger}s)+\chi(d^{\dagger}\tilde{d})^{(2)}$ the number operator for quadrupole bosons, and the quadrupole operator, respectively, and $c$ is a scaling factor. The above Hamiltonian encompasses the entire IBM symmetry triangle~\cite{casten00,feng81,iachello98}, along with the U(5)--O(6)--SU(3) dynamical symmetry limits of the IBM, by making use of two structural parameters, $\zeta$ and $\chi$. The parameters $(\zeta,\chi)$ describing a nucleus can be placed in the IBM symmetry triangle by converting them into polar coordinates, through the relations~\cite{mccutchan04,zyriliou22}:
\begin{equation}
\label{eq:polar}
    \rho = \frac{\sqrt{3}\zeta}{\sqrt{3}\cos(\theta_{\chi})-\sin(\theta_{\chi})},~~ \theta = \frac{\pi}{3}+\theta_{\chi},
\end{equation}
with $\theta_{\chi}=(2/\sqrt{7})(\pi/3)\chi$.

Finally, the coherent state formalism~\cite{arima87,ginocchio80,dieperink80} of the IBM is employed to extract the following expression for the energy surface, $E(\beta,\gamma)$, corresponding to the Hamiltonian of Eq.~(\ref{eq:Hecqf})~\cite{mccutchan06}:
\begin{eqnarray}
\label{eq:ibm-pes}
    E(\beta,\gamma)=\frac{cN_B\beta^2}{1+\beta^2}\Biggl[(1-\zeta)-(\chi^2+1)\frac{\zeta}{4N_B}\Biggr]\nonumber\\
    -\frac{5c\zeta}{4(1+\beta^2)}
    -\frac{c\zeta(N_B-1)}{4(1+\beta^2)^2}\nonumber\\
    \times\Biggl[4\beta^2-4\sqrt{\frac{2}{7}}\chi\beta^3\cos{3\gamma}+\frac{2}{7}\chi^2\beta^4 \Biggr].
\end{eqnarray}
Eq.~(\ref{eq:ibm-pes}) relates the structural parameters $(\chi,\zeta)$ of the ECQF Hamiltonian with the $(\beta,\gamma)$ classical coordinates linked to the Bohr geometrical variables~\cite{iachello87,bohr52,bohr_mot98b}. More specifically, $\zeta$ is related to the the axial quadrupole deformation parameter, $\beta$, while $\chi$ is associated with the triaxiality parameter, $\gamma$, which regulates the degree of triaxial deformation of a nucleus.  

It should be noted that the deformation parameter of the boson system is not identical to that of the geometrical model, since the former is derived by taking into account only the valence bosons, in contrast to the latter, for which the entirety of the shell-model space is considered~\cite{ginocchio80a}. Thus, the IBM deformation parameter, $\beta$ is always larger than the corresponding fermionic one, $\beta_F$, and, to a good approximation, one can assume $\beta \propto \beta_F$~\cite{ginocchio80a}. One can then go on to write:
\begin{equation}
    \label{eq:beta_cond}
    \beta=C_{\beta}\beta_F,
\end{equation}
with $C_{\beta}(>1)$ being the proportionality coefficient for the $\beta$--deformation~\cite{nomura08,rudigier15}.

Regarding the triaxiality parameter of the bosonic PEC, as a first approach we assume axial deformations, and equate:
\begin{equation}
	\label{eq:gamma_cond}
	\gamma_F=\gamma=0^{\circ}.  
\end{equation}

By imposing the conditions~(\ref{eq:gamma_cond}) and (\ref{eq:beta_cond}) in Eq.~(\ref{eq:ibm-pes}), we obtain the boson potential energy curves for each of the studied isotopes, as functions of the axial quadrupole deformation, $\beta$, with parameters $\chi, \zeta, c$ and $C_{\beta}$, namely:
\begin{equation}
    \label{eq:ibm-pec}
    E(\beta,\gamma=0)\rightarrow E(\beta)\equiv E(\beta_F;C_{\beta},\chi,\zeta,c,N_B).
\end{equation}
The procedure followed for the mapping of the IBM PECs to the EDF calculated ones, and the subsequent determination of the optimal set of IBM-1 parameters for each isotope, is outlined below.

The intervals $\chi=[-\sqrt{7}/2,0]$ and $\zeta=[0,1]$ are divided in steps, with fixed step-sizes $d\chi$ and $d\zeta$, and the same is done for $C_{\beta}$, for which the selected interval is $[1,10]$. The choice of the lower limit for $C_{\beta}$ has already been justified by the discussion in the preceding paragraphs, while for the upper limit of this range, a sufficiently large value has been used, such that the $C_{\beta}^{best}$ optimal value is enclosed in the defined interval. The choice of $C_{\beta}^{max}=10$ seems to be sufficient, based on our results, as well as the results of similar studies on \isotope[166-180]{Hf}~\cite{rudigier15} and \isotope[182-194]{Hf}~\cite{nomura11a}, where an IBM-2 Hamiltonian was mapped to Hartree--Fock--Bogolyubov (HFB) EDFs for calculations. 

\begin{figure*}[!htbp]
\centering
\includegraphics[width=\textwidth]{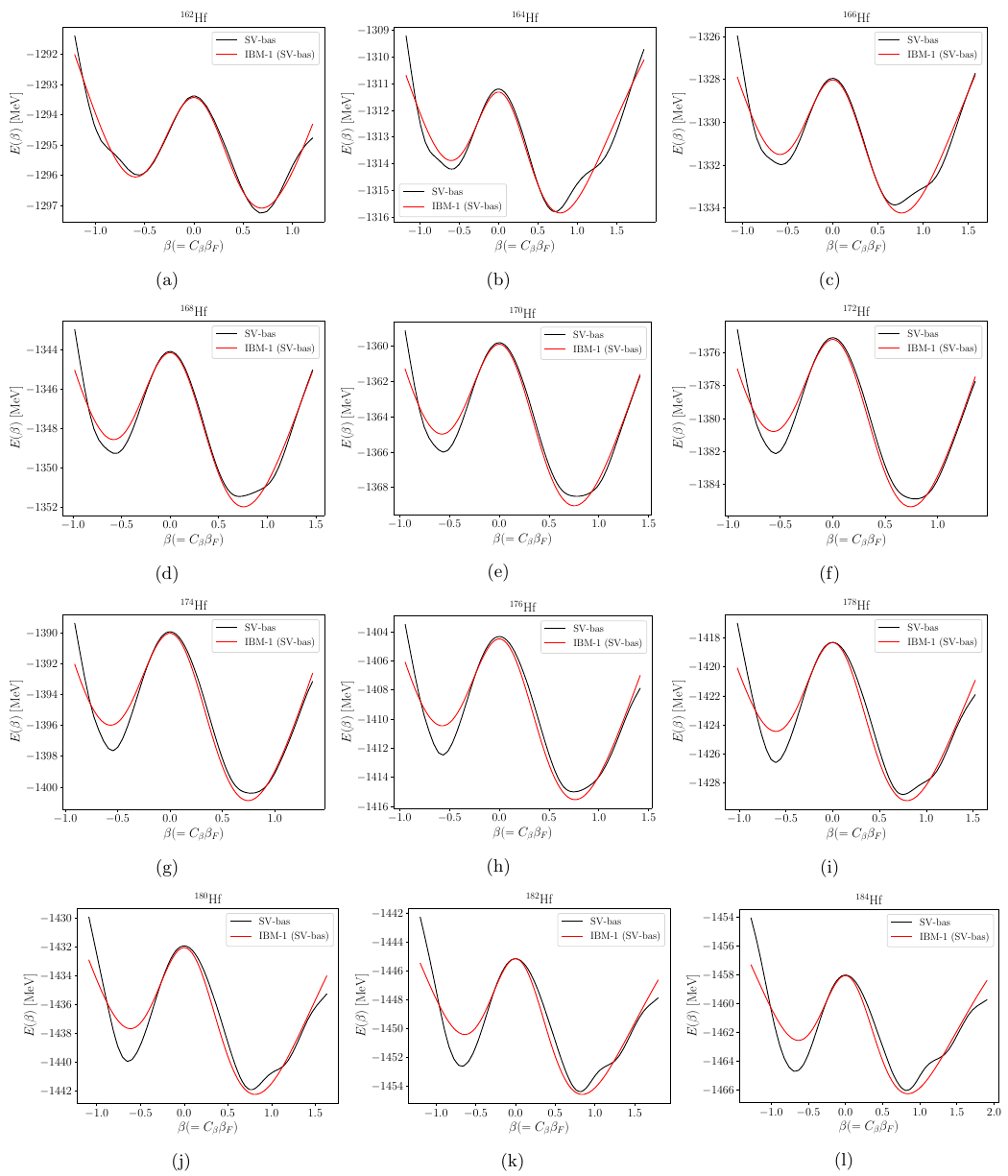}
\caption{\label{fig:Hf-PECs}SV-bas EDF potential energy curves (black) vs. the corresponding IBM-1 ones (red), for \isotope[162-184]{Hf}. The IBM-1 PECs resulted from the mapping process outlined in Section~\ref{sec:models}, with the incorporation of proxy-SU(3) irreps.}
\end{figure*}

\begin{figure*}[!htbp]
\centering
\includegraphics[width=\textwidth]{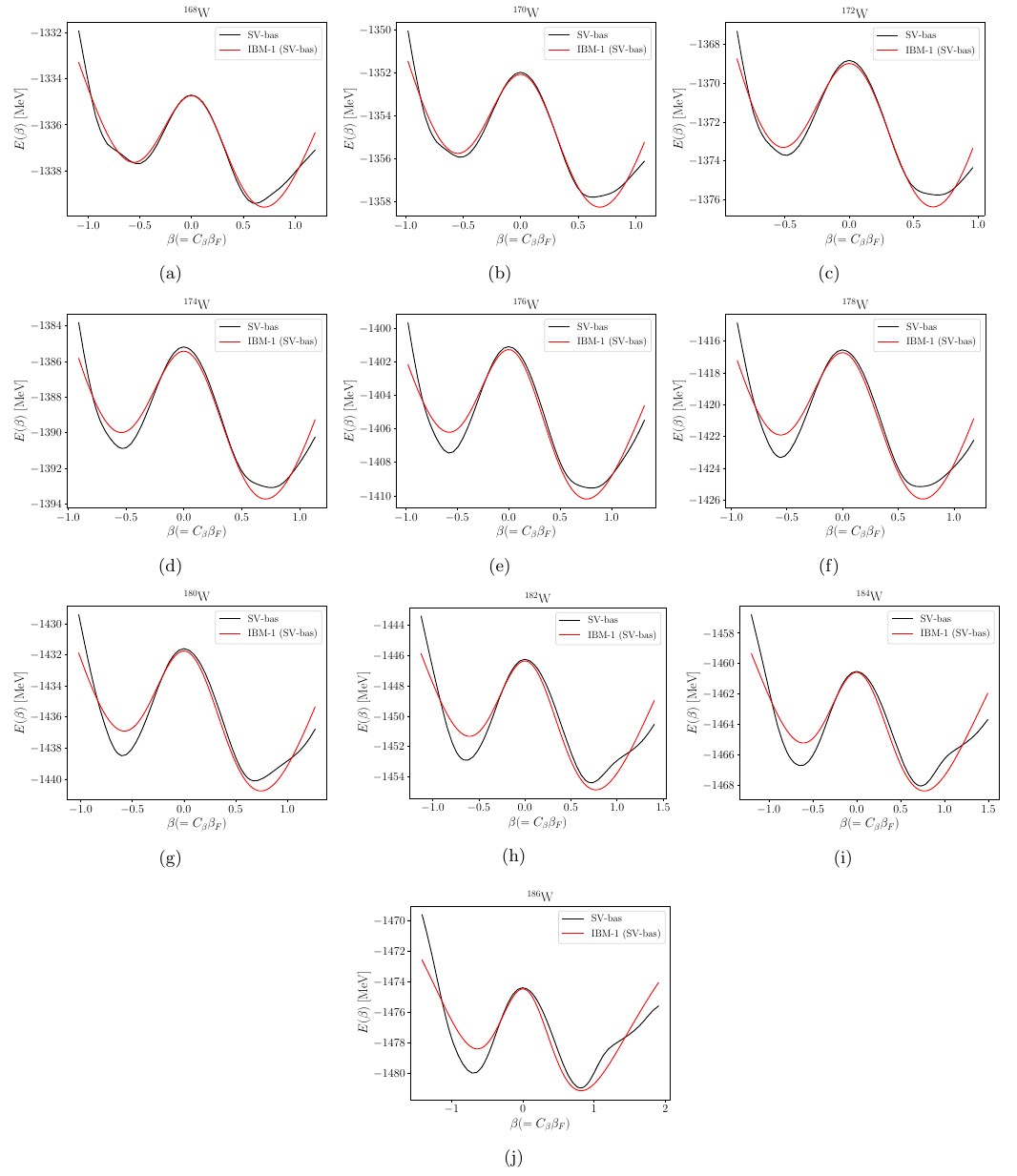}
\caption{\label{fig:W-PECs}Same as in Fig.~\ref{fig:Hf-PECs}, for \isotope[168-186]{W}.}
\end{figure*}

For each set of $(\chi^i,\zeta^i,C_{\beta}^i)$, an IBM PEC is generated through Eqs.~(\ref{eq:ibm-pes}) and (\ref{eq:ibm-pec}), and compared to the HF PEC. The set of optimal $(\chi,\zeta,C_{\beta})$ parameters are chosen such that the best reproduction of the overall shape and curvature of the HF PEC by the IBM PEC, up to a range of a few MeV from the absolute minimum of the microscopic PEC is achieved (Figs.~\ref{fig:Hf-PECs} and \ref{fig:W-PECs}).

The resulting $\chi$ and $\zeta$ values, along with the valence boson number, $N_B$, are then used as inputs for the diagonilization of the IBM-1 ECQF Hamiltonian of Eq.~(\ref{eq:Hecqf}), carried out with the IBAR code by Casperson~\cite{casperson12}. The scale, $c$, entering Eq.~(\ref{eq:Hecqf}) to obtain quantitative results for the calculated energy levels, is chosen so as to reproduce the experimental $E(2^+_1)$ for each isotope.
\begin{table*}[!htbp]
	\caption{\label{tab:ibm-params}%
		The parameters of the IBM-1 Hamiltonian of Eq.~(\ref{eq:Hecqf}), derived from the mapping process described in Section~\ref{sec:models}, for the case of $\gamma=0^{\circ}$. The $N_B,\chi,\zeta,c$ parameters were used as IBAR code inputs for the calculation of energy levels for the ground-, $\beta$- and $\gamma$-bands in the studied Hf and W isotopes. Tabulated are also the effective charges, $e_{B}$, proportionality coefficients for $\beta$ deformation, $C_{\beta}$, and the experimental (exp.) and calculated (th.) $R_{4/2}$ energy ratios.
	}
	\begin{ruledtabular}
		\begin{tabular}{cdddddddd}
			Isotope 
			& \multicolumn{1}{c}{$N_B$} 
			& \multicolumn{1}{c}{$\chi$} 
			& \multicolumn{1}{c}{$\zeta$} 
			& \multicolumn{1}{c}{$c$~[MeV]}
			& \multicolumn{1}{c}{$e_B$~[$e$fm$^2$]} 
			& \multicolumn{1}{c}{$C_{\beta}$}
			& \multicolumn{1}{c}{$R_{4/2}$~(exp.)} 
			& \multicolumn{1}{c}{$R_{4/2}$~(th.)} \\
			\colrule
\isotope[162]{Hf} &  9  & -0.185 & 0.770 & 2.080 & 14.7 & 3.070 & 2.56 & 2.58 \\
\isotope[164]{Hf} & 10	& -0.344 & 0.820 & 2.405 & 16.0 & 3.070 & 2.79 & 2.95 \\
\isotope[166]{Hf} & 11	& -0.370 & 0.790 & 1.956 & 16.9 & 2.620 & 2.97 & 2.97 \\
\isotope[168]{Hf} & 12	& -0.357 & 0.780 & 1.652 & 17.0 & 2.440 & 3.11 & 3.00 \\
\isotope[170]{Hf} & 13	& -0.357 & 0.760 & 1.503 & 16.6 & 2.260 & 3.19 & 3.00 \\
\isotope[172]{Hf} & 14	& -0.357 & 0.750 & 1.473 & 16.2 & 2.170 & 3.25 & 3.02 \\
\isotope[174]{Hf} & 15	& -0.370 & 0.760 & 1.572 & 15.2 & 2.260 & 3.27 & 3.11 \\
\isotope[176]{Hf} & 16	& -0.370 & 0.770 & 1.723 & 14.3 & 2.350 & 3.28 & 3.18 \\
\isotope[178]{Hf} & 15	& -0.384 & 0.790 & 1.790 & 14.7 & 2.530 & 3.29 & 3.19 \\
\isotope[180]{Hf} & 14	& -0.384 & 0.810 & 1.723 & 14.7 & 2.710 & 3.31 & 3.19 \\
\isotope[182]{Hf} & 13	& -0.397 & 0.830 & 1.733 & 14.6 & 2.980 & 3.29 & 3.18 \\
\isotope[184]{Hf} & 12	& -0.410 & 0.830 & 1.725 & 15.2 & 3.160 & 3.26 & 3.17 \\
\hline
\isotope[168]{W}  & 10	& -0.304 & 0.770 & 1.870 & 17.7 & 2.710 & 2.82 & 2.77 \\
\isotope[170]{W}  & 11	& -0.291 & 0.730 & 1.423 & 18.0 & 2.260 & 2.95 & 2.71 \\
\isotope[172]{W}  & 12	& -0.278 & 0.700 & 1.084 & 18.1 & 1.990 & 3.06 & 2.65 \\
\isotope[174]{W}  & 13	& -0.344 & 0.760 & 1.549 & 17.0 & 2.350 & 3.15 & 2.98 \\
\isotope[176]{W}  & 14	& -0.370 & 0.790 & 1.871 & 15.8 & 2.530 & 3.22 & 3.13 \\
\isotope[178]{W}  & 15	& -0.344 & 0.750 & 1.713 & 15.1 & 2.350 & 3.24 & 3.05 \\
\isotope[180]{W}  & 14	& -0.344 & 0.780 & 1.689 & 15.1 & 2.620 & 3.26 & 3.08 \\
\isotope[182]{W}  & 13	& -0.331 & 0.780 & 1.448 & 15.3 & 2.710 & 3.29 & 3.01 \\
\isotope[184]{W}  & 12	& -0.331 & 0.800 & 1.529 & 15.4 & 2.980 & 3.27 & 3.01 \\
\isotope[186]{W}  & 11	& -0.344 & 0.840 & 1.685 & 15.3 & 3.520 & 3.23 & 3.05 \\
\end{tabular}
\end{ruledtabular}
\end{table*}

\begin{figure}[!htbp]
	\includegraphics[width=0.45\textwidth]{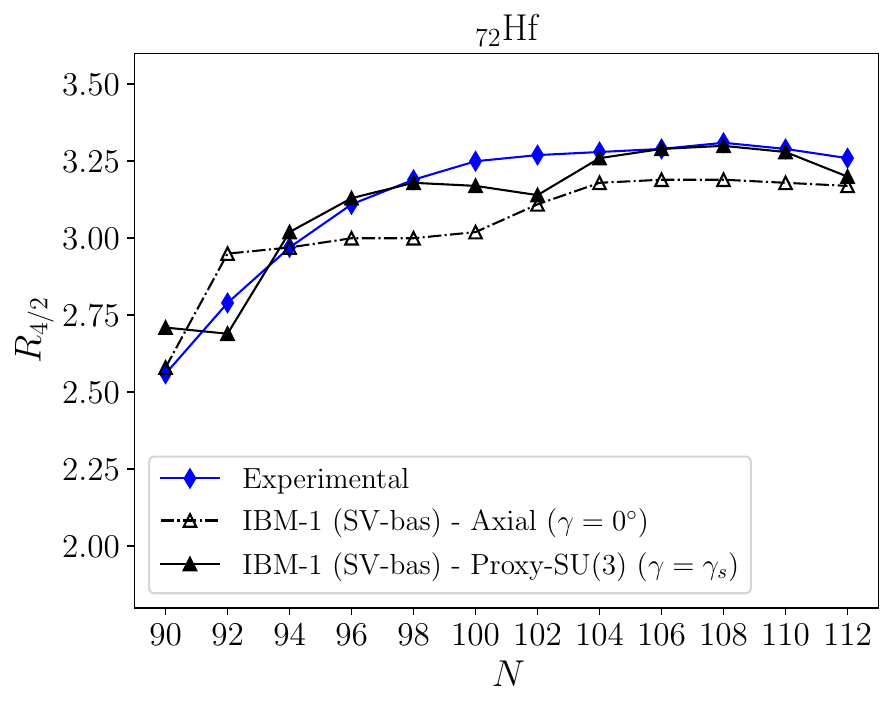}
	\caption{\label{fig:hf_r42} Experimental (blue diamonds) vs. calculated $R_{4/2}$ ratios for \isotope[162-184]{Hf}. Results for the case of ``pure'' axial deformations ($\gamma = 0^{\circ}$) are shown in empty black triangles, while the predictions made with the use of proxy-SU(3) irreps are plotted in black color with solid symbols (see Section~\ref{sec:models} for details on the calculations).}
\end{figure}

\begin{figure}[!htbp]
	\includegraphics[width=0.45\textwidth]{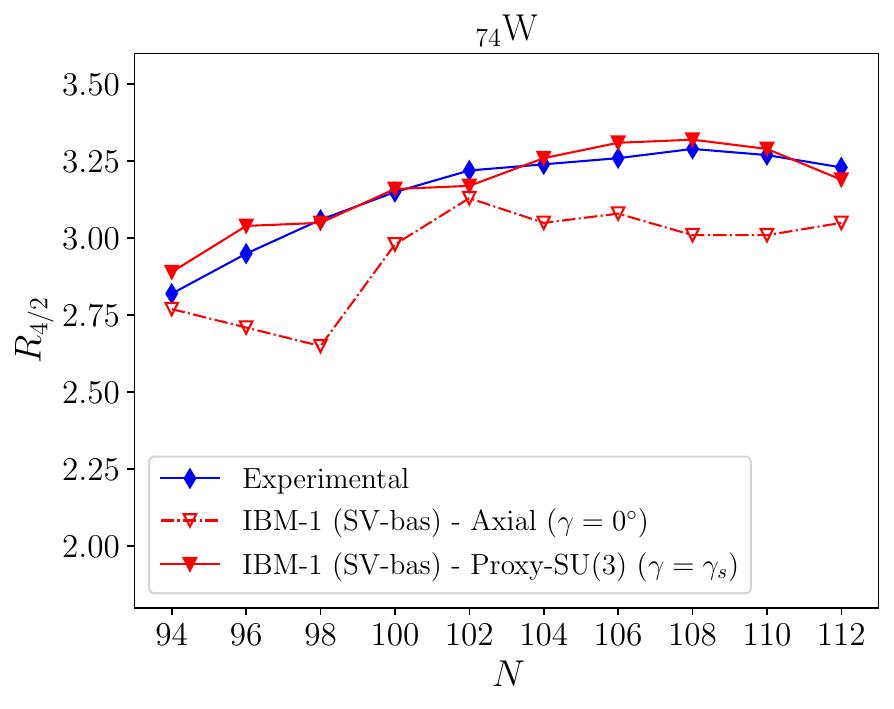}
	\caption{\label{fig:w_r42} Same as in Fig.~\ref{fig:w_r42}, for \isotope[168-186]{W} (experimental values shown in blue diamonds, theoretical ones shown in red triangles).}
\end{figure}

\begin{figure*}[!htbp]
\centering
\includegraphics[width=0.75\textwidth]{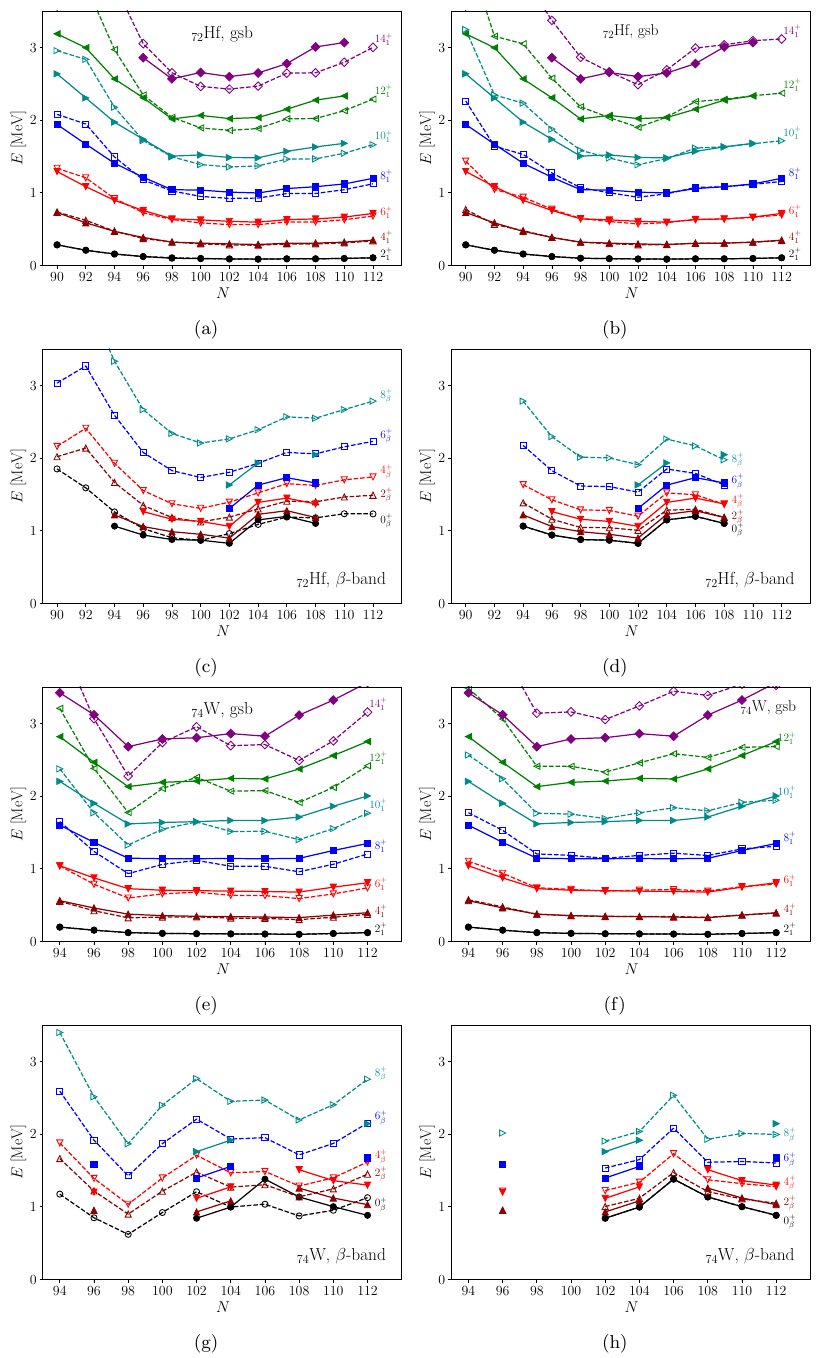}
\caption{\label{fig:Hf-W-gs-beta} Experimental (solid lines and symbols) vs. theoretical (dashed lines with empty symbols) g.s. and $\beta$-band energies for the examined Hf and W isotopes. Calculations with $\gamma=0^{\circ}$ are shown in the left column (panels (a), (c), (e), (g)), while calculations performed with $\gamma=\gamma_s$ and the use of the $\alpha_{\beta}$, $\alpha_{\gamma}$ mass coefficients are shown on the right (panels (b), (d), (f), (h)).}
\end{figure*}

\begin{figure*}[!htbp]
\centering
\includegraphics[width=0.95\textwidth]{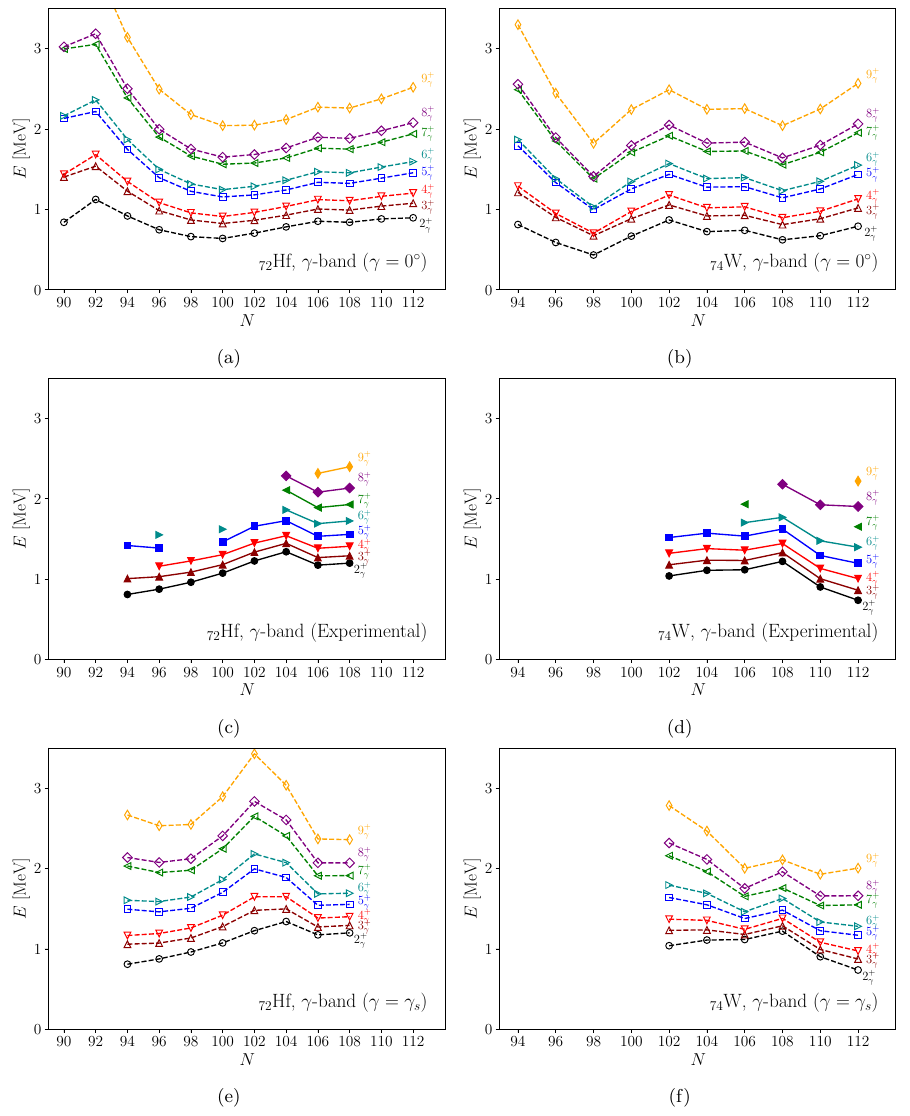}
\caption{\label{fig:Hf-W-gamma} Similar to Fig.~\ref{fig:Hf-W-gs-beta}, but for the $\gamma$-bands. Theoretical calculations and experimental data are split into separate panels, to improve readability.}
\end{figure*}

\begin{figure}[!htbp]
\centering
\includegraphics[width=0.48\textwidth]{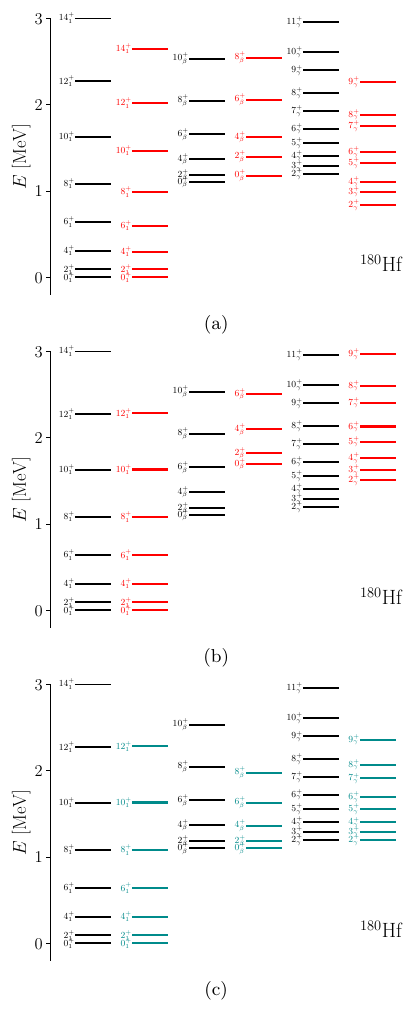}
\caption{\label{fig:180Hf}Experimental (black) vs. calculated (red) g.s., $\beta$- and $\gamma$-band levels of \isotope[180]{Hf}, for the case of $\gamma=0^{\circ}$ (a) and $\gamma=\gamma_s$ (b). Levels calculated for $\gamma=\gamma_s$, with the use of mass coefficients, are plotted in darkcyan color and compared with the experimental ones (in black) in panel (c) (see Section~\ref{sec:models} for details).}
\end{figure}

The assumption of ``pure'' axial deformations ($\gamma=0^{\circ}$) is enough to sufficiently reproduce the first few low-lying energy levels of the ground-state band (gsb) of the studied nuclei (up to the $8^+_1$ level, in most cases). However, the following discrepancies still persist:
\begin{enumerate}
	\item[(i)] The $R_{4/2} \equiv E(4^+_1)/E(2^+_1)$ ratios (see Table~\ref{tab:ibm-params}, and Figs.~\ref{fig:hf_r42}, \ref{fig:w_r42}) are generally underestimated, compared to the experimental ones.
	\item[(ii)] The calculated gsb energy levels begin to diverge from the experimental ones, with increasing values of spin (Figs.~\ref{fig:Hf-W-gs-beta}a, \ref{fig:Hf-W-gs-beta}e).
	\item[(iii)] The band spacings of the members of the $\beta$-band are not accurately reproduced (Figs.~\ref{fig:Hf-W-gs-beta}c, \ref{fig:Hf-W-gs-beta}g).
	\item[(iv)] The energy values for the $\gamma$ band members are systematically underpredicted, while there is a pronounced odd--even staggering (i.e. $2^+_{\gamma}$, $(3^+_{\gamma},4^+_{\gamma})$, $(5^+_{\gamma},6^+_{\gamma})$,...) within the $\gamma$ bands (Figs.~\ref{fig:Hf-W-gamma}a, \ref{fig:Hf-W-gamma}b). 
\end{enumerate}
The resulting band levels for \isotope[162-184]{Hf} and \isotope[168-186]{W} are shown in Figs.~\ref{fig:Hf-W-gs-beta}a, \ref{fig:Hf-W-gs-beta}c, \ref{fig:Hf-W-gs-beta}e, \ref{fig:Hf-W-gs-beta}g and \ref{fig:Hf-W-gamma}a--\ref{fig:Hf-W-gamma}d (see also Fig.~\ref{fig:180Hf}a for an indicative level scheme). It is evident that this initial approach cannot provide a sufficiently accurate simultaneous description of the ground-, $\beta$- and $\gamma$-bands in the investigated members of the Hf and W isotopic chains.

Taking a closer look at the empirical values for the triaxiality parameter, $\gamma$, spanning the entire rare-earth region of the nuclear chart, it is observed that even for the most axially-symmetric nuclei, low, yet non-zero values of $\gamma \approx 8^{\circ}$ appear (this is discussed in \cite{casten00,ginocchio80a,davydov58,esser97,casten84}, see also Fig.~5 of Ref.~\cite{bonatsos17b}). Thus, the incorporation of a degree of triaxiality, and the impact on the various bands follows as the next step in our calculations.

In the framework of the IBM-1, triaxiality can be achieved through the inclusion of higher order terms, such as three- and four-body terms of the form $(d^{\dagger}\times d^{\dagger}\times d^{\dagger})\cdot(\tilde{d}\times\tilde{d}\times\tilde{d})$~\cite{pvisacker81,heyde84}, $(\hat{Q}\times\hat{Q}\times\hat{Q})$, $(\hat{L}\times\hat{Q}\times\hat{L})$ and $(\hat{L}\times\hat{Q})\cdot(\hat{L}\times\hat{Q})$~\cite{zhang14,zhang22}. An alternate path to triaxiality was undertaken in recent years, through the implementation of group-theoretical methods, utilizing SU(3) irreps (see e.g.~\cite{zhang14,zhang22,leschber87,naqvi95,smirnov00}, and references therein).

For our calculations, we make use of the proxy-SU(3) highest-weight (h.w.) irreps~\cite{bonatsos17b,martinou21a}, to derive a value for the triaxiality parameter~\cite{castanos88,draayer89}:
\begin{equation}
	\label{eq:gamma_s}
	\gamma_s = \arctan\bigg(\frac{\sqrt{3}(\mu+1)}{2\lambda+\mu+3}\bigg).
\end{equation} 
We then go on to substitute the above expression, along with Eq.~(\ref{eq:beta_cond}), into Eq.~(\ref{eq:ibm-pes}), obtaining the following IBM-1 PECs:
\begin{equation}
	\label{eq:ibm-pec-proxy}
	E(\beta,\gamma=\gamma_s)\rightarrow E(\beta)\equiv E(\beta_F;C_{\beta},\chi,\zeta,c,N_B).
\end{equation}
The same procedure as in the case of $\gamma=0^{\circ}$ is then undertaken for the determination of the new values of $(\chi,\zeta,C_{\beta})$, for each of the studied Hf and W isotopes. The parameters are tabulated in Table~\ref{tab:ibm-params2} (see also Fig.~\ref{fig:180Hf}b for an indicative level scheme).
\begin{table*}[!htbp]
	\caption{\label{tab:ibm-params2}%
		The parameters of the IBM-1 Hamiltonian of Eq.~(\ref{eq:Hecqf}), derived from the mapping process described in Section~\ref{sec:models}, for the case of $\gamma_s$. The $N_B,\chi,\zeta,c$ parameters were used as IBAR code inputs for the calculation of energy levels for the ground-, $\beta$- and $\gamma$-bands in the studied Hf and W isotopes. Tabulated are also the effective charges, $e_{B}$, proportionality coefficients for $\beta$ deformation, $C_{\beta}$, the intrinsic $\gamma$ deformation parameter, $\gamma_s$, the experimental (exp.) and calculated (th.) $R_{4/2}$ energy ratios.
	}
	\begin{ruledtabular}
		\begin{tabular}{cddddddddd}
Isotope 
& \multicolumn{1}{c}{$N_B$} 
& \multicolumn{1}{c}{$\chi$} 
& \multicolumn{1}{c}{$\zeta$} 
& \multicolumn{1}{c}{$c$~[MeV]} 
& \multicolumn{1}{c}{$e_B$~[$e$fm$^2$]} 
& \multicolumn{1}{c}{$C_{\beta}$}
& \multicolumn{1}{c}{$\gamma_s$} 
& \multicolumn{1}{c}{$R_{4/2}$~(exp.)} 
& \multicolumn{1}{c}{$R_{4/2}$~(th.)} \\
\colrule
\isotope[162]{Hf} &  9 & -0.265 & 0.790 & 2.370 & 14.6 & 3.340 & 13.923 & 2.56 & 2.71 \\
\isotope[164]{Hf} & 10 & -0.291 & 0.740 & 1.759 & 16.6 & 3.070 & 12.834 & 2.79 & 2.69 \\
\isotope[166]{Hf} & 11 & -0.397 & 0.790 & 2.015 & 16.8 & 2.620 &  8.308 & 2.97 & 3.02 \\
\isotope[168]{Hf} & 12 & -0.463 & 0.780 & 1.841 & 16.5 & 2.440 & 13.407 & 3.11 & 3.13 \\
\isotope[170]{Hf} & 13 & -0.503 & 0.770 & 1.672 & 15.9 & 2.350 & 14.840 & 3.19 & 3.18 \\
\isotope[172]{Hf} & 14 & -0.463 & 0.760 & 1.638 & 15.6 & 2.260 & 12.949 & 3.25 & 3.17 \\
\isotope[174]{Hf} & 15 & -0.397 & 0.760 & 1.616 & 15.1 & 2.260 &  7.735 & 3.27 & 3.14 \\
\isotope[176]{Hf} & 16 & -0.529 & 0.760 & 1.893 & 13.7 & 2.350 & 14.840 & 3.28 & 3.26 \\
\isotope[178]{Hf} & 15 & -0.661 & 0.780 & 2.066 & 13.6 & 2.530 & 18.793 & 3.29 & 3.29 \\
\isotope[180]{Hf} & 14 & -0.728 & 0.790 & 1.998 & 13.4 & 2.710 & 19.423 & 3.31 & 3.30 \\
\isotope[182]{Hf} & 13 & -0.609 & 0.820 & 1.925 & 13.9 & 2.980 & 16.558 & 3.29 & 3.28 \\
\isotope[184]{Hf} & 12 & -0.463 & 0.830 & 1.796 & 15.0 & 3.160 &  9.339 & 3.26 & 3.20 \\
\hline
\isotope[168]{W}  & 10 & -0.384 & 0.770 & 2.048 & 17.5 & 2.710 & 12.834 & 2.82 & 2.89 \\
\isotope[170]{W}  & 11 & -0.503 & 0.750 & 1.945 & 17.0 & 2.440 & 18.048 & 2.95 & 3.04 \\
\isotope[172]{W}  & 12 & -0.542 & 0.720 & 1.600 & 16.8 & 2.170 & 19.423 & 3.06 & 3.05 \\
\isotope[174]{W}  & 13 & -0.556 & 0.740 & 1.785 & 16.2 & 2.260 & 17.418 & 3.15 & 3.16 \\
\isotope[176]{W}  & 14 & -0.450 & 0.770 & 1.899 & 15.6 & 2.440 & 11.860 & 3.22 & 3.17 \\
\isotope[178]{W}  & 15 & -0.635 & 0.740 & 2.122 & 14.0 & 2.350 & 19.423 & 3.24 & 3.26 \\
\isotope[180]{W}  & 14 & -0.979 & 0.740 & 2.230 & 12.8 & 2.530 & 23.606 & 3.26 & 3.31 \\
\isotope[182]{W}  & 13 & -1.111 & 0.750 & 2.094 & 12.4 & 2.800 & 24.523 & 3.29 & 3.32 \\
\isotope[184]{W}  & 12 & -0.767 & 0.780 & 1.994 & 13.8 & 2.980 & 21.772 & 3.27 & 3.29 \\
\isotope[186]{W}  & 11 & -0.489 & 0.830 & 1.871 & 14.8 & 3.520 & 14.496 & 3.23 & 3.19 \\
		\end{tabular}
	\end{ruledtabular}
\end{table*}

In Fig.~2 of the paper~\cite{martinou17} the proxy-SU(3) predictions for the deformation variable $\gamma$ of the Hf and W isotopes under study have been compared to the Gogny D1S predictions of Ref.~\cite{delaroche10}, as well as to empirical values, with good agreement seen. This fact adds reliability to the proxy-SU(3) predictions, making them appropriate for use in the present work.

The inclusion of an intrinsic $\gamma$ deformation, in the form of $\gamma_s$, resulting from the use of proxy-SU(3) h.w. irreps has minimal impact on the curvature and overall shape of the calculated IBM-1 PECs, which are (almost) indistinguishable to the ones obtained in the $\gamma = 0^{\circ}$ case, with the use of Eq.~(\ref{eq:ibm-pec}). This is also reflected in the $C_{\beta}$ proportionality coefficients tabulated in Tables~\ref{tab:ibm-params} and \ref{tab:ibm-params2}, which are (nearly/) identical in both cases. Thus, only the IBM-1 PECs obtained from Eq.~(\ref{eq:ibm-pec-proxy}) are plotted in Figs.~\ref{fig:Hf-PECs}a--\ref{fig:Hf-PECs}l and \ref{fig:W-PECs}a--\ref{fig:W-PECs}j, together with the SV-bas EDF PECs, shown for comparison. 
 
As it can be seen from Figs.~\ref{fig:hf_r42}, \ref{fig:w_r42}, the inclusion of an intrinsic $\gamma$ deformation, through the use of proxy-SU(3) h.w. irreps leads to a significant improvement in the predicted $R_{4/2}$ ratios for the studied isotopes. Additionally, there is a significantly improved agreement for the higher-spin levels of the gsb, which can be attributed to the contribution to the corresponding moment of inertia via the second order Casimir operator of SU(3)~\cite{elliott58a,elliott58b,elliott63,elliott68}:
\begin{equation}
	\label{eq:c2su3}
	\hat{C}_2[\mathrm{SU(3)}] = 2\hat{Q}\cdot\hat{Q}+\frac{3}{4}\hat{L}^2,
\end{equation}
entering implicitly in the calculations via the $(\lambda,\mu)$ irreps used to derive $\gamma_s$ through Eq.~(\ref{eq:gamma_s}). Furthermore, the level spacings for the $\beta$ bands are much improved, while there is also a vast improvement in the predicted behavior for the staggering inside the $\gamma$ bands.

For the final step in our calculations, we perform a rescaling of the predicted energy levels for the $\beta$- and $\gamma$-bands to the respective experimental band-heads, where available (see Table~\ref{tab:ibm-params3}). This rescaling, which will be discussed in the next section, does not affect the energy ratios or the staggering within individual bands. However, it is necessary in order to obtain quantitative results for the energy levels. The final band levels are presented for the even-even \isotope[166-180]{Hf} and \isotope[170,176-186]{W} in Figs.~\ref{fig:Hf-W-gs-beta}b, \ref{fig:Hf-W-gs-beta}d, \ref{fig:Hf-W-gs-beta}f, \ref{fig:Hf-W-gs-beta}h and \ref{fig:Hf-W-gamma}e--\ref{fig:Hf-W-gamma}f (see also Fig.~\ref{fig:180Hf}c for an indicative level scheme).


Level schemes, similar to those presented for \isotope[180]{Hf} in Fig.~\ref{fig:180Hf}, are provided for all of the studied Hf and W isotopes in the Supplemental Material of this paper.

\section{\label{sec:disc}Results and Discussion}
\subsection{Parameter systematics}
\label{ssec:systematics}

\begin{figure*}[!htbp]
\centering
\includegraphics[width=0.90\textwidth]{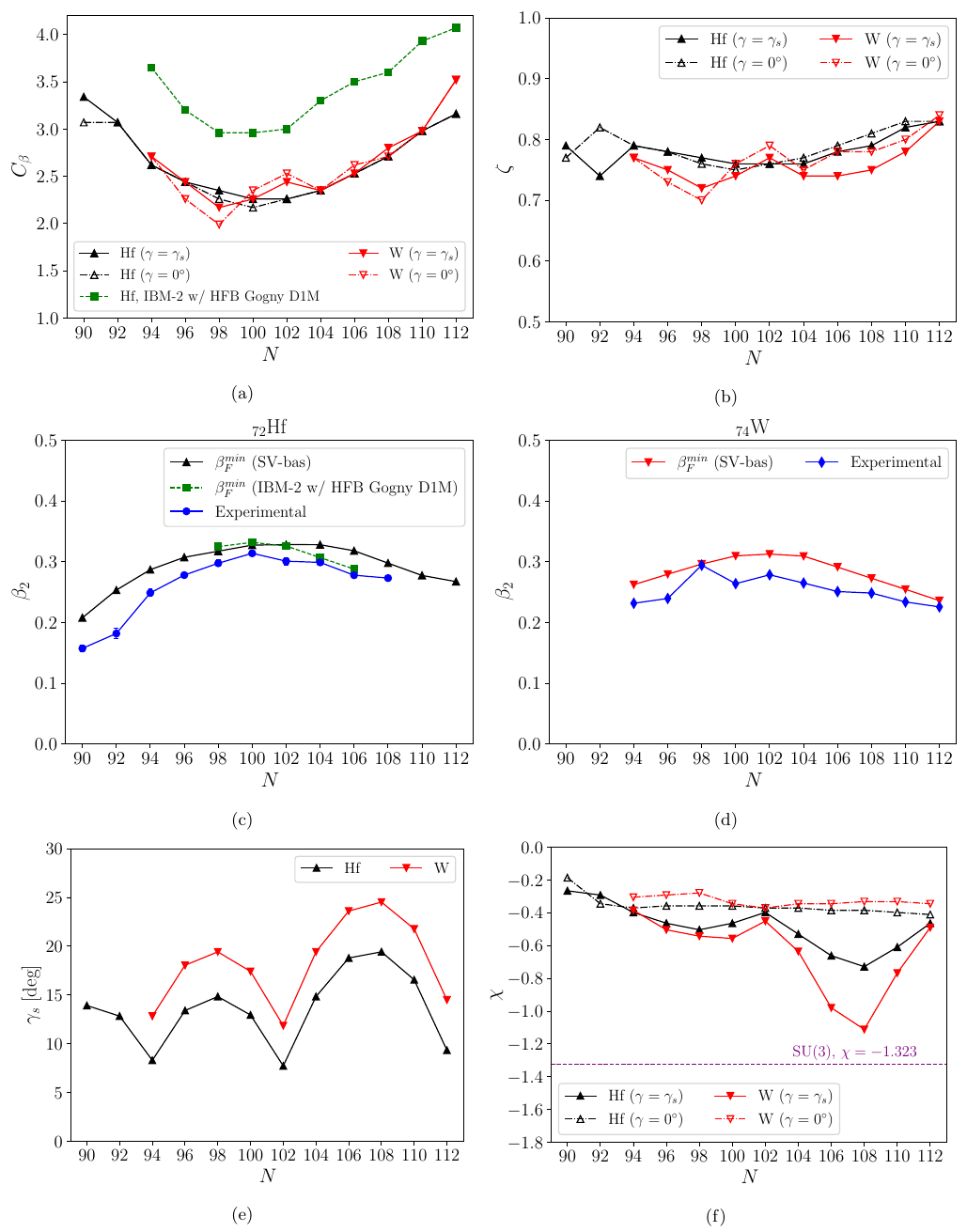}
\caption{\label{fig:systematics} Systematics for the IBM parameters as functions of the neutron number, N. Experimental quadrupole deformation parameters, $\beta_2$, taken from~\cite{pritychenko16}.}
\end{figure*}

The parameter systematics tabulated in Tables~\ref{tab:ibm-params} and \ref{tab:ibm-params2} are plotted in Fig.~\ref{fig:systematics} for the Hf and W isotopes studied in this work. The same notation is used for all of the panels of Fig.~\ref{fig:systematics}, i.e., dot-dashed lines with empty symbols for the case of $\gamma=0^{\circ}$, solid lines with filled symbols for the case of $\gamma=\gamma_s$, black color for Hf and red for W.

One can immediately notice the relation between the proportionality coefficient, $C_{\beta}$ (Fig.~\ref{fig:systematics}a) and the $\zeta$ parameter of the IBM-1 Hamiltonian of Eq.~(\ref{eq:Hecqf}) (Fig.~\ref{fig:systematics}b), related to axial quadrupole deformation. Both parameters follow the same trend, exhibiting minima around the middle of the major shell where the PEC shows the largest deformation. This behavior is in agreement with the results of earlier works of Nomura {\em et al.} on Xe and Ba isotopes~\cite{nomura10}, situated in the $N=50-82$ major neutron shell, as well as \isotope[166-194]{Hf}~\cite{nomura11a,rudigier15}, for the $N=82-126$ shell. In the former, an IBM-2 Hamiltonian was mapped to a microscopic PES, derived from Hartree-Fock+BCS calculations employing SLy4~\cite{chabanat98} and SkM*~\cite{bartel82} effective interactions, while for the latter, a microscopic PES obtained from Hartree-Fock-Bogolyubov calculations with Gogny D1S~\cite{berger84} and D1M~\cite{goriely09} EDFs was used for mapping to an IBM-2 Hamiltonian. The $C_{\beta}$ values of~\cite{nomura11a,rudigier15} are also plotted in Fig.~\ref{fig:systematics}a, for comparison (green dashed curves with solid squares).  

Despite the use of different IBM models (IBM-1 vs IBM-2), pairing schemes (HF+BCS vs HFB), and effective interactions (Skyrme type SV-bas vs Gogny D1M/D1S), there is a very good qualitative agreement between the results presented in this work, and the ones of~\cite{nomura11a,rudigier15}, for the Hf isotopes. The quantitative differences for the proportionality coefficients, $C_{\beta}$, can be explained by the use of different effective interactions, leading to larger $\beta_F$ values for the microscopic PECs/PESs (see e.g. Table~VI of~\cite{rudigier15} for a comparison between Gogny D1M and D1S energy density functionals). Furthermore, depending on the employed EDF, the maximum deformation is observed at either $N=100$ or $102$, 4 or 2 neutrons away from the $N=104$ midshell, respectively (see also discussion and Fig.~3a of~\cite{vasileiou23} for a comparison of quadrupole deformation parameters between various nuclear models). The calculated $\beta_F^{min}$ values are plotted as functions of $N$, together with the experimental data~\cite{pritychenko16}, in Figs.~\ref{fig:systematics}c and \ref{fig:systematics}d.

It can be seen from Figs.~\ref{fig:systematics}a and \ref{fig:systematics}b that the inclusion of an intrinsic deformation, $\gamma_s$, resulting from the use of proxy-SU(3) h.w. irreps, has only a minor quantitative effect on $C_{\beta}$, and only in some of the considered Hf and W isotopes.

However, this is not the case for $\gamma_s$ and $\chi$, the IBM-1 Hamiltonian parameter associated with the degree of triaxiality in the nucleus. As it can be seen from Figs.~\ref{fig:systematics}e and \ref{fig:systematics}f, the inclusion of $\gamma_s$ leads to a drastically different picture for $\chi$, compared to the $\gamma=0^{\circ}$ case. Again, the trend of $\gamma_s$ translates well into $\chi$, with larger values of $\gamma_s$ leading to larger deformations. These values are more realistic, since they get closer to the SU(3) limit of $\chi=-\sqrt{7}/2=-1.323$, as is typically the case for well-deformed nuclei. Furthermore, the local minima of $\gamma_s$ for $N=102,112$, which correspond to proxy-SU(3) neutron irreps with $\mu=0$~\cite{bonatsos17b}, are translated into local maxima for $\chi$, in qualitative agreement.

\begin{table}[!htbp]
	\caption{\label{tab:ibm-params3}%
		Values for the quadrupole deformation parameter at the energy minimum of the HF PECs, denoted as $\beta_F^{min}$, for each of the studied Hf and W isotopes. The mass coefficients for the $\beta$- ($\alpha_{\beta}$) and $\gamma$- ($\alpha_{\gamma}$) bands are also tabulated.
	}
	\begin{ruledtabular}
		\begin{tabular}{cddd}
			Isotope 
			& \multicolumn{1}{c}{$\beta_F^{min}$} 
			& \multicolumn{1}{c}{$\alpha_{\beta}$} 
			& \multicolumn{1}{c}{$\alpha_{\gamma}$} \\
			\colrule
			\isotope[162]{Hf} & 0.208 & \multicolumn{1}{c}{---} & \multicolumn{1}{c}{---} \\
			\isotope[164]{Hf} & 0.253 & \multicolumn{1}{c}{---} & \multicolumn{1}{c}{---} \\
			\isotope[166]{Hf} & 0.287 & 0.803 & 0.821 \\
			\isotope[168]{Hf} & 0.307 & 0.750 & 0.893 \\
			\isotope[170]{Hf} & 0.317 & 0.756 & 1.031 \\
			\isotope[172]{Hf} & 0.327 & 0.799 & 1.261 \\
			\isotope[174]{Hf} & 0.328 & 0.813 & 1.616 \\
			\isotope[176]{Hf} & 0.328 & 0.844 & 1.231 \\
			\isotope[178]{Hf} & 0.318 & 0.716 & 0.818 \\
			\isotope[180]{Hf} & 0.298 & 0.650 & 0.795 \\
			\isotope[182]{Hf} & 0.278 & \multicolumn{1}{c}{---} & \multicolumn{1}{c}{---} \\
			\isotope[184]{Hf} & 0.267 & \multicolumn{1}{c}{---} & \multicolumn{1}{c}{---} \\
			\hline
			\isotope[168]{W}  & 0.262 & \multicolumn{1}{c}{---} & \multicolumn{1}{c}{---} \\
			\isotope[170]{W}  & 0.280 & 0.578\footnote{Calculated by replacing $0^+_{\beta}$ with $2^+_{\beta}$ in Eq.~(\ref{eq:mass_coeffs2}), due to the absence of an experimental value for $0^+_{\beta}$ in~\cite{nndc}.} & 0.883  \\
			\isotope[172]{W}  & 0.296 & \multicolumn{1}{c}{---} & \multicolumn{1}{c}{---} \\
			\isotope[174]{W}  & 0.310 & \multicolumn{1}{c}{---} & \multicolumn{1}{c}{---} \\
			\isotope[176]{W}  & 0.313 & 0.659 & 1.058  \\
			\isotope[178]{W}  & 0.309 & 0.652 & 0.819  \\
			\isotope[180]{W}  & 0.291 & 0.740 & 0.553  \\
			\isotope[182]{W}  & 0.273 & 0.605 & 0.576  \\
			\isotope[184]{W}  & 0.255 & 0.609 & 0.585  \\
			\isotope[186]{W}  & 0.236 & 0.619 & 0.673  \\
		\end{tabular}
	\end{ruledtabular}
\end{table}

\subsection{Energy spectra}
\label{ssec:spectra}

We now proceed to a more thorough examination of the calculated level schemes for the Hf and W isotopes, building upon the first observations made at Section~\ref{sec:models}. 

Regarding the gsb levels of the studied isotopes, the calculations performed with $\gamma=0^{\circ}$ can sufficiently reproduce the energies of the first few low-lying states (up to $8^+_1$), with divergences from experimental data making their appearance for higher spins. These divergences have been associated with the moment of inertia, and for well-deformed rotational nuclei, they have been remedied through the inclusion of an additional $(\hat{L}\cdot\hat{L})$ term to the employed IBM Hamiltonian, which provided the experimentally observed $L(L+1)$ level energy dependence~\cite{nomura11}. In our calculations, this contribution to the moment of inertia is taken into account implicitly, through the use of h.w. proxy-SU(3) irreps, thus overcoming the need to add extra terms to the IBM-1 ECQF Hamiltonian of Eq.~(\ref{eq:Hecqf}).

On the subject of the first $K=0^+$ excited bands, which are associated with the $\beta$-- bands in this mass region, these are formed by the calculated $0^+_2$, $2^+_3$, $4^+_3$, $6^+_3$ and $8^+_3$ states, both in the $\gamma=0^{\circ}$ and in the $\gamma=\gamma_s$ cases. However, their structure differs significantly between the two cases, with a behavior assimilating a $\Delta L=2$ staggering presenting itself for $\gamma=0^{\circ}$. Such a behavior has been recently observed also in \isotope[154]{Gd}, in the framework of the {\em sdg}-IBM-1 (see Fig.~3 of~\cite{lotina24a}), and appears to arise from the assumption of ``pure'' axial deformations (i.e. axial symmetry at a mean-field level, without any intrinsic triaxial deformation). The inclusion of the intrinsic deformation, $\gamma_s$, stemming from the proxy-SU(3) h.w. irreps, is enough to remedy this picture, leading to a good qualitative agreement with the experimental data, while preserving the axial symmetry of the IBM potential energy surface, which is dictated by the $\cos{3\gamma}$ term in Eq.~(\ref{eq:ibm-pes}). 

The effect of the proxy-SU(3) irreps is even more impactful on the $\gamma$ bands, which are formed by the $2^+_2$, $3^+_1$, $4^+_2$, $5^+_1$, ..., $9^+_1$ excited states. These bands exhibit a pronounced odd--even staggering, when only axial deformations are assumed (Figs.~\ref{fig:Hf-W-gamma}a, \ref{fig:Hf-W-gamma}b, \ref{fig:180Hf}a). Such a behavior has been observed in the past, in calculations using the mapping method for \isotope[190]{Os}, in the framework of {\em sd}-IBM-2~\cite{nomura12}. The inclusion of a {\em g} boson ($L=4$) does not seem to resolve this staggering, which also appeared in \isotope[154]{Gd}, in the framework of the {\em sdg}-IBM-1~\cite{lotina24a}. The path followed in~\cite{nomura12} to remedy this behavior for the case of \isotope[190]{Os} involved the inclusion of a 3-body term in the IBM-2 Hamiltonian, of the form:
\begin{equation}
	\label{eq:h3b-ibm2}
H_{3B} = \sum_{\rho\neq\rho^{\prime}}\theta^{\rho}[d_{\rho}^{\dagger}d_{\rho}^{\dagger}d_{\rho^{\prime}}^{\dagger}]^{(3)}\cdot[\tilde{d}_{\rho^{\prime}}\tilde{d}_{\rho}\tilde{d}_{\rho}]^{(3)},~~\rho = \pi, \nu.
\end{equation}
The above term is associated with a $\cos^23\gamma$ term, capable of producing triaxial minima in the IBM potential energy surface.

In this work, the introduction of the intrinsic triaxial deformation, $\gamma_s$, resulting from the proxy-SU(3) h.w. irreps is sufficient to achieve a qualitative agreement with the experimental staggering for the $\gamma$-bands (Fig.~\ref{fig:180Hf}b, see also Supplementary Material), thus avoiding the need for additional terms in the IBM Hamiltonian, and the associated computational complexity arising from the addition of extra parameters.

A final step towards obtaining quantitative results for the energy levels of the $\beta$- and $\gamma$-bands in the studied isotopes is undertaken with the inclusion of different mass coefficients for these bands, determined via a rescaling with respect to the experimental bandheads, i.e.:
\begin{equation}
	\label{eq:mass_coeffs1}
	E_{\beta(\gamma)}(J) \rightarrow \tilde{E}_{\beta(\gamma)}(J)=\alpha_{\beta(\gamma)}E_{\beta}(J),
\end{equation}
where:
\begin{equation}
	\label{eq:mass_coeffs2}
	\alpha_{\beta}=E^{exp.}(0^+_{\beta})/E(0^+_{\beta}),~~
	\alpha_{\gamma}=E^{exp.}(2^+_{\gamma})/E(2^+_{\gamma}).
\end{equation}
The need for the introduction of different mass coefficients in phenomenological models, associated with the varying moments of inertia for different modes of excitation (ground band rotational motion, $\beta$ and $\gamma$ vibrations), has been outlined and discussed in the works of Jolos {\em et al.} (see e.g.~\cite{jolos05a,jolos05b,jolos06,jolos07,jolos21}, and references therein).

However, it should be stressed that the inclusion of different mass coefficients for the $\beta$ and $\gamma$ bands does not alter the qualitative picture of the calculations, but rather affects the calculated energy values on a quantitative level. The resulting band levels are presented for \isotope[166-180]{Hf} and \isotope[170,176-186]{W} in Figs.~\ref{fig:Hf-W-gs-beta}b,\ref{fig:Hf-W-gs-beta}d, \ref{fig:Hf-W-gs-beta}f, \ref{fig:Hf-W-gs-beta}h  and Figs.~\ref{fig:Hf-W-gamma}c--\ref{fig:Hf-W-gamma}f.

Overall, the inclusion of an intrinsic triaxial deformation, resulting from the use of proxy-SU(3) irreps, leads to a significant improvement on the qualitative description of the gs-, $\beta$- and $\gamma$-bands, further improved on the quantitative level with the introduction of different mass coefficents for these bands. The unusually low $\gamma_s$ predictions of proxy-SU(3), for $N=94,102,112$ (neutron irreps with $\mu=0$) are also reflected in the IBM-1 calculations. These small $\gamma_s$ values, while sufficient for the qualitative description of ground state and $\beta$-bands, are not able to completely remedy the $\gamma$-band staggering, leading to a qualitative picture which is closer to the $\gamma=0^{\circ}$ case.

\subsection{E2 transitions}

We move on to calculate the $B(E2)$ values for some low-lying gsb states. In the ECQF formalism, the relevant transition operator is defined as~\cite{casperson12}:
\begin{equation}
	\hat{T}(E2) = e_B\hat{Q}^{\chi}\cdot\hat{Q}^{\chi},
\end{equation} 
where $\hat{Q}^{\chi}$ is the quadrupole boson creation operator of Eq.~(\ref{eq:Hecqf}), and $e_B$ is the effective charge. One option for the determination of the effective charge is the use of a fixed value across the isotopic chain studied, chosen such that it reproduces the experimental $B(E2;2^+_1\rightarrow 0^+_1)$ for a specific isotope. However, as shown in previous IBM fitting calculations in the W, Os~\cite{rudigier10} and Hf~\cite{rudigier15} isotopic chains, the assumption of a constant value for $e_B$ leads to a maximization of the $B(E2;J\rightarrow J-2)$ values at the midshell $N=104$ (\isotope[176]{Hf}, \isotope[178]{W}), in contrast with the experimentally observed maxima at $N=98$ for W and $N=100$ for Hf~\cite{pritychenko16}.

An alternative for the derivation of the effective charges is the assumption of a mass dependence, with two possible options; the first one is the choice of $e_B$ individually for each nucleus, fitted to the corresponding experimental $B(E2;2^+_1\rightarrow 0^+_1)$ values. This is a commonly used approach amongst IBM calculations (see e.g.~\cite{zyriliou22,mccutchan04,mccutchan05a,mccutchan05b,mccutchan06,harter22})

The second one is the choice of $e_B$ individually for each nucleus, such that the intrinsic quadrupole deformation parameter, $\beta_t(2^+_1\rightarrow 0^+_1)$ of the EDF-IBM model equals the $\beta_F^{min}$ minimum of the mean-field PEC~\cite{rudigier15}. More specifically, in the framework of the nuclear collective model, the intrinsic quadrupole deformation parameter, $\beta_t(J\rightarrow J')$ is related to the transition quadrupole moment, $Q_t(J\rightarrow J')$, and the associated $B(E2;J\rightarrow J')$ matrix element, through the relations~\cite{bohr_mot98b}:
\begin{equation}
\label{eq:Q_t}
 Q_t(J\rightarrow J')=\sqrt{\frac{16\pi}{5}\frac{B(E2;J\rightarrow J')}{(J200|J^{\prime}0)^2}}
\end{equation}
and
\begin{equation}
	\label{eq:b_t}
	\beta_t(J\rightarrow J')=\frac{\sqrt{5\pi}}{3ZR^2}Q_t(J\rightarrow J'),
\end{equation}
where $J$ and $J^{\prime}$ the spins of the initial and final states, $(J200|J^{\prime}0)$ the relevant Clebsch--Gordan coefficient, $Z$ the atomic number and $R$ the nuclear radius. By equating $\beta_t(2^+_1\rightarrow 0^+_1)=\beta_F^{min}$, one obtains the effective charge, $e_B$, for each isotope under investigation.

Since the HF+BCS calculations with the SV-bas EDF give a satisfactory reproduction of the systematics of the $\beta_2$ values, and, by extension, the $B(E2;2^+_1\rightarrow 0^+_1)$ for the Hf and W isotopic chains, the latter approach is adopted for the derivation of $e_B$. This choice comes with the advantage of avoiding the addition of extra fitting inputs, albeit at the cost of some small quantitative deviations from the experimental $B(E2)$ values.

\begin{figure*}[!htbp]
\centering
\includegraphics[width=0.90\textwidth]{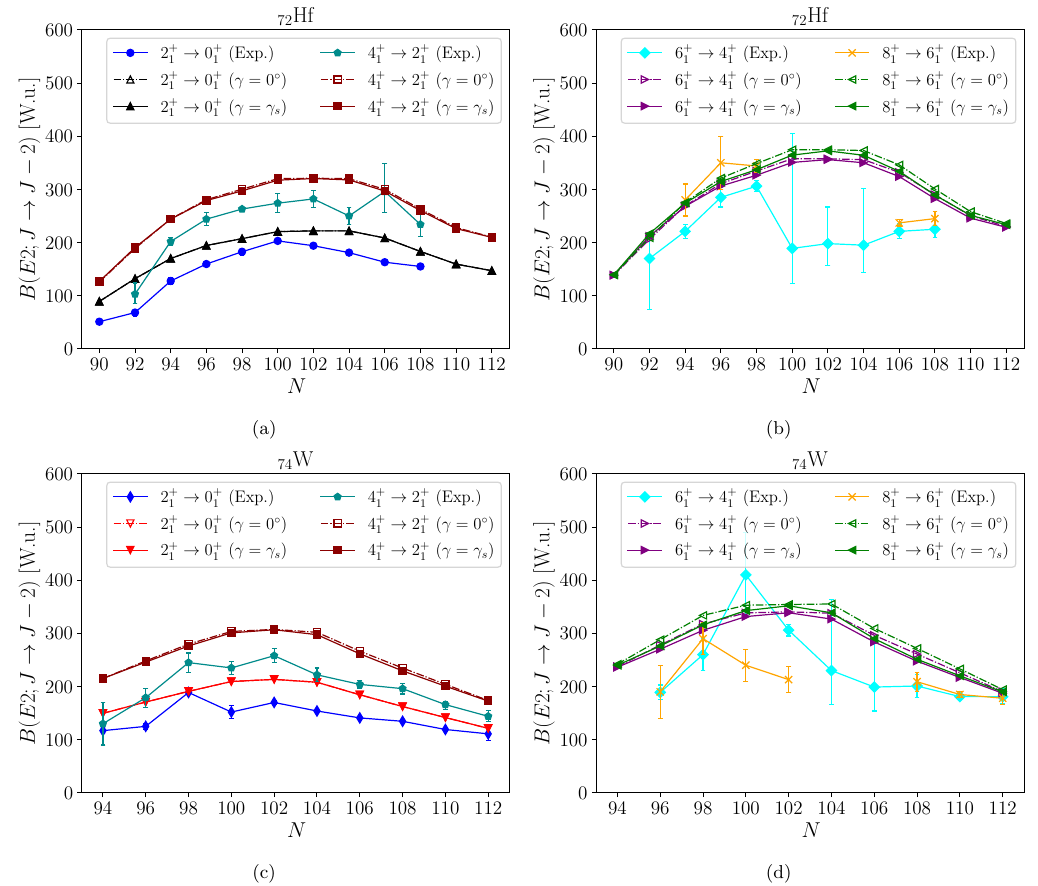}
\caption{\label{fig:be2s} Experimental vs. calculated $B(E2;J\rightarrow J-2)$ values for the Hf and W isotopes studied in this work. Calculations for $\gamma=0^{\circ}$ are shown as dot--dashed lines with open symbols, while solid lines and symbols, of the same color, correspond to calculations for the case of $\gamma=\gamma_s$. Literature data were taken from~\cite{nndc,pritychenko16,wiederhold19} for Hf, and~\cite{nndc,pritychenko16,harter22} for W.}
\end{figure*}

For each isotopic chain, two separate sets of $e_B$ are calculated, one for the case of $\gamma=0^{\circ}$ and one for $\gamma=\gamma_s$. The resulting $B(E2)$ values are plotted, together with the experimental data existing in the literature~\cite{nndc,pritychenko16,wiederhold19,harter22}, in Fig.~\ref{fig:be2s}.

A reasonable agreement with the experimental values, on both qualitative and quantitative levels is observed for the $B(E2;2^+_1\rightarrow 0^+_1)$ and $B(E2;4^+_1\rightarrow 2^+_1)$ transition strengths. This seems to hold true also for $B(E2;6^+_1\rightarrow 4^+_1)$ and $B(E2;8^+_1\rightarrow 6^+_1)$, however, the lack of experimental data, and the large uncertainties accompanying the existing ones, prevent a clear comparison between trends for these quantities.

The maximal $B(E2)$ values are obtained for $N=102$, instead of $N=100$ for Hf and $N=98$ for W, which can be attributed to the particular choice of EDF (see also earlier discussion in Subsection~\ref{ssec:systematics}). This displacement of the maxima with respect to $N=104$ indicates a saturation of collectivity on the neutron-deficient side, as one moves towards the neutron mid-shell in the examined isotopic chains. The pre-mid-shell saturation of the $B(E2)$ transition strengths, which gets more pronounced with increasing proton numbers, was partially attributed in~\cite{zamfir95} to the influence of a hexadecapole deformation, $\beta_4$, entering the expression for the transitional quadrupole moment, $Q_t$. In the {\em sd}-IBM framework, such an effect, caused by the renormalization of {\em g}-bosons, could be incorporated to an extent via an expansion of the expression for the quadrupole operator of Eq.~(\ref{eq:Hecqf}), made to include higher-order (two-body) terms~\cite{rudigier15,zamfir95}. A more extensive study on the influence of hexadecapole deformations on $B(E2;2^+_1\rightarrow 0^+_1)$ was undertaken for the even-even nuclei of the Er-Yb-Hf-W mass region in~\cite{wiederhold19}, however, it could not offer a complete explanation of the early $B(E2)$ maxima. 

The $B(E2)$ systematics calculated in this work are consistent with earlier findings of Ref.~\cite{rudigier15,wiederhold19} on \isotope[166-180]{Hf} and Ref.~\cite{harter22,wiederhold19} on \isotope[172-182]{W}, while at the same time expanding the area of study towards the more neutron deficient side of these isotopic chains.  

It should be noted that the inclusion of an intrinsic triaxial deformation, $\gamma_s$, generated through the proxy-SU(3) irreps, has minimal influence on the $B(E2)$ values for the gsb transitions examined. More specifically, it leads to a lowering of the $B(E2)$ predictions, which gets slightly more pronounced for higher-spin states (Fig.~\ref{fig:be2s}). Overall, the $B(E2)$ strengths appear to exhibit similar sensitivity to the inclusion of $\gamma_s$ compared to $C_{\beta}$, which is not unexpected, given their connection to the quadrupole deformation parameters (see also Eqs.~(\ref{eq:Q_t}), (\ref{eq:b_t})).

\section{\label{sec:conclusions}Summary and Outlook}
In view of recent findings, which indicate the presence of some degree of triaxiality all over the nuclear chart~\cite{tsunoda21,otsuka23,rouoof24}, an effort is made to include triaxiality in the framework of the standard IBM-1, in which only one- and two-body terms are taken into account, and no distinction between protons and neutrons is made. The aim of this effort is to provide an easy tool (the existing IBM code~\cite{casperson12}) for calculating spectra and B(E2) transition rates for many medium-mass and heavy nuclei. 

Along the path taken, potential energy curves are calculated using a self-consistent mean-field approach employing a Skyrme energy density functional, namely the axial Hartree-Fock+BCS code SkyAx~\cite{skyax}. The PECs derived by IBM-1 are then fitted to the microscopic PECs, in order to have the IBM-1 parameters determined. 
Once this is done, spectra and B(E2) transition rates for the nuclei under study are readily obtained through the IBAR code~\cite{casperson12}. 

Two sets of calculations have been performed, one assuming axial symmetry, as it is the case in the original IBM-1 with only one- and two-body terms included in the Hamiltonian, and an additional one, in which an intrinsic triaxial deformation has been added to the potential energy curve corresponding to the classical limit of IBM-1. The intrinsic triaxial deformation has not been treated as a free parameter; on the contrary, it has been obtained from the proxy-SU(3) approximation to the shell model~\cite{bonatsos17a,bonatsos17b,bonatsos23} in a parameter-free way, taking into account the Pauli principle and the short-range nature of the nucleon-nucleon interaction.

Significantly improved results have been obtained in the latter case, providing an \textit{a posteriori} justification for the inclusion of a microscopically derived intrinsic triaxial deformation in the potential energy curve corresponding to the classical limit of IBM-1. 

In conclusion, the preponderance of triaxial shapes over most of the nuclear chart is predicted in a parameter-free way by the proxy-SU(3) approximation to the shell model, while detailed predictions for spectra and $B(E2)$ transition rates for specific nuclei can be readily calculated through the IBM-1 code IBAR~\cite{casperson12} after including an intrinsic triaxial deformation to the potential energy curve of the IBM-1 in its classical limit, and determining the IBM-1 parameters through fitting of the resulting potential energy curve to the one derived through the axial Hartree-Fock+BCS code SkyAx~\cite{skyax}. 

In the present work, the Hf and W series of isotopes have been used as the test-ground of the new approach. A region of obvious interest for further calculations consists of the Os and Pt series of isotopes, for which sufficient experimental data exist for chains of isotopes ranging from moderate to strong quadrupole deformation, extending beyond $N=116$, where a prolate to oblate shape/phase transition is expected to take place~\cite{jolie03,bonatsos04,zhang12,bonatsos17b,wang23}.

It would be interesting to examine if and how much the results of the present approach would be influenced by fitting the IBM-1 PEC with intrinsic triaxial deformation to a microscopic PEC derived by a self-consistent mean-field approach including triaxiality~\cite{maruhn14,schuetrumpf18}. Recent calculations in the Ce isotopes~\cite{alexa22} have indicated that triaxiality appears only for $N<82$, but the validity of this result has to be checked at higher $Z$.

Regarding the Hf and W isotopes studied in the present work, triaxial mean-field calculations performed in the framework of the HFB formalism with the Gogny D1S EDF~\cite{delaroche10} result in axially-symmetric PESs, with minima along the $\gamma=0^{\circ}$ line of the $\beta$-$\gamma$ plane. These calculations are available in the AMEDEE database~\cite{amedee1} (see column 5 in the Table of~\cite{amedee2}). The $\gamma$ deformation at the HFB energy minimum is $0^{\circ}$ for all of the examined Hf and W isotopes (with the exception of \isotope[180]{W}, for which $\gamma=1^{\circ}\simeq 0^{\circ}$). This is consistent with the results of similar calculations found in Refs.~\cite{nomura11a,rudigier15}, which were carried out on the $\beta-\gamma$ plane with the HFB + Gogny D1S and D1M EDFs, and the results of the present work.

Still, many of the constrained mean-field calculations performed with triaxial quadrupole deformations give energy surfaces that have a triaxial minimum or are considerably
$\gamma$-soft. This might play a role as one moves on to examine more axially asymmetric nuclei (e.g. Os and Pt), making for an interesting future study subject.


\nocite{*}

\bibliography{Hf}

\begin{thebibliography}{121}%
\makeatletter
\providecommand \@ifxundefined [1]{%
 \@ifx{#1\undefined}
}%
\providecommand \@ifnum [1]{%
 \ifnum #1\expandafter \@firstoftwo
 \else \expandafter \@secondoftwo
 \fi
}%
\providecommand \@ifx [1]{%
 \ifx #1\expandafter \@firstoftwo
 \else \expandafter \@secondoftwo
 \fi
}%
\providecommand \natexlab [1]{#1}%
\providecommand \enquote  [1]{``#1''}%
\providecommand \bibnamefont  [1]{#1}%
\providecommand \bibfnamefont [1]{#1}%
\providecommand \citenamefont [1]{#1}%
\providecommand \href@noop [0]{\@secondoftwo}%
\providecommand \href [0]{\begingroup \@sanitize@url \@href}%
\providecommand \@href[1]{\@@startlink{#1}\@@href}%
\providecommand \@@href[1]{\endgroup#1\@@endlink}%
\providecommand \@sanitize@url [0]{\catcode `\\12\catcode `\$12\catcode
  `\&12\catcode `\#12\catcode `\^12\catcode `\_12\catcode `\%12\relax}%
\providecommand \@@startlink[1]{}%
\providecommand \@@endlink[0]{}%
\providecommand \url  [0]{\begingroup\@sanitize@url \@url }%
\providecommand \@url [1]{\endgroup\@href {#1}{\urlprefix }}%
\providecommand \urlprefix  [0]{URL }%
\providecommand \Eprint [0]{\href }%
\providecommand \doibase [0]{https://doi.org/}%
\providecommand \selectlanguage [0]{\@gobble}%
\providecommand \bibinfo  [0]{\@secondoftwo}%
\providecommand \bibfield  [0]{\@secondoftwo}%
\providecommand \translation [1]{[#1]}%
\providecommand \BibitemOpen [0]{}%
\providecommand \bibitemStop [0]{}%
\providecommand \bibitemNoStop [0]{.\EOS\space}%
\providecommand \EOS [0]{\spacefactor3000\relax}%
\providecommand \BibitemShut  [1]{\csname bibitem#1\endcsname}%
\let\auto@bib@innerbib\@empty
\bibitem [{\citenamefont {Mayer}\ and\ \citenamefont {Jensen}(1955)}]{mayer55}%
  \BibitemOpen
  \bibfield  {author} {\bibinfo {author} {\bibfnamefont {M.~G.}\ \bibnamefont
  {Mayer}}\ and\ \bibinfo {author} {\bibfnamefont {J.~H.~D.}\ \bibnamefont
  {Jensen}},\ }\href@noop {} {\emph {\bibinfo {title} {{Elementary Theory of
  Nuclear Shell Structure}}}}\ (\bibinfo  {publisher} {Wiley: New York},\
  \bibinfo {year} {1955})\BibitemShut {NoStop}%
\bibitem [{\citenamefont {Heyde}(1990)}]{heyde90}%
  \BibitemOpen
  \bibfield  {author} {\bibinfo {author} {\bibfnamefont {K.~L.~G.}\
  \bibnamefont {Heyde}},\ }\href@noop {} {\emph {\bibinfo {title} {{The Nuclear
  Shell Model}}}}\ (\bibinfo  {publisher} {Springer: Berlin},\ \bibinfo {year}
  {1990})\BibitemShut {NoStop}%
\bibitem [{\citenamefont {Talmi}(1993)}]{talmi93}%
  \BibitemOpen
  \bibfield  {author} {\bibinfo {author} {\bibfnamefont {I.}~\bibnamefont
  {Talmi}},\ }\href@noop {} {\emph {\bibinfo {title} {{Simple Models of Complex
  Nuclei: The Shell Model and the Interacting Boson Model}}}}\ (\bibinfo
  {publisher} {Harwood: Chur},\ \bibinfo {year} {1993})\BibitemShut {NoStop}%
\bibitem [{\citenamefont {Wybourne}(1974)}]{wybourne74}%
  \BibitemOpen
  \bibfield  {author} {\bibinfo {author} {\bibfnamefont {B.~G.}\ \bibnamefont
  {Wybourne}},\ }\href@noop {} {\emph {\bibinfo {title} {{Classical Groups for
  Physicists}}}}\ (\bibinfo  {publisher} {Wiley: New York, NY, USA},\ \bibinfo
  {year} {1974})\BibitemShut {NoStop}%
\bibitem [{\citenamefont {Moshinsky}\ and\ \citenamefont
  {Smirnov}(1996)}]{moshinsky96}%
  \BibitemOpen
  \bibfield  {author} {\bibinfo {author} {\bibfnamefont {M.}~\bibnamefont
  {Moshinsky}}\ and\ \bibinfo {author} {\bibfnamefont {Y.~F.}\ \bibnamefont
  {Smirnov}},\ }\href@noop {} {\emph {\bibinfo {title} {{The Harmonic
  Oscillator in Modern Physics}}}}\ (\bibinfo  {publisher} {Harwood:
  Amsterdam},\ \bibinfo {year} {1996})\BibitemShut {NoStop}%
\bibitem [{\citenamefont {Iachello}(2006)}]{iachello06}%
  \BibitemOpen
  \bibfield  {author} {\bibinfo {author} {\bibfnamefont {F.}~\bibnamefont
  {Iachello}},\ }\href@noop {} {\emph {\bibinfo {title} {{Lie Algebras and
  Applications}}}}\ (\bibinfo  {publisher} {Springer: Berlin},\ \bibinfo {year}
  {2006})\BibitemShut {NoStop}%
\bibitem [{\citenamefont {Navr\'atil}\ \emph {et~al.}(2000)\citenamefont
  {Navr\'atil}, \citenamefont {Vary},\ and\ \citenamefont
  {Barrett}}]{navratil00}%
  \BibitemOpen
  \bibfield  {author} {\bibinfo {author} {\bibfnamefont {P.}~\bibnamefont
  {Navr\'atil}}, \bibinfo {author} {\bibfnamefont {J.~P.}\ \bibnamefont
  {Vary}},\ and\ \bibinfo {author} {\bibfnamefont {B.~R.}\ \bibnamefont
  {Barrett}},\ }\bibfield  {title} {\bibinfo {title} {{Large-basis ab initio
  no-core shell model and its application to ${}^{12}\mathbf{C}$}},\ }\href
  {https://doi.org/10.1103/PhysRevC.62.054311} {\bibfield  {journal} {\bibinfo
  {journal} {Phys. Rev. C}\ }\textbf {\bibinfo {volume} {62}},\ \bibinfo
  {pages} {054311} (\bibinfo {year} {2000})}\BibitemShut {NoStop}%
\bibitem [{\citenamefont {Dytrych}\ \emph {et~al.}(2008)\citenamefont
  {Dytrych}, \citenamefont {Sviratcheva}, \citenamefont {Draayer},
  \citenamefont {Bahri},\ and\ \citenamefont {Vary}}]{dytrych08}%
  \BibitemOpen
  \bibfield  {author} {\bibinfo {author} {\bibfnamefont {T.}~\bibnamefont
  {Dytrych}}, \bibinfo {author} {\bibfnamefont {K.~D.}\ \bibnamefont
  {Sviratcheva}}, \bibinfo {author} {\bibfnamefont {J.~P.}\ \bibnamefont
  {Draayer}}, \bibinfo {author} {\bibfnamefont {C.}~\bibnamefont {Bahri}},\
  and\ \bibinfo {author} {\bibfnamefont {J.~P.}\ \bibnamefont {Vary}},\
  }\bibfield  {title} {\bibinfo {title} {{Ab initio symplectic no-core shell
  model}},\ }\href {https://doi.org/10.1088/0954-3899/35/12/123101} {\bibfield
  {journal} {\bibinfo  {journal} {Journal of Physics G: Nuclear and Particle
  Physics}\ }\textbf {\bibinfo {volume} {35}},\ \bibinfo {pages} {123101}
  (\bibinfo {year} {2008})}\BibitemShut {NoStop}%
\bibitem [{\citenamefont {Launey}\ \emph {et~al.}(2020)\citenamefont {Launey},
  \citenamefont {Dytrych}, \citenamefont {Sargsyan}, \citenamefont {Baker},\
  and\ \citenamefont {Draayer}}]{launey20}%
  \BibitemOpen
  \bibfield  {author} {\bibinfo {author} {\bibfnamefont {K.~D.}\ \bibnamefont
  {Launey}}, \bibinfo {author} {\bibfnamefont {T.}~\bibnamefont {Dytrych}},
  \bibinfo {author} {\bibfnamefont {G.~H.}\ \bibnamefont {Sargsyan}}, \bibinfo
  {author} {\bibfnamefont {R.~B.}\ \bibnamefont {Baker}},\ and\ \bibinfo
  {author} {\bibfnamefont {J.~P.}\ \bibnamefont {Draayer}},\ }\bibfield
  {title} {\bibinfo {title} {{Emergent symplectic symmetry in atomic nuclei: Ab
  initio symmetry-adapted no-core shell model}},\ }\href
  {https://doi.org/10.1140/epjst/e2020-000178-3} {\bibfield  {journal}
  {\bibinfo  {journal} {Eur. Phys. J. Special Topics}\ }\textbf {\bibinfo
  {volume} {229}},\ \bibinfo {pages} {2429} (\bibinfo {year}
  {2020})}\BibitemShut {NoStop}%
\bibitem [{\citenamefont {Bohr}(1952)}]{bohr52}%
  \BibitemOpen
  \bibfield  {author} {\bibinfo {author} {\bibfnamefont {A.}~\bibnamefont
  {Bohr}},\ }\bibfield  {title} {\bibinfo {title} {{The coupling of nuclear
  surface oscillations to the motion of individual nucleons}},\ }\href@noop {}
  {\bibfield  {journal} {\bibinfo  {journal} {Dan. Mat. Fys. Medd.}\ }\textbf
  {\bibinfo {volume} {26}},\ \bibinfo {pages} {14} (\bibinfo {year}
  {1952})}\BibitemShut {NoStop}%
\bibitem [{\citenamefont {{Aage Bohr and Ben R.
  Mottelson}}(1998{\natexlab{a}})}]{bohr_mot98a}%
  \BibitemOpen
  \bibfield  {author} {\bibinfo {author} {\bibnamefont {{Aage Bohr and Ben R.
  Mottelson}}},\ }\href@noop {} {\emph {\bibinfo {title} {{Nuclear
  Structure}}}},\ Vol.\ \bibinfo {volume} {I: Single-Particle Motion}\
  (\bibinfo  {publisher} {World Scientific Publishing},\ \bibinfo {year}
  {1998})\BibitemShut {NoStop}%
\bibitem [{\citenamefont {{Aage Bohr and Ben R.
  Mottelson}}(1998{\natexlab{b}})}]{bohr_mot98b}%
  \BibitemOpen
  \bibfield  {author} {\bibinfo {author} {\bibnamefont {{Aage Bohr and Ben R.
  Mottelson}}},\ }\href@noop {} {\emph {\bibinfo {title} {{Nuclear
  Structure}}}},\ Vol.\ \bibinfo {volume} {II: Nuclear Deformations}\ (\bibinfo
   {publisher} {World Scientific Publishing},\ \bibinfo {year}
  {1998})\BibitemShut {NoStop}%
\bibitem [{\citenamefont {{F. Iachello and A.
  Arima}}(1987{\natexlab{a}})}]{iachello87}%
  \BibitemOpen
  \bibfield  {author} {\bibinfo {author} {\bibnamefont {{F. Iachello and A.
  Arima}}},\ }\href@noop {} {\emph {\bibinfo {title} {{The Interacting Boson
  Model}}}},\ Cambridge Monographs on Mathematical Physics\ (\bibinfo
  {publisher} {Cambridge University Press},\ \bibinfo {year}
  {1987})\BibitemShut {NoStop}%
\bibitem [{\citenamefont {{A. Arima and F. Iachello}}(1976)}]{arima76}%
  \BibitemOpen
  \bibfield  {author} {\bibinfo {author} {\bibnamefont {{A. Arima and F.
  Iachello}}},\ }\bibfield  {title} {\bibinfo {title} {{Interacting boson model
  of collective states I. The vibrational limit}},\ }\href
  {https://doi.org/https://doi.org/10.1016/0003-4916(76)90097-X} {\bibfield
  {journal} {\bibinfo  {journal} {Annals of Physics (NY)}\ }\textbf {\bibinfo
  {volume} {99}},\ \bibinfo {pages} {253} (\bibinfo {year} {1976})}\BibitemShut
  {NoStop}%
\bibitem [{\citenamefont {{A. Arima and F. Iachello}}(1978)}]{arima78}%
  \BibitemOpen
  \bibfield  {author} {\bibinfo {author} {\bibnamefont {{A. Arima and F.
  Iachello}}},\ }\bibfield  {title} {\bibinfo {title} {{Interacting boson model
  of collective nuclear states II. The rotational limit}},\ }\href
  {https://doi.org/https://doi.org/10.1016/0003-4916(78)90228-2} {\bibfield
  {journal} {\bibinfo  {journal} {Annals of Physics (NY)}\ }\textbf {\bibinfo
  {volume} {111}},\ \bibinfo {pages} {201} (\bibinfo {year}
  {1978})}\BibitemShut {NoStop}%
\bibitem [{\citenamefont {{A. Arima and F. Iachello}}(1979)}]{arima79}%
  \BibitemOpen
  \bibfield  {author} {\bibinfo {author} {\bibnamefont {{A. Arima and F.
  Iachello}}},\ }\bibfield  {title} {\bibinfo {title} {{Interacting boson model
  of collective nuclear states IV. The O(6) limit}},\ }\href
  {https://doi.org/https://doi.org/10.1016/0003-4916(79)90347-6} {\bibfield
  {journal} {\bibinfo  {journal} {Annals of Physics (NY)}\ }\textbf {\bibinfo
  {volume} {123}},\ \bibinfo {pages} {468} (\bibinfo {year}
  {1979})}\BibitemShut {NoStop}%
\bibitem [{\citenamefont {Ginocchio}\ and\ \citenamefont
  {Kirson}(1980{\natexlab{a}})}]{ginocchio80}%
  \BibitemOpen
  \bibfield  {author} {\bibinfo {author} {\bibfnamefont {J.~N.}\ \bibnamefont
  {Ginocchio}}\ and\ \bibinfo {author} {\bibfnamefont {M.~W.}\ \bibnamefont
  {Kirson}},\ }\bibfield  {title} {\bibinfo {title} {{Relationship between the
  Bohr Collective Hamiltonian and the Interacting-Boson Model}},\ }\href
  {https://doi.org/10.1103/PhysRevLett.44.1744} {\bibfield  {journal} {\bibinfo
   {journal} {Phys. Rev. Lett.}\ }\textbf {\bibinfo {volume} {44}},\ \bibinfo
  {pages} {1744} (\bibinfo {year} {1980}{\natexlab{a}})}\BibitemShut {NoStop}%
\bibitem [{\citenamefont {Ginocchio}\ and\ \citenamefont
  {Kirson}(1980{\natexlab{b}})}]{ginocchio80a}%
  \BibitemOpen
  \bibfield  {author} {\bibinfo {author} {\bibfnamefont {J.}~\bibnamefont
  {Ginocchio}}\ and\ \bibinfo {author} {\bibfnamefont {M.}~\bibnamefont
  {Kirson}},\ }\bibfield  {title} {\bibinfo {title} {{An intrinsic state for
  the interacting boson model and its relationship to the Bohr-Mottelson
  model}},\ }\href
  {https://doi.org/https://doi.org/10.1016/0375-9474(80)90387-5} {\bibfield
  {journal} {\bibinfo  {journal} {Nuclear Physics A}\ }\textbf {\bibinfo
  {volume} {350}},\ \bibinfo {pages} {31} (\bibinfo {year}
  {1980}{\natexlab{b}})}\BibitemShut {NoStop}%
\bibitem [{\citenamefont {Dieperink}\ \emph {et~al.}(1980)\citenamefont
  {Dieperink}, \citenamefont {Scholten},\ and\ \citenamefont
  {Iachello}}]{dieperink80}%
  \BibitemOpen
  \bibfield  {author} {\bibinfo {author} {\bibfnamefont {A.~E.~L.}\
  \bibnamefont {Dieperink}}, \bibinfo {author} {\bibfnamefont {O.}~\bibnamefont
  {Scholten}},\ and\ \bibinfo {author} {\bibfnamefont {F.}~\bibnamefont
  {Iachello}},\ }\bibfield  {title} {\bibinfo {title} {{Classical Limit of the
  Interacting-Boson Model}},\ }\href
  {https://doi.org/10.1103/PhysRevLett.44.1747} {\bibfield  {journal} {\bibinfo
   {journal} {Phys. Rev. Lett.}\ }\textbf {\bibinfo {volume} {44}},\ \bibinfo
  {pages} {1747} (\bibinfo {year} {1980})}\BibitemShut {NoStop}%
\bibitem [{\citenamefont {Bender}\ \emph {et~al.}(2003)\citenamefont {Bender},
  \citenamefont {Heenen},\ and\ \citenamefont {Reinhard}}]{bender03}%
  \BibitemOpen
  \bibfield  {author} {\bibinfo {author} {\bibfnamefont {M.}~\bibnamefont
  {Bender}}, \bibinfo {author} {\bibfnamefont {P.-H.}\ \bibnamefont {Heenen}},\
  and\ \bibinfo {author} {\bibfnamefont {P.-G.}\ \bibnamefont {Reinhard}},\
  }\bibfield  {title} {\bibinfo {title} {{Self-consistent mean-field models for
  nuclear structure}},\ }\href {https://doi.org/10.1103/RevModPhys.75.121}
  {\bibfield  {journal} {\bibinfo  {journal} {Rev. Mod. Phys.}\ }\textbf
  {\bibinfo {volume} {75}},\ \bibinfo {pages} {121} (\bibinfo {year}
  {2003})}\BibitemShut {NoStop}%
\bibitem [{\citenamefont {Delaroche}\ \emph {et~al.}(2010)\citenamefont
  {Delaroche}, \citenamefont {Girod}, \citenamefont {Libert}, \citenamefont
  {Goutte}, \citenamefont {Hilaire}, \citenamefont {P\'eru}, \citenamefont
  {Pillet},\ and\ \citenamefont {Bertsch}}]{delaroche10}%
  \BibitemOpen
  \bibfield  {author} {\bibinfo {author} {\bibfnamefont {J.~P.}\ \bibnamefont
  {Delaroche}}, \bibinfo {author} {\bibfnamefont {M.}~\bibnamefont {Girod}},
  \bibinfo {author} {\bibfnamefont {J.}~\bibnamefont {Libert}}, \bibinfo
  {author} {\bibfnamefont {H.}~\bibnamefont {Goutte}}, \bibinfo {author}
  {\bibfnamefont {S.}~\bibnamefont {Hilaire}}, \bibinfo {author} {\bibfnamefont
  {S.}~\bibnamefont {P\'eru}}, \bibinfo {author} {\bibfnamefont
  {N.}~\bibnamefont {Pillet}},\ and\ \bibinfo {author} {\bibfnamefont {G.~F.}\
  \bibnamefont {Bertsch}},\ }\bibfield  {title} {\bibinfo {title} {{Structure
  of even-even nuclei using a mapped collective Hamiltonian and the D1S Gogny
  interaction}},\ }\href {https://doi.org/10.1103/PhysRevC.81.014303}
  {\bibfield  {journal} {\bibinfo  {journal} {Phys. Rev. C}\ }\textbf {\bibinfo
  {volume} {81}},\ \bibinfo {pages} {014303} (\bibinfo {year}
  {2010})}\BibitemShut {NoStop}%
\bibitem [{\citenamefont {Erler}\ \emph {et~al.}(2011)\citenamefont {Erler},
  \citenamefont {Kl\"upfel},\ and\ \citenamefont {Reinhard}}]{erler11}%
  \BibitemOpen
  \bibfield  {author} {\bibinfo {author} {\bibfnamefont {J.}~\bibnamefont
  {Erler}}, \bibinfo {author} {\bibnamefont {Kl\"upfel}},\ and\ \bibinfo
  {author} {\bibfnamefont {P.~G.}\ \bibnamefont {Reinhard}},\ }\bibfield
  {title} {\bibinfo {title} {{Self-consistent nuclear mean-field models:
  example Skyrme–Hartree–Fock}},\ }\href
  {https://doi.org/10.1088/0954-3899/38/3/033101} {\bibfield  {journal}
  {\bibinfo  {journal} {Journal of Physics G: Nuclear and Particle Physics}\
  }\textbf {\bibinfo {volume} {38}},\ \bibinfo {pages} {033101} (\bibinfo
  {year} {2011})}\BibitemShut {NoStop}%
\bibitem [{\citenamefont {Lalazissis}\ \emph {et~al.}(1997)\citenamefont
  {Lalazissis}, \citenamefont {K\"onig},\ and\ \citenamefont
  {Ring}}]{lalazissis97}%
  \BibitemOpen
  \bibfield  {author} {\bibinfo {author} {\bibfnamefont {G.~A.}\ \bibnamefont
  {Lalazissis}}, \bibinfo {author} {\bibfnamefont {J.}~\bibnamefont
  {K\"onig}},\ and\ \bibinfo {author} {\bibfnamefont {P.}~\bibnamefont
  {Ring}},\ }\bibfield  {title} {\bibinfo {title} {{New parametrization for the
  Lagrangian density of relativistic mean field theory}},\ }\href
  {https://doi.org/10.1103/PhysRevC.55.540} {\bibfield  {journal} {\bibinfo
  {journal} {Phys. Rev. C}\ }\textbf {\bibinfo {volume} {55}},\ \bibinfo
  {pages} {540} (\bibinfo {year} {1997})}\BibitemShut {NoStop}%
\bibitem [{\citenamefont {Vretenar}\ \emph {et~al.}(2005)\citenamefont
  {Vretenar}, \citenamefont {Afanasjev}, \citenamefont {Lalazissis},\ and\
  \citenamefont {Ring}}]{vretenar05}%
  \BibitemOpen
  \bibfield  {author} {\bibinfo {author} {\bibfnamefont {D.}~\bibnamefont
  {Vretenar}}, \bibinfo {author} {\bibfnamefont {A.~V.}\ \bibnamefont
  {Afanasjev}}, \bibinfo {author} {\bibfnamefont {G.~A.}\ \bibnamefont
  {Lalazissis}},\ and\ \bibinfo {author} {\bibfnamefont {P.}~\bibnamefont
  {Ring}},\ }\bibfield  {title} {\bibinfo {title} {{Relativistic
  Hartree–Bogoliubov theory: static and dynamic aspects of exotic nuclear
  structure}},\ }\href
  {https://doi.org/https://doi.org/10.1016/j.physrep.2004.10.001} {\bibfield
  {journal} {\bibinfo  {journal} {Physics Reports}\ }\textbf {\bibinfo {volume}
  {409}},\ \bibinfo {pages} {101} (\bibinfo {year} {2005})}\BibitemShut
  {NoStop}%
\bibitem [{\citenamefont {Nikšić}\ \emph {et~al.}(2014)\citenamefont
  {Nikšić}, \citenamefont {Paar}, \citenamefont {Vretenar},\ and\
  \citenamefont {Ring}}]{niksic14}%
  \BibitemOpen
  \bibfield  {author} {\bibinfo {author} {\bibfnamefont {T.}~\bibnamefont
  {Nikšić}}, \bibinfo {author} {\bibfnamefont {N.}~\bibnamefont {Paar}},
  \bibinfo {author} {\bibfnamefont {D.}~\bibnamefont {Vretenar}},\ and\
  \bibinfo {author} {\bibfnamefont {P.}~\bibnamefont {Ring}},\ }\bibfield
  {title} {\bibinfo {title} {{DIRHB--A relativistic self-consistent mean-field
  framework for atomic nuclei}},\ }\href
  {https://doi.org/https://doi.org/10.1016/j.cpc.2014.02.027} {\bibfield
  {journal} {\bibinfo  {journal} {Computer Physics Communications}\ }\textbf
  {\bibinfo {volume} {185}},\ \bibinfo {pages} {1808} (\bibinfo {year}
  {2014})}\BibitemShut {NoStop}%
\bibitem [{\citenamefont {Nomura}\ \emph {et~al.}(2008)\citenamefont {Nomura},
  \citenamefont {Shimizu},\ and\ \citenamefont {Otsuka}}]{nomura08}%
  \BibitemOpen
  \bibfield  {author} {\bibinfo {author} {\bibfnamefont {K.}~\bibnamefont
  {Nomura}}, \bibinfo {author} {\bibfnamefont {N.}~\bibnamefont {Shimizu}},\
  and\ \bibinfo {author} {\bibfnamefont {T.}~\bibnamefont {Otsuka}},\
  }\bibfield  {title} {\bibinfo {title} {{Mean-Field Derivation of the
  Interacting Boson Model Hamiltonian and Exotic Nuclei}},\ }\href
  {https://doi.org/10.1103/PhysRevLett.101.142501} {\bibfield  {journal}
  {\bibinfo  {journal} {Phys. Rev. Lett.}\ }\textbf {\bibinfo {volume} {101}},\
  \bibinfo {pages} {142501} (\bibinfo {year} {2008})}\BibitemShut {NoStop}%
\bibitem [{\citenamefont {Nomura}\ \emph {et~al.}(2010)\citenamefont {Nomura},
  \citenamefont {Shimizu},\ and\ \citenamefont {Otsuka}}]{nomura10}%
  \BibitemOpen
  \bibfield  {author} {\bibinfo {author} {\bibfnamefont {K.}~\bibnamefont
  {Nomura}}, \bibinfo {author} {\bibfnamefont {N.}~\bibnamefont {Shimizu}},\
  and\ \bibinfo {author} {\bibfnamefont {T.}~\bibnamefont {Otsuka}},\
  }\bibfield  {title} {\bibinfo {title} {{Formulating the interacting boson
  model by mean-field methods}},\ }\href
  {https://doi.org/10.1103/PhysRevC.81.044307} {\bibfield  {journal} {\bibinfo
  {journal} {Phys. Rev. C}\ }\textbf {\bibinfo {volume} {81}},\ \bibinfo
  {pages} {044307} (\bibinfo {year} {2010})}\BibitemShut {NoStop}%
\bibitem [{\citenamefont {Nomura}\ \emph
  {et~al.}(2011{\natexlab{a}})\citenamefont {Nomura}, \citenamefont {Otsuka},
  \citenamefont {Shimizu},\ and\ \citenamefont {Guo}}]{nomura11}%
  \BibitemOpen
  \bibfield  {author} {\bibinfo {author} {\bibfnamefont {K.}~\bibnamefont
  {Nomura}}, \bibinfo {author} {\bibfnamefont {T.}~\bibnamefont {Otsuka}},
  \bibinfo {author} {\bibfnamefont {N.}~\bibnamefont {Shimizu}},\ and\ \bibinfo
  {author} {\bibfnamefont {L.}~\bibnamefont {Guo}},\ }\bibfield  {title}
  {\bibinfo {title} {{Microscopic formulation of the interacting boson model
  for rotational nuclei}},\ }\href {https://doi.org/10.1103/PhysRevC.83.041302}
  {\bibfield  {journal} {\bibinfo  {journal} {Phys. Rev. C}\ }\textbf {\bibinfo
  {volume} {83}},\ \bibinfo {pages} {041302(R)} (\bibinfo {year}
  {2011}{\natexlab{a}})}\BibitemShut {NoStop}%
\bibitem [{\citenamefont {Casperson}(2012)}]{casperson12}%
  \BibitemOpen
  \bibfield  {author} {\bibinfo {author} {\bibfnamefont {R.}~\bibnamefont
  {Casperson}},\ }\bibfield  {title} {\bibinfo {title} {{IBAR: Interacting
  boson model calculations for large system sizes}},\ }\href
  {https://doi.org/https://doi.org/10.1016/j.cpc.2011.12.024} {\bibfield
  {journal} {\bibinfo  {journal} {Computer Physics Communications}\ }\textbf
  {\bibinfo {volume} {183}},\ \bibinfo {pages} {1029} (\bibinfo {year}
  {2012})}\BibitemShut {NoStop}%
\bibitem [{\citenamefont {Davydov}\ and\ \citenamefont
  {Filippov}(1958)}]{davydov58}%
  \BibitemOpen
  \bibfield  {author} {\bibinfo {author} {\bibfnamefont {A.~S.}\ \bibnamefont
  {Davydov}}\ and\ \bibinfo {author} {\bibfnamefont {G.~F.}\ \bibnamefont
  {Filippov}},\ }\bibfield  {title} {\bibinfo {title} {{Rotational states in
  even atomic nuclei}},\ }\href
  {https://doi.org/https://doi.org/10.1016/0029-5582(58)90153-6} {\bibfield
  {journal} {\bibinfo  {journal} {Nuclear Physics}\ }\textbf {\bibinfo {volume}
  {8}},\ \bibinfo {pages} {237} (\bibinfo {year} {1958})}\BibitemShut {NoStop}%
\bibitem [{\citenamefont {Davydov}\ and\ \citenamefont
  {Rostovsky}(1959)}]{davydov59}%
  \BibitemOpen
  \bibfield  {author} {\bibinfo {author} {\bibfnamefont {A.~S.}\ \bibnamefont
  {Davydov}}\ and\ \bibinfo {author} {\bibfnamefont {V.~S.}\ \bibnamefont
  {Rostovsky}},\ }\bibfield  {title} {\bibinfo {title} {{Relative transition
  probabilities between rotational levels of non-axial nuclei}},\ }\href
  {https://doi.org/https://doi.org/10.1016/0029-5582(59)90127-0} {\bibfield
  {journal} {\bibinfo  {journal} {Nuclear Physics}\ }\textbf {\bibinfo {volume}
  {12}},\ \bibinfo {pages} {58} (\bibinfo {year} {1959})}\BibitemShut {NoStop}%
\bibitem [{\citenamefont {Meyer-Ter-Vehn}(1975)}]{meyertervehn75}%
  \BibitemOpen
  \bibfield  {author} {\bibinfo {author} {\bibfnamefont {J.}~\bibnamefont
  {Meyer-Ter-Vehn}},\ }\bibfield  {title} {\bibinfo {title} {{Collective model
  description of transitional odd-A nuclei: (I). The
  triaxial-rotor-plus-particle model}},\ }\href
  {https://doi.org/https://doi.org/10.1016/0375-9474(75)90095-0} {\bibfield
  {journal} {\bibinfo  {journal} {Nuclear Physics A}\ }\textbf {\bibinfo
  {volume} {249}},\ \bibinfo {pages} {111} (\bibinfo {year}
  {1975})}\BibitemShut {NoStop}%
\bibitem [{\citenamefont {Zamfir}\ and\ \citenamefont
  {Casten}(1991)}]{zamfir91}%
  \BibitemOpen
  \bibfield  {author} {\bibinfo {author} {\bibfnamefont {N.~V.}\ \bibnamefont
  {Zamfir}}\ and\ \bibinfo {author} {\bibfnamefont {R.~F.}\ \bibnamefont
  {Casten}},\ }\bibfield  {title} {\bibinfo {title} {{Signatures of $\gamma$
  softness or triaxiality in low energy nuclear spectra}},\ }\href
  {https://doi.org/https://doi.org/10.1016/0370-2693(91)91610-8} {\bibfield
  {journal} {\bibinfo  {journal} {Physics Letters B}\ }\textbf {\bibinfo
  {volume} {260}},\ \bibinfo {pages} {265} (\bibinfo {year}
  {1991})}\BibitemShut {NoStop}%
\bibitem [{\citenamefont {McCutchan}\ \emph {et~al.}(2007)\citenamefont
  {McCutchan}, \citenamefont {Bonatsos}, \citenamefont {Zamfir},\ and\
  \citenamefont {Casten}}]{mccutchan07}%
  \BibitemOpen
  \bibfield  {author} {\bibinfo {author} {\bibfnamefont {E.~A.}\ \bibnamefont
  {McCutchan}}, \bibinfo {author} {\bibfnamefont {D.}~\bibnamefont {Bonatsos}},
  \bibinfo {author} {\bibfnamefont {N.~V.}\ \bibnamefont {Zamfir}},\ and\
  \bibinfo {author} {\bibfnamefont {R.~F.}\ \bibnamefont {Casten}},\ }\bibfield
   {title} {\bibinfo {title} {{Staggering in \ensuremath{\gamma}-band energies
  and the transition between different structural symmetries in nuclei}},\
  }\href {https://doi.org/10.1103/PhysRevC.76.024306} {\bibfield  {journal}
  {\bibinfo  {journal} {Phys. Rev. C}\ }\textbf {\bibinfo {volume} {76}},\
  \bibinfo {pages} {024306} (\bibinfo {year} {2007})}\BibitemShut {NoStop}%
\bibitem [{\citenamefont {Tsunoda}\ and\ \citenamefont
  {Otsuka}(2021)}]{tsunoda21}%
  \BibitemOpen
  \bibfield  {author} {\bibinfo {author} {\bibfnamefont {Y.}~\bibnamefont
  {Tsunoda}}\ and\ \bibinfo {author} {\bibfnamefont {T.}~\bibnamefont
  {Otsuka}},\ }\bibfield  {title} {\bibinfo {title} {{Triaxial rigidity of
  $^{166}\mathrm{Er}$ and its Bohr-model realization}},\ }\href
  {https://doi.org/10.1103/PhysRevC.103.L021303} {\bibfield  {journal}
  {\bibinfo  {journal} {Phys. Rev. C}\ }\textbf {\bibinfo {volume} {103}},\
  \bibinfo {pages} {L021303} (\bibinfo {year} {2021})}\BibitemShut {NoStop}%
\bibitem [{\citenamefont {Otsuka}\ \emph {et~al.}(2023)\citenamefont {Otsuka},
  \citenamefont {Tsunoda}, \citenamefont {Utsuno}, \citenamefont {Shimizu},
  \citenamefont {Abe},\ and\ \citenamefont {Ueno}}]{otsuka23}%
  \BibitemOpen
  \bibfield  {author} {\bibinfo {author} {\bibfnamefont {T.}~\bibnamefont
  {Otsuka}}, \bibinfo {author} {\bibfnamefont {Y.}~\bibnamefont {Tsunoda}},
  \bibinfo {author} {\bibfnamefont {Y.}~\bibnamefont {Utsuno}}, \bibinfo
  {author} {\bibfnamefont {N.}~\bibnamefont {Shimizu}}, \bibinfo {author}
  {\bibfnamefont {T.}~\bibnamefont {Abe}},\ and\ \bibinfo {author}
  {\bibfnamefont {H.}~\bibnamefont {Ueno}},\ }\href@noop {} {\bibinfo {title}
  {{Prevailing Triaxial Shapes in Heavy Nuclei Driven by Nuclear Tensor
  Force}}} (\bibinfo {year} {2023}),\ \Eprint
  {https://arxiv.org/abs/2303.11299} {arXiv:2303.11299 [nucl-th]} \BibitemShut
  {NoStop}%
\bibitem [{\citenamefont {Rouoof}\ \emph {et~al.}(2024)\citenamefont {Rouoof},
  \citenamefont {Nazir}, \citenamefont {Jehangir}, \citenamefont {Bhat},
  \citenamefont {Sheikh}, \citenamefont {Rather},\ and\ \citenamefont
  {Frauendorf}}]{rouoof24}%
  \BibitemOpen
  \bibfield  {author} {\bibinfo {author} {\bibfnamefont {S.}~\bibnamefont
  {Rouoof}}, \bibinfo {author} {\bibfnamefont {N.}~\bibnamefont {Nazir}},
  \bibinfo {author} {\bibfnamefont {S.}~\bibnamefont {Jehangir}}, \bibinfo
  {author} {\bibfnamefont {G.~H.}\ \bibnamefont {Bhat}}, \bibinfo {author}
  {\bibfnamefont {J.~A.}\ \bibnamefont {Sheikh}}, \bibinfo {author}
  {\bibfnamefont {N.}~\bibnamefont {Rather}},\ and\ \bibinfo {author}
  {\bibfnamefont {S.}~\bibnamefont {Frauendorf}},\ }\bibfield  {title}
  {\bibinfo {title} {{Fingerprints of the triaxial deformation from energies
  and B(E2) transition probabilities of $\gamma$-bands in transitional and
  deformed nuclei}},\ }\href {https://doi.org/10.1140/epja/s10050-024-01257-y}
  {\bibfield  {journal} {\bibinfo  {journal} {Eur. Phys. J. A}\ }\textbf
  {\bibinfo {volume} {60}},\ \bibinfo {pages} {40} (\bibinfo {year}
  {2024})}\BibitemShut {NoStop}%
\bibitem [{\citenamefont {Van~Isacker}\ and\ \citenamefont
  {Chen}(1981)}]{pvisacker81}%
  \BibitemOpen
  \bibfield  {author} {\bibinfo {author} {\bibfnamefont {P.}~\bibnamefont
  {Van~Isacker}}\ and\ \bibinfo {author} {\bibfnamefont {J.-Q.}\ \bibnamefont
  {Chen}},\ }\bibfield  {title} {\bibinfo {title} {{Classical limit of the
  interacting boson Hamiltonian}},\ }\href
  {https://doi.org/10.1103/PhysRevC.24.684} {\bibfield  {journal} {\bibinfo
  {journal} {Phys. Rev. C}\ }\textbf {\bibinfo {volume} {24}},\ \bibinfo
  {pages} {684} (\bibinfo {year} {1981})}\BibitemShut {NoStop}%
\bibitem [{\citenamefont {Heyde}\ \emph {et~al.}(1984)\citenamefont {Heyde},
  \citenamefont {Van~Isacker}, \citenamefont {Waroquier},\ and\ \citenamefont
  {Moreau}}]{heyde84}%
  \BibitemOpen
  \bibfield  {author} {\bibinfo {author} {\bibfnamefont {K.}~\bibnamefont
  {Heyde}}, \bibinfo {author} {\bibfnamefont {P.}~\bibnamefont {Van~Isacker}},
  \bibinfo {author} {\bibfnamefont {M.}~\bibnamefont {Waroquier}},\ and\
  \bibinfo {author} {\bibfnamefont {J.}~\bibnamefont {Moreau}},\ }\bibfield
  {title} {\bibinfo {title} {{Triaxial shapes in the interacting boson
  model}},\ }\href {https://doi.org/10.1103/PhysRevC.29.1420} {\bibfield
  {journal} {\bibinfo  {journal} {Phys. Rev. C}\ }\textbf {\bibinfo {volume}
  {29}},\ \bibinfo {pages} {1420} (\bibinfo {year} {1984})}\BibitemShut
  {NoStop}%
\bibitem [{\citenamefont {Thiamova}(2010)}]{thiamova10}%
  \BibitemOpen
  \bibfield  {author} {\bibinfo {author} {\bibfnamefont {G.}~\bibnamefont
  {Thiamova}},\ }\bibfield  {title} {\bibinfo {title} {{The IBM description of
  triaxial nuclei}},\ }\href {https://doi.org/10.1140/epja/i2010-10982-2}
  {\bibfield  {journal} {\bibinfo  {journal} {Eur. Phys. J. A}\ }\textbf
  {\bibinfo {volume} {45}},\ \bibinfo {pages} {81} (\bibinfo {year}
  {2010})}\BibitemShut {NoStop}%
\bibitem [{\citenamefont {Fortunato}\ \emph {et~al.}(2011)\citenamefont
  {Fortunato}, \citenamefont {Alonso}, \citenamefont {Arias}, \citenamefont
  {Garc\'{\i}a-Ramos},\ and\ \citenamefont {Vitturi}}]{fortunato11}%
  \BibitemOpen
  \bibfield  {author} {\bibinfo {author} {\bibfnamefont {L.}~\bibnamefont
  {Fortunato}}, \bibinfo {author} {\bibfnamefont {C.~E.}\ \bibnamefont
  {Alonso}}, \bibinfo {author} {\bibfnamefont {J.~M.}\ \bibnamefont {Arias}},
  \bibinfo {author} {\bibfnamefont {J.~E.}\ \bibnamefont {Garc\'{\i}a-Ramos}},\
  and\ \bibinfo {author} {\bibfnamefont {A.}~\bibnamefont {Vitturi}},\
  }\bibfield  {title} {\bibinfo {title} {{Phase diagram for a cubic-$Q$
  interacting boson model Hamiltonian: Signs of triaxiality}},\ }\href
  {https://doi.org/10.1103/PhysRevC.84.014326} {\bibfield  {journal} {\bibinfo
  {journal} {Phys. Rev. C}\ }\textbf {\bibinfo {volume} {84}},\ \bibinfo
  {pages} {014326} (\bibinfo {year} {2011})}\BibitemShut {NoStop}%
\bibitem [{\citenamefont {Dieperink}\ and\ \citenamefont
  {Bijker}(1982)}]{dieperink82}%
  \BibitemOpen
  \bibfield  {author} {\bibinfo {author} {\bibfnamefont {A.~E.~L.}\
  \bibnamefont {Dieperink}}\ and\ \bibinfo {author} {\bibfnamefont
  {R.}~\bibnamefont {Bijker}},\ }\bibfield  {title} {\bibinfo {title} {{On
  triaxial features in the neutron-proton IBA}},\ }\href
  {https://doi.org/https://doi.org/10.1016/0370-2693(82)90979-0} {\bibfield
  {journal} {\bibinfo  {journal} {Physics Letters B}\ }\textbf {\bibinfo
  {volume} {116}},\ \bibinfo {pages} {77} (\bibinfo {year} {1982})}\BibitemShut
  {NoStop}%
\bibitem [{\citenamefont {Walet}\ and\ \citenamefont
  {Brussaard}(1987)}]{walet87}%
  \BibitemOpen
  \bibfield  {author} {\bibinfo {author} {\bibfnamefont {N.~R.}\ \bibnamefont
  {Walet}}\ and\ \bibinfo {author} {\bibfnamefont {P.~J.}\ \bibnamefont
  {Brussaard}},\ }\bibfield  {title} {\bibinfo {title} {{A study of the SU(3)*
  limit of IBM-2}},\ }\href
  {https://doi.org/https://doi.org/10.1016/0375-9474(87)90194-1} {\bibfield
  {journal} {\bibinfo  {journal} {Nuclear Physics A}\ }\textbf {\bibinfo
  {volume} {474}},\ \bibinfo {pages} {61} (\bibinfo {year} {1987})}\BibitemShut
  {NoStop}%
\bibitem [{\citenamefont {Sevrin}\ \emph {et~al.}(1987)\citenamefont {Sevrin},
  \citenamefont {Heyde},\ and\ \citenamefont {Jolie}}]{sevrin87}%
  \BibitemOpen
  \bibfield  {author} {\bibinfo {author} {\bibfnamefont {A.}~\bibnamefont
  {Sevrin}}, \bibinfo {author} {\bibfnamefont {K.}~\bibnamefont {Heyde}},\ and\
  \bibinfo {author} {\bibfnamefont {J.}~\bibnamefont {Jolie}},\ }\bibfield
  {title} {\bibinfo {title} {{Triaxiality in the proton-neutron interacting
  boson model: Systematic study of perturbations in the
  ${\mathrm{SU}}^{/\mathrm{e}\mathrm{m}\mathrm{p}\mathrm{h}>}$(3) limit}},\
  }\href {https://doi.org/10.1103/PhysRevC.36.2621} {\bibfield  {journal}
  {\bibinfo  {journal} {Phys. Rev. C}\ }\textbf {\bibinfo {volume} {36}},\
  \bibinfo {pages} {2621} (\bibinfo {year} {1987})}\BibitemShut {NoStop}%
\bibitem [{\citenamefont {Bonatsos}\ \emph
  {et~al.}(2017{\natexlab{a}})\citenamefont {Bonatsos}, \citenamefont
  {Assimakis}, \citenamefont {Minkov}, \citenamefont {Martinou}, \citenamefont
  {Cakirli}, \citenamefont {Casten},\ and\ \citenamefont
  {Blaum}}]{bonatsos17a}%
  \BibitemOpen
  \bibfield  {author} {\bibinfo {author} {\bibfnamefont {D.}~\bibnamefont
  {Bonatsos}}, \bibinfo {author} {\bibfnamefont {I.~E.}\ \bibnamefont
  {Assimakis}}, \bibinfo {author} {\bibfnamefont {N.}~\bibnamefont {Minkov}},
  \bibinfo {author} {\bibfnamefont {A.}~\bibnamefont {Martinou}}, \bibinfo
  {author} {\bibfnamefont {R.~B.}\ \bibnamefont {Cakirli}}, \bibinfo {author}
  {\bibfnamefont {R.~F.}\ \bibnamefont {Casten}},\ and\ \bibinfo {author}
  {\bibfnamefont {K.}~\bibnamefont {Blaum}},\ }\bibfield  {title} {\bibinfo
  {title} {{Proxy-SU(3) symmetry in heavy deformed nuclei}},\ }\href
  {https://doi.org/10.1103/PhysRevC.95.064325} {\bibfield  {journal} {\bibinfo
  {journal} {Phys. Rev. C}\ }\textbf {\bibinfo {volume} {95}},\ \bibinfo
  {pages} {064325} (\bibinfo {year} {2017}{\natexlab{a}})}\BibitemShut
  {NoStop}%
\bibitem [{\citenamefont {Bonatsos}\ \emph
  {et~al.}(2017{\natexlab{b}})\citenamefont {Bonatsos}, \citenamefont
  {Assimakis}, \citenamefont {Minkov}, \citenamefont {Martinou}, \citenamefont
  {Sarantopoulou}, \citenamefont {Cakirli}, \citenamefont {Casten},\ and\
  \citenamefont {Blaum}}]{bonatsos17b}%
  \BibitemOpen
  \bibfield  {author} {\bibinfo {author} {\bibfnamefont {D.}~\bibnamefont
  {Bonatsos}}, \bibinfo {author} {\bibfnamefont {I.~E.}\ \bibnamefont
  {Assimakis}}, \bibinfo {author} {\bibfnamefont {N.}~\bibnamefont {Minkov}},
  \bibinfo {author} {\bibfnamefont {A.}~\bibnamefont {Martinou}}, \bibinfo
  {author} {\bibfnamefont {S.}~\bibnamefont {Sarantopoulou}}, \bibinfo {author}
  {\bibfnamefont {R.~B.}\ \bibnamefont {Cakirli}}, \bibinfo {author}
  {\bibfnamefont {R.~F.}\ \bibnamefont {Casten}},\ and\ \bibinfo {author}
  {\bibfnamefont {K.}~\bibnamefont {Blaum}},\ }\bibfield  {title} {\bibinfo
  {title} {{Analytic predictions for nuclear shapes, prolate dominance, and the
  prolate-oblate shape transition in the proxy-SU(3) model}},\ }\href
  {https://doi.org/10.1103/PhysRevC.95.064326} {\bibfield  {journal} {\bibinfo
  {journal} {Phys. Rev. C}\ }\textbf {\bibinfo {volume} {95}},\ \bibinfo
  {pages} {064326} (\bibinfo {year} {2017}{\natexlab{b}})}\BibitemShut
  {NoStop}%
\bibitem [{\citenamefont {Bonatsos}\ \emph {et~al.}(2023)\citenamefont
  {Bonatsos}, \citenamefont {Martinou}, \citenamefont {Peroulis}, \citenamefont
  {Mertzimekis},\ and\ \citenamefont {Minkov}}]{bonatsos23}%
  \BibitemOpen
  \bibfield  {author} {\bibinfo {author} {\bibfnamefont {D.}~\bibnamefont
  {Bonatsos}}, \bibinfo {author} {\bibfnamefont {A.}~\bibnamefont {Martinou}},
  \bibinfo {author} {\bibfnamefont {S.~K.}\ \bibnamefont {Peroulis}}, \bibinfo
  {author} {\bibfnamefont {T.~J.}\ \bibnamefont {Mertzimekis}},\ and\ \bibinfo
  {author} {\bibfnamefont {N.}~\bibnamefont {Minkov}},\ }\bibfield  {title}
  {\bibinfo {title} {{The Proxy-SU(3) Symmetry in Atomic Nuclei}},\ }\href
  {https://doi.org/10.3390/sym15010169} {\bibfield  {journal} {\bibinfo
  {journal} {Symmetry}\ }\textbf {\bibinfo {volume} {15}},\ \bibinfo {pages}
  {169} (\bibinfo {year} {2023})}\BibitemShut {NoStop}%
\bibitem [{\citenamefont {Reinhard}\ \emph {et~al.}(2021)\citenamefont
  {Reinhard}, \citenamefont {Schuetrumpf},\ and\ \citenamefont
  {Maruhn}}]{skyax}%
  \BibitemOpen
  \bibfield  {author} {\bibinfo {author} {\bibfnamefont {P.-G.}\ \bibnamefont
  {Reinhard}}, \bibinfo {author} {\bibfnamefont {B.}~\bibnamefont
  {Schuetrumpf}},\ and\ \bibinfo {author} {\bibfnamefont {J.}~\bibnamefont
  {Maruhn}},\ }\bibfield  {title} {\bibinfo {title} {{The Axial Hartree–Fock
  + BCS Code SkyAx}},\ }\href
  {https://doi.org/https://doi.org/10.1016/j.cpc.2020.107603} {\bibfield
  {journal} {\bibinfo  {journal} {Computer Physics Communications}\ }\textbf
  {\bibinfo {volume} {258}},\ \bibinfo {pages} {107603} (\bibinfo {year}
  {2021})}\BibitemShut {NoStop}%
\bibitem [{\citenamefont {Nomura}\ \emph
  {et~al.}(2011{\natexlab{b}})\citenamefont {Nomura}, \citenamefont {Otsuka},
  \citenamefont {Rodr\'{\i}guez-Guzm\'an}, \citenamefont {Robledo},\ and\
  \citenamefont {Sarriguren}}]{nomura11a}%
  \BibitemOpen
  \bibfield  {author} {\bibinfo {author} {\bibfnamefont {K.}~\bibnamefont
  {Nomura}}, \bibinfo {author} {\bibfnamefont {T.}~\bibnamefont {Otsuka}},
  \bibinfo {author} {\bibfnamefont {R.}~\bibnamefont
  {Rodr\'{\i}guez-Guzm\'an}}, \bibinfo {author} {\bibfnamefont {L.~M.}\
  \bibnamefont {Robledo}},\ and\ \bibinfo {author} {\bibfnamefont
  {P.}~\bibnamefont {Sarriguren}},\ }\bibfield  {title} {\bibinfo {title}
  {{Collective structural evolution in neutron-rich Yb, Hf, W, Os, and Pt
  isotopes}},\ }\href {https://doi.org/10.1103/PhysRevC.84.054316} {\bibfield
  {journal} {\bibinfo  {journal} {Phys. Rev. C}\ }\textbf {\bibinfo {volume}
  {84}},\ \bibinfo {pages} {054316} (\bibinfo {year}
  {2011}{\natexlab{b}})}\BibitemShut {NoStop}%
\bibitem [{\citenamefont {Nomura}\ \emph
  {et~al.}(2011{\natexlab{c}})\citenamefont {Nomura}, \citenamefont {Otsuka},
  \citenamefont {Rodr\'{\i}guez-Guzm\'an}, \citenamefont {Robledo},
  \citenamefont {Sarriguren}, \citenamefont {Regan}, \citenamefont
  {Stevenson},\ and\ \citenamefont {Podoly\'ak}}]{nomura11b}%
  \BibitemOpen
  \bibfield  {author} {\bibinfo {author} {\bibfnamefont {K.}~\bibnamefont
  {Nomura}}, \bibinfo {author} {\bibfnamefont {T.}~\bibnamefont {Otsuka}},
  \bibinfo {author} {\bibfnamefont {R.}~\bibnamefont
  {Rodr\'{\i}guez-Guzm\'an}}, \bibinfo {author} {\bibfnamefont {L.~M.}\
  \bibnamefont {Robledo}}, \bibinfo {author} {\bibfnamefont {P.}~\bibnamefont
  {Sarriguren}}, \bibinfo {author} {\bibfnamefont {P.~H.}\ \bibnamefont
  {Regan}}, \bibinfo {author} {\bibfnamefont {P.~D.}\ \bibnamefont
  {Stevenson}},\ and\ \bibinfo {author} {\bibfnamefont {Z.}~\bibnamefont
  {Podoly\'ak}},\ }\bibfield  {title} {\bibinfo {title} {{Spectroscopic
  calculations of the low-lying structure in exotic Os and W isotopes}},\
  }\href {https://doi.org/10.1103/PhysRevC.83.054303} {\bibfield  {journal}
  {\bibinfo  {journal} {Phys. Rev. C}\ }\textbf {\bibinfo {volume} {83}},\
  \bibinfo {pages} {054303} (\bibinfo {year} {2011}{\natexlab{c}})}\BibitemShut
  {NoStop}%
\bibitem [{\citenamefont {Rudigier}\ \emph {et~al.}(2015)\citenamefont
  {Rudigier}, \citenamefont {Nomura}, \citenamefont {Dannhoff}, \citenamefont
  {Gerst}, \citenamefont {Jolie}, \citenamefont {Saed-Samii}, \citenamefont
  {Stegemann}, \citenamefont {R\'egis}, \citenamefont {Robledo}, \citenamefont
  {Rodr\'{\i}guez-Guzm\'an}, \citenamefont {Blazhev}, \citenamefont {Fransen},
  \citenamefont {Warr},\ and\ \citenamefont {Zell}}]{rudigier15}%
  \BibitemOpen
  \bibfield  {author} {\bibinfo {author} {\bibfnamefont {M.}~\bibnamefont
  {Rudigier}}, \bibinfo {author} {\bibfnamefont {K.}~\bibnamefont {Nomura}},
  \bibinfo {author} {\bibfnamefont {M.}~\bibnamefont {Dannhoff}}, \bibinfo
  {author} {\bibfnamefont {R.-B.}\ \bibnamefont {Gerst}}, \bibinfo {author}
  {\bibfnamefont {J.}~\bibnamefont {Jolie}}, \bibinfo {author} {\bibfnamefont
  {N.}~\bibnamefont {Saed-Samii}}, \bibinfo {author} {\bibfnamefont
  {S.}~\bibnamefont {Stegemann}}, \bibinfo {author} {\bibfnamefont {J.-M.}\
  \bibnamefont {R\'egis}}, \bibinfo {author} {\bibfnamefont {L.~M.}\
  \bibnamefont {Robledo}}, \bibinfo {author} {\bibfnamefont {R.}~\bibnamefont
  {Rodr\'{\i}guez-Guzm\'an}}, \bibinfo {author} {\bibfnamefont
  {A.}~\bibnamefont {Blazhev}}, \bibinfo {author} {\bibfnamefont
  {C.}~\bibnamefont {Fransen}}, \bibinfo {author} {\bibfnamefont
  {N.}~\bibnamefont {Warr}},\ and\ \bibinfo {author} {\bibfnamefont {K.~O.}\
  \bibnamefont {Zell}},\ }\bibfield  {title} {\bibinfo {title} {{Evolution of
  $E2$ transition strength in deformed hafnium isotopes from new measurements
  on $^{172}\mathrm{Hf},\phantom{\rule{0.16em}{0ex}}^{174}\mathrm{Hf}$, and
  $^{176}\mathrm{Hf}$}},\ }\href {https://doi.org/10.1103/PhysRevC.91.044301}
  {\bibfield  {journal} {\bibinfo  {journal} {Phys. Rev. C}\ }\textbf {\bibinfo
  {volume} {91}},\ \bibinfo {pages} {044301} (\bibinfo {year}
  {2015})}\BibitemShut {NoStop}%
\bibitem [{\citenamefont {Elliott}(1958{\natexlab{a}})}]{elliott58a}%
  \BibitemOpen
  \bibfield  {author} {\bibinfo {author} {\bibfnamefont {J.~P.}\ \bibnamefont
  {Elliott}},\ }\bibfield  {title} {\bibinfo {title} {{Collective motion in the
  nuclear shell model. I. Classification schemes for states of mixed
  configurations}},\ }\href {https://doi.org/10.1098/rspa.1958.0072} {\bibfield
   {journal} {\bibinfo  {journal} {Proceedings of the Royal Society of London.
  Series A. Mathematical and Physical Sciences}\ }\textbf {\bibinfo {volume}
  {245}},\ \bibinfo {pages} {128} (\bibinfo {year}
  {1958}{\natexlab{a}})}\BibitemShut {NoStop}%
\bibitem [{\citenamefont {Elliott}(1958{\natexlab{b}})}]{elliott58b}%
  \BibitemOpen
  \bibfield  {author} {\bibinfo {author} {\bibfnamefont {J.~P.}\ \bibnamefont
  {Elliott}},\ }\bibfield  {title} {\bibinfo {title} {{Collective motion in the
  nuclear shell model II. The introduction of intrinsic wave-functions}},\
  }\href {https://doi.org/10.1098/rspa.1958.0101} {\bibfield  {journal}
  {\bibinfo  {journal} {Proceedings of the Royal Society of London. Series A.
  Mathematical and Physical Sciences}\ }\textbf {\bibinfo {volume} {245}},\
  \bibinfo {pages} {562} (\bibinfo {year} {1958}{\natexlab{b}})}\BibitemShut
  {NoStop}%
\bibitem [{\citenamefont {Elliott}\ and\ \citenamefont
  {Harvey}(1963)}]{elliott63}%
  \BibitemOpen
  \bibfield  {author} {\bibinfo {author} {\bibfnamefont {J.~P.}\ \bibnamefont
  {Elliott}}\ and\ \bibinfo {author} {\bibfnamefont {M.}~\bibnamefont
  {Harvey}},\ }\bibfield  {title} {\bibinfo {title} {{Collective motion in the
  nuclear shell model III. The calculation of spectra}},\ }\href
  {https://doi.org/10.1098/rspa.1963.0071} {\bibfield  {journal} {\bibinfo
  {journal} {Proceedings of the Royal Society of London. Series A. Mathematical
  and Physical Sciences}\ }\textbf {\bibinfo {volume} {272}},\ \bibinfo {pages}
  {557} (\bibinfo {year} {1963})}\BibitemShut {NoStop}%
\bibitem [{\citenamefont {Martinou}\ \emph {et~al.}(2020)\citenamefont
  {Martinou}, \citenamefont {Bonatsos}, \citenamefont {Minkov}, \citenamefont
  {Assimakis}, \citenamefont {Peroulis}, \citenamefont {Sarantopoulou},\ and\
  \citenamefont {Cseh}}]{martinou20}%
  \BibitemOpen
  \bibfield  {author} {\bibinfo {author} {\bibfnamefont {A.}~\bibnamefont
  {Martinou}}, \bibinfo {author} {\bibfnamefont {D.}~\bibnamefont {Bonatsos}},
  \bibinfo {author} {\bibfnamefont {N.}~\bibnamefont {Minkov}}, \bibinfo
  {author} {\bibfnamefont {I.~E.}\ \bibnamefont {Assimakis}}, \bibinfo {author}
  {\bibfnamefont {S.~K.}\ \bibnamefont {Peroulis}}, \bibinfo {author}
  {\bibfnamefont {S.}~\bibnamefont {Sarantopoulou}},\ and\ \bibinfo {author}
  {\bibfnamefont {J.}~\bibnamefont {Cseh}},\ }\bibfield  {title} {\bibinfo
  {title} {{Proxy-SU(3) symmetry in the shell model basis}},\ }\href
  {https://doi.org/10.1140/epja/s10050-020-00239-0} {\bibfield  {journal}
  {\bibinfo  {journal} {Eur. Phys. J. A}\ }\textbf {\bibinfo {volume} {56}},\
  \bibinfo {pages} {239} (\bibinfo {year} {2020})}\BibitemShut {NoStop}%
\bibitem [{\citenamefont {Bonatsos}\ \emph {et~al.}(2020)\citenamefont
  {Bonatsos}, \citenamefont {Sobhani},\ and\ \citenamefont
  {Hassanabadi}}]{bonatsos20}%
  \BibitemOpen
  \bibfield  {author} {\bibinfo {author} {\bibfnamefont {D.}~\bibnamefont
  {Bonatsos}}, \bibinfo {author} {\bibfnamefont {H.}~\bibnamefont {Sobhani}},\
  and\ \bibinfo {author} {\bibfnamefont {H.}~\bibnamefont {Hassanabadi}},\
  }\bibfield  {title} {\bibinfo {title} {{Shell model structure of proxy-SU(3)
  pairs of orbitals}},\ }\href
  {https://doi.org/10.1140/epjp/s13360-020-00749-2} {\bibfield  {journal}
  {\bibinfo  {journal} {Eur. Phys. J. Plus}\ }\textbf {\bibinfo {volume}
  {135}},\ \bibinfo {pages} {710} (\bibinfo {year} {2020})}\BibitemShut
  {NoStop}%
\bibitem [{\citenamefont {Nilson}(1955)}]{nilsson55}%
  \BibitemOpen
  \bibfield  {author} {\bibinfo {author} {\bibfnamefont {S.~G.}\ \bibnamefont
  {Nilson}},\ }\bibfield  {title} {\bibinfo {title} {{Binding states of
  individual nucleons in strongly deformed nuclei}},\ }\href@noop {} {\bibfield
   {journal} {\bibinfo  {journal} {Dan. Mat. Fys. Medd.}\ }\textbf {\bibinfo
  {volume} {29}},\ \bibinfo {pages} {16} (\bibinfo {year} {1955})}\BibitemShut
  {NoStop}%
\bibitem [{\citenamefont {Nilson}\ and\ \citenamefont
  {Ragnarsson}(1995)}]{nilsson95}%
  \BibitemOpen
  \bibfield  {author} {\bibinfo {author} {\bibfnamefont {S.~G.}\ \bibnamefont
  {Nilson}}\ and\ \bibinfo {author} {\bibfnamefont {I.}~\bibnamefont
  {Ragnarsson}},\ }\href@noop {} {\emph {\bibinfo {title} {{Shapes and Shells
  in Nuclear Structure}}}}\ (\bibinfo  {publisher} {Cambridge University Press:
  Cambridge},\ \bibinfo {year} {1995})\BibitemShut {NoStop}%
\bibitem [{\citenamefont {Martinou}\ \emph
  {et~al.}(2021{\natexlab{a}})\citenamefont {Martinou}, \citenamefont
  {Bonatsos}, \citenamefont {Karakatsanis}, \citenamefont {Sarantopoulou},
  \citenamefont {Assimakis}, \citenamefont {Peroulis},\ and\ \citenamefont
  {Minkov}}]{martinou21b}%
  \BibitemOpen
  \bibfield  {author} {\bibinfo {author} {\bibfnamefont {A.}~\bibnamefont
  {Martinou}}, \bibinfo {author} {\bibfnamefont {D.}~\bibnamefont {Bonatsos}},
  \bibinfo {author} {\bibfnamefont {K.~E.}\ \bibnamefont {Karakatsanis}},
  \bibinfo {author} {\bibfnamefont {S.}~\bibnamefont {Sarantopoulou}}, \bibinfo
  {author} {\bibfnamefont {I.~E.}\ \bibnamefont {Assimakis}}, \bibinfo {author}
  {\bibfnamefont {S.~K.}\ \bibnamefont {Peroulis}},\ and\ \bibinfo {author}
  {\bibfnamefont {N.}~\bibnamefont {Minkov}},\ }\bibfield  {title} {\bibinfo
  {title} {{Why nuclear forces favor the highest weight irreducible
  representations of the fermionic SU(3) symmetry}},\ }\href
  {https://doi.org/10.1140/epja/s10050-021-00395-x} {\bibfield  {journal}
  {\bibinfo  {journal} {Eur. Phys. J. A}\ }\textbf {\bibinfo {volume} {57}},\
  \bibinfo {pages} {83} (\bibinfo {year} {2021}{\natexlab{a}})}\BibitemShut
  {NoStop}%
\bibitem [{\citenamefont {Casta\~nos}\ \emph {et~al.}(1988)\citenamefont
  {Casta\~nos}, \citenamefont {Draayer},\ and\ \citenamefont
  {Leschber}}]{castanos88}%
  \BibitemOpen
  \bibfield  {author} {\bibinfo {author} {\bibfnamefont {O.}~\bibnamefont
  {Casta\~nos}}, \bibinfo {author} {\bibfnamefont {J.~P.}\ \bibnamefont
  {Draayer}},\ and\ \bibinfo {author} {\bibfnamefont {Y.}~\bibnamefont
  {Leschber}},\ }\bibfield  {title} {\bibinfo {title} {{Shape variables and the
  shell model}},\ }\href@noop {} {\bibfield  {journal} {\bibinfo  {journal} {Z.
  Phys. A}\ }\textbf {\bibinfo {volume} {329}},\ \bibinfo {pages} {33}
  (\bibinfo {year} {1988})}\BibitemShut {NoStop}%
\bibitem [{\citenamefont {Draayer}\ \emph {et~al.}(1989)\citenamefont
  {Draayer}, \citenamefont {Park},\ and\ \citenamefont
  {Casta\~nos}}]{draayer89}%
  \BibitemOpen
  \bibfield  {author} {\bibinfo {author} {\bibfnamefont {J.~P.}\ \bibnamefont
  {Draayer}}, \bibinfo {author} {\bibfnamefont {S.~C.}\ \bibnamefont {Park}},\
  and\ \bibinfo {author} {\bibfnamefont {O.}~\bibnamefont {Casta\~nos}},\
  }\bibfield  {title} {\bibinfo {title} {{Shell-Model Interpretation of the
  Collective-Model Potential-Energy Surface}},\ }\href
  {https://doi.org/10.1103/PhysRevLett.62.20} {\bibfield  {journal} {\bibinfo
  {journal} {Phys. Rev. Lett.}\ }\textbf {\bibinfo {volume} {62}},\ \bibinfo
  {pages} {20} (\bibinfo {year} {1989})}\BibitemShut {NoStop}%
\bibitem [{\citenamefont {Hamamoto}\ and\ \citenamefont
  {Mottelson}(2009)}]{hamamoto09}%
  \BibitemOpen
  \bibfield  {author} {\bibinfo {author} {\bibfnamefont {I.}~\bibnamefont
  {Hamamoto}}\ and\ \bibinfo {author} {\bibfnamefont {B.~R.}\ \bibnamefont
  {Mottelson}},\ }\bibfield  {title} {\bibinfo {title} {{Further examination of
  prolate-shape dominance in nuclear deformation}},\ }\href
  {https://doi.org/10.1103/PhysRevC.79.034317} {\bibfield  {journal} {\bibinfo
  {journal} {Phys. Rev. C}\ }\textbf {\bibinfo {volume} {79}},\ \bibinfo
  {pages} {034317} (\bibinfo {year} {2009})}\BibitemShut {NoStop}%
\bibitem [{\citenamefont {Hamamoto}\ and\ \citenamefont
  {Mottelson}(2012)}]{hamamoto12}%
  \BibitemOpen
  \bibfield  {author} {\bibinfo {author} {\bibfnamefont {I.}~\bibnamefont
  {Hamamoto}}\ and\ \bibinfo {author} {\bibfnamefont {B.~R.}\ \bibnamefont
  {Mottelson}},\ }\bibfield  {title} {\bibinfo {title} {{Shape deformations in
  atomic nuclei}},\ }\href@noop {} {\bibfield  {journal} {\bibinfo  {journal}
  {Scholarpedia}\ }\textbf {\bibinfo {volume} {7}},\ \bibinfo {pages} {10693}
  (\bibinfo {year} {2012})}\BibitemShut {NoStop}%
\bibitem [{ens()}]{ensdf}%
  \BibitemOpen
  \href@noop {} {\bibinfo {title} {{ENSDF}}},\ \bibinfo {howpublished}
  {{https://www.nndc.bnl.gov/ensdf}}\BibitemShut {NoStop}%
\bibitem [{\citenamefont {{R. F. Casten}}(2000)}]{casten00}%
  \BibitemOpen
  \bibfield  {author} {\bibinfo {author} {\bibnamefont {{R. F. Casten}}},\
  }\href@noop {} {\emph {\bibinfo {title} {{Nuclear Structure from a Simple
  Perspective}}}}\ (\bibinfo  {publisher} {Oxford University Press},\ \bibinfo
  {year} {2000})\BibitemShut {NoStop}%
\bibitem [{\citenamefont {Esser}\ \emph {et~al.}(1997)\citenamefont {Esser},
  \citenamefont {Neuneyer}, \citenamefont {Casten},\ and\ \citenamefont {von
  Brentano}}]{esser97}%
  \BibitemOpen
  \bibfield  {author} {\bibinfo {author} {\bibfnamefont {L.}~\bibnamefont
  {Esser}}, \bibinfo {author} {\bibfnamefont {U.}~\bibnamefont {Neuneyer}},
  \bibinfo {author} {\bibfnamefont {R.~F.}\ \bibnamefont {Casten}},\ and\
  \bibinfo {author} {\bibfnamefont {P.}~\bibnamefont {von Brentano}},\
  }\bibfield  {title} {\bibinfo {title} {{Correlations of the deformation
  variables $\ensuremath{\beta}$ and $\ensuremath{\gamma}$ in even-even Hf, W,
  Os, Pt, and Hg nuclei}},\ }\href {https://doi.org/10.1103/PhysRevC.55.206}
  {\bibfield  {journal} {\bibinfo  {journal} {Phys. Rev. C}\ }\textbf {\bibinfo
  {volume} {55}},\ \bibinfo {pages} {206} (\bibinfo {year} {1997})}\BibitemShut
  {NoStop}%
\bibitem [{\citenamefont {Casten}\ and\ \citenamefont
  {Warner}(1988)}]{casten88}%
  \BibitemOpen
  \bibfield  {author} {\bibinfo {author} {\bibfnamefont {R.~F.}\ \bibnamefont
  {Casten}}\ and\ \bibinfo {author} {\bibfnamefont {D.~D.}\ \bibnamefont
  {Warner}},\ }\bibfield  {title} {\bibinfo {title} {{The interacting boson
  approximation}},\ }\href {https://doi.org/10.1103/RevModPhys.60.389}
  {\bibfield  {journal} {\bibinfo  {journal} {Rev. Mod. Phys.}\ }\textbf
  {\bibinfo {volume} {60}},\ \bibinfo {pages} {389} (\bibinfo {year}
  {1988})}\BibitemShut {NoStop}%
\bibitem [{\citenamefont {Ratna~Raju}\ \emph {et~al.}(1973)\citenamefont
  {Ratna~Raju}, \citenamefont {Draayer},\ and\ \citenamefont
  {Hecht}}]{ratnaraju73}%
  \BibitemOpen
  \bibfield  {author} {\bibinfo {author} {\bibfnamefont {R.~D.}\ \bibnamefont
  {Ratna~Raju}}, \bibinfo {author} {\bibfnamefont {J.~P.}\ \bibnamefont
  {Draayer}},\ and\ \bibinfo {author} {\bibfnamefont {K.~T.}\ \bibnamefont
  {Hecht}},\ }\bibfield  {title} {\bibinfo {title} {{Search for a coupling
  scheme in heavy deformed nuclei: The pseudo SU(3) model}},\ }\href
  {https://doi.org/https://doi.org/10.1016/0375-9474(73)90635-0} {\bibfield
  {journal} {\bibinfo  {journal} {Nuclear Physics A}\ }\textbf {\bibinfo
  {volume} {202}},\ \bibinfo {pages} {433} (\bibinfo {year}
  {1973})}\BibitemShut {NoStop}%
\bibitem [{\citenamefont {Draayer}\ \emph {et~al.}(1982)\citenamefont
  {Draayer}, \citenamefont {Weeks},\ and\ \citenamefont {Hecht}}]{draayer82}%
  \BibitemOpen
  \bibfield  {author} {\bibinfo {author} {\bibfnamefont {J.~P.}\ \bibnamefont
  {Draayer}}, \bibinfo {author} {\bibfnamefont {K.~J.}\ \bibnamefont {Weeks}},\
  and\ \bibinfo {author} {\bibfnamefont {K.~T.}\ \bibnamefont {Hecht}},\
  }\bibfield  {title} {\bibinfo {title} {{Strength of the $Q_{\pi}\cdot
  Q_{\nu}$ interaction and the strong-coupled pseudo-SU(3) limit}},\ }\href
  {https://doi.org/https://doi.org/10.1016/0375-9474(82)90497-3} {\bibfield
  {journal} {\bibinfo  {journal} {Nuclear Physics A}\ }\textbf {\bibinfo
  {volume} {381}},\ \bibinfo {pages} {1} (\bibinfo {year} {1982})}\BibitemShut
  {NoStop}%
\bibitem [{\citenamefont {Draayer}\ and\ \citenamefont
  {Weeks}(1984)}]{draayer84}%
  \BibitemOpen
  \bibfield  {author} {\bibinfo {author} {\bibfnamefont {J.~P.}\ \bibnamefont
  {Draayer}}\ and\ \bibinfo {author} {\bibfnamefont {K.~J.}\ \bibnamefont
  {Weeks}},\ }\bibfield  {title} {\bibinfo {title} {{Towards a shell model
  description of the low-energy structure of deformed nuclei I. Even-even
  systems}},\ }\href
  {https://doi.org/https://doi.org/10.1016/0003-4916(84)90210-0} {\bibfield
  {journal} {\bibinfo  {journal} {Annals of Physics (NY)}\ }\textbf {\bibinfo
  {volume} {156}},\ \bibinfo {pages} {41} (\bibinfo {year} {1984})}\BibitemShut
  {NoStop}%
\bibitem [{\citenamefont {Kl\"upfel}\ \emph {et~al.}(2009)\citenamefont
  {Kl\"upfel}, \citenamefont {Reinhard}, \citenamefont {B\"urvenich},\ and\
  \citenamefont {Maruhn}}]{klupfel08}%
  \BibitemOpen
  \bibfield  {author} {\bibinfo {author} {\bibfnamefont {P.}~\bibnamefont
  {Kl\"upfel}}, \bibinfo {author} {\bibfnamefont {P.-G.}\ \bibnamefont
  {Reinhard}}, \bibinfo {author} {\bibfnamefont {T.~J.}\ \bibnamefont
  {B\"urvenich}},\ and\ \bibinfo {author} {\bibfnamefont {J.~A.}\ \bibnamefont
  {Maruhn}},\ }\bibfield  {title} {\bibinfo {title} {{Variations on a theme by
  Skyrme: A systematic study of adjustments of model parameters}},\ }\href
  {https://doi.org/10.1103/PhysRevC.79.034310} {\bibfield  {journal} {\bibinfo
  {journal} {Phys. Rev. C}\ }\textbf {\bibinfo {volume} {79}},\ \bibinfo
  {pages} {034310} (\bibinfo {year} {2009})}\BibitemShut {NoStop}%
\bibitem [{\citenamefont {Abolghasem}\ and\ \citenamefont
  {Alexa}(2024)}]{abolghasem24}%
  \BibitemOpen
  \bibfield  {author} {\bibinfo {author} {\bibfnamefont {M.}~\bibnamefont
  {Abolghasem}}\ and\ \bibinfo {author} {\bibfnamefont {P.}~\bibnamefont
  {Alexa}},\ }\bibfield  {title} {\bibinfo {title} {{Microscopic study of
  ground-state binding energies in $Z=52-70$ neutron-rich nuclei}},\ }\href
  {https://doi.org/https://doi.org/10.1016/j.nuclphysa.2024.122841} {\bibfield
  {journal} {\bibinfo  {journal} {Nuclear Physics A}\ }\textbf {\bibinfo
  {volume} {1044}},\ \bibinfo {pages} {122841} (\bibinfo {year}
  {2024})}\BibitemShut {NoStop}%
\bibitem [{\citenamefont {Lipas}\ \emph {et~al.}(1985)\citenamefont {Lipas},
  \citenamefont {Toivonen},\ and\ \citenamefont {Warner}}]{lipas85}%
  \BibitemOpen
  \bibfield  {author} {\bibinfo {author} {\bibfnamefont {P.}~\bibnamefont
  {Lipas}}, \bibinfo {author} {\bibfnamefont {P.}~\bibnamefont {Toivonen}},\
  and\ \bibinfo {author} {\bibfnamefont {D.}~\bibnamefont {Warner}},\
  }\bibfield  {title} {\bibinfo {title} {{IBA consistent-Q formalism extended
  to the vibrational region}},\ }\href
  {https://doi.org/https://doi.org/10.1016/0370-2693(85)91573-4} {\bibfield
  {journal} {\bibinfo  {journal} {Physics Letters B}\ }\textbf {\bibinfo
  {volume} {155}},\ \bibinfo {pages} {295} (\bibinfo {year}
  {1985})}\BibitemShut {NoStop}%
\bibitem [{\citenamefont {Warner}\ and\ \citenamefont
  {Casten}(1982)}]{warner82}%
  \BibitemOpen
  \bibfield  {author} {\bibinfo {author} {\bibfnamefont {D.~D.}\ \bibnamefont
  {Warner}}\ and\ \bibinfo {author} {\bibfnamefont {R.~F.}\ \bibnamefont
  {Casten}},\ }\bibfield  {title} {\bibinfo {title} {{Revised Formulation of
  the Phenomenological Interacting Boson Approximation}},\ }\href
  {https://doi.org/10.1103/PhysRevLett.48.1385} {\bibfield  {journal} {\bibinfo
   {journal} {Phys. Rev. Lett.}\ }\textbf {\bibinfo {volume} {48}},\ \bibinfo
  {pages} {1385} (\bibinfo {year} {1982})}\BibitemShut {NoStop}%
\bibitem [{\citenamefont {Warner}\ and\ \citenamefont
  {Casten}(1983)}]{warner83}%
  \BibitemOpen
  \bibfield  {author} {\bibinfo {author} {\bibfnamefont {D.~D.}\ \bibnamefont
  {Warner}}\ and\ \bibinfo {author} {\bibfnamefont {R.~F.}\ \bibnamefont
  {Casten}},\ }\bibfield  {title} {\bibinfo {title} {{Predictions of the
  interacting boson approximation in a consistent $Q$ framework}},\ }\href
  {https://doi.org/10.1103/PhysRevC.28.1798} {\bibfield  {journal} {\bibinfo
  {journal} {Phys. Rev. C}\ }\textbf {\bibinfo {volume} {28}},\ \bibinfo
  {pages} {1798} (\bibinfo {year} {1983})}\BibitemShut {NoStop}%
\bibitem [{\citenamefont {Zamfir}\ \emph {et~al.}(2002)\citenamefont {Zamfir},
  \citenamefont {von Brentano}, \citenamefont {Casten},\ and\ \citenamefont
  {Jolie}}]{zamfir02}%
  \BibitemOpen
  \bibfield  {author} {\bibinfo {author} {\bibfnamefont {N.~V.}\ \bibnamefont
  {Zamfir}}, \bibinfo {author} {\bibfnamefont {P.}~\bibnamefont {von
  Brentano}}, \bibinfo {author} {\bibfnamefont {R.~F.}\ \bibnamefont
  {Casten}},\ and\ \bibinfo {author} {\bibfnamefont {J.}~\bibnamefont
  {Jolie}},\ }\bibfield  {title} {\bibinfo {title} {{Test of two-level crossing
  at the $N=90$ spherical-deformed critical point}},\ }\href
  {https://doi.org/10.1103/PhysRevC.66.021304} {\bibfield  {journal} {\bibinfo
  {journal} {Phys. Rev. C}\ }\textbf {\bibinfo {volume} {66}},\ \bibinfo
  {pages} {021304(R)} (\bibinfo {year} {2002})}\BibitemShut {NoStop}%
\bibitem [{\citenamefont {Werner}\ \emph {et~al.}(2005)\citenamefont {Werner},
  \citenamefont {von Brentano}, \citenamefont {Casten}, \citenamefont {Scholl},
  \citenamefont {McCutchan}, \citenamefont {Kr\"ucken},\ and\ \citenamefont
  {Jolie}}]{werner05}%
  \BibitemOpen
  \bibfield  {author} {\bibinfo {author} {\bibfnamefont {V.}~\bibnamefont
  {Werner}}, \bibinfo {author} {\bibfnamefont {P.}~\bibnamefont {von
  Brentano}}, \bibinfo {author} {\bibfnamefont {R.}~\bibnamefont {Casten}},
  \bibinfo {author} {\bibfnamefont {C.}~\bibnamefont {Scholl}}, \bibinfo
  {author} {\bibfnamefont {E.}~\bibnamefont {McCutchan}}, \bibinfo {author}
  {\bibfnamefont {R.}~\bibnamefont {Kr\"ucken}},\ and\ \bibinfo {author}
  {\bibfnamefont {J.}~\bibnamefont {Jolie}},\ }\bibfield  {title} {\bibinfo
  {title} {{Alternative interpretation of $E0$ strengths in transitional
  regions}},\ }\href@noop {} {\bibfield  {journal} {\bibinfo  {journal} {Eur.
  Phys. J. A}\ }\textbf {\bibinfo {volume} {25}},\ \bibinfo {pages} {455}
  (\bibinfo {year} {2005})}\BibitemShut {NoStop}%
\bibitem [{\citenamefont {Feng}\ \emph {et~al.}(1981)\citenamefont {Feng},
  \citenamefont {Gilmore},\ and\ \citenamefont {Deans}}]{feng81}%
  \BibitemOpen
  \bibfield  {author} {\bibinfo {author} {\bibfnamefont {D.~H.}\ \bibnamefont
  {Feng}}, \bibinfo {author} {\bibfnamefont {R.}~\bibnamefont {Gilmore}},\ and\
  \bibinfo {author} {\bibfnamefont {S.~R.}\ \bibnamefont {Deans}},\ }\bibfield
  {title} {\bibinfo {title} {{Phase transitions and the geometric properties of
  the interacting boson model}},\ }\href
  {https://doi.org/10.1103/PhysRevC.23.1254} {\bibfield  {journal} {\bibinfo
  {journal} {Phys. Rev. C}\ }\textbf {\bibinfo {volume} {23}},\ \bibinfo
  {pages} {1254} (\bibinfo {year} {1981})}\BibitemShut {NoStop}%
\bibitem [{\citenamefont {Iachello}\ \emph {et~al.}(1998)\citenamefont
  {Iachello}, \citenamefont {Zamfir},\ and\ \citenamefont
  {Casten}}]{iachello98}%
  \BibitemOpen
  \bibfield  {author} {\bibinfo {author} {\bibfnamefont {F.}~\bibnamefont
  {Iachello}}, \bibinfo {author} {\bibfnamefont {N.~V.}\ \bibnamefont
  {Zamfir}},\ and\ \bibinfo {author} {\bibfnamefont {R.~F.}\ \bibnamefont
  {Casten}},\ }\bibfield  {title} {\bibinfo {title} {{Phase Coexistence in
  Transitional Nuclei and the Interacting-Boson Model}},\ }\href
  {https://doi.org/10.1103/PhysRevLett.81.1191} {\bibfield  {journal} {\bibinfo
   {journal} {Phys. Rev. Lett.}\ }\textbf {\bibinfo {volume} {81}},\ \bibinfo
  {pages} {1191} (\bibinfo {year} {1998})}\BibitemShut {NoStop}%
\bibitem [{\citenamefont {McCutchan}\ \emph {et~al.}(2004)\citenamefont
  {McCutchan}, \citenamefont {Zamfir},\ and\ \citenamefont
  {Casten}}]{mccutchan04}%
  \BibitemOpen
  \bibfield  {author} {\bibinfo {author} {\bibfnamefont {E.~A.}\ \bibnamefont
  {McCutchan}}, \bibinfo {author} {\bibfnamefont {N.~V.}\ \bibnamefont
  {Zamfir}},\ and\ \bibinfo {author} {\bibfnamefont {R.~F.}\ \bibnamefont
  {Casten}},\ }\bibfield  {title} {\bibinfo {title} {{Mapping the interacting
  boson approximation symmetry triangle: New trajectories of structural
  evolution of rare-earth nuclei}},\ }\href
  {https://doi.org/10.1103/PhysRevC.69.064306} {\bibfield  {journal} {\bibinfo
  {journal} {Phys. Rev. C}\ }\textbf {\bibinfo {volume} {69}},\ \bibinfo
  {pages} {064306} (\bibinfo {year} {2004})}\BibitemShut {NoStop}%
\bibitem [{\citenamefont {Zyriliou}\ \emph {et~al.}(2022)\citenamefont
  {Zyriliou}, \citenamefont {Mertzimekis}, \citenamefont {Chalil},
  \citenamefont {Vasileiou}, \citenamefont {Mavrommatis}, \citenamefont
  {Bonatsos}, \citenamefont {Martinou}, \citenamefont {Peroulis},\ and\
  \citenamefont {Minkov}}]{zyriliou22}%
  \BibitemOpen
  \bibfield  {author} {\bibinfo {author} {\bibfnamefont {A.}~\bibnamefont
  {Zyriliou}}, \bibinfo {author} {\bibfnamefont {T.~J.}\ \bibnamefont
  {Mertzimekis}}, \bibinfo {author} {\bibfnamefont {A.}~\bibnamefont {Chalil}},
  \bibinfo {author} {\bibfnamefont {P.}~\bibnamefont {Vasileiou}}, \bibinfo
  {author} {\bibfnamefont {E.}~\bibnamefont {Mavrommatis}}, \bibinfo {author}
  {\bibfnamefont {D.}~\bibnamefont {Bonatsos}}, \bibinfo {author}
  {\bibfnamefont {A.}~\bibnamefont {Martinou}}, \bibinfo {author}
  {\bibfnamefont {S.}~\bibnamefont {Peroulis}},\ and\ \bibinfo {author}
  {\bibfnamefont {N.}~\bibnamefont {Minkov}},\ }\bibfield  {title} {\bibinfo
  {title} {{A study of some aspects of the nuclear structure in the even--even
  Yb isotopes}},\ }\href {https://doi.org/10.1140/epjp/s13360-022-02414-2}
  {\bibfield  {journal} {\bibinfo  {journal} {Eur. Phys. J. Plus}\ }\textbf
  {\bibinfo {volume} {137}},\ \bibinfo {pages} {352} (\bibinfo {year}
  {2022})}\BibitemShut {NoStop}%
\bibitem [{\citenamefont {{F. Iachello and A.
  Arima}}(1987{\natexlab{b}})}]{arima87}%
  \BibitemOpen
  \bibfield  {author} {\bibinfo {author} {\bibnamefont {{F. Iachello and A.
  Arima}}},\ }\href@noop {} {\emph {\bibinfo {title} {{The Interacting Boson
  Model}}}},\ Cambridge Monographs on Mathematical Physics\ (\bibinfo
  {publisher} {Cambridge University Press},\ \bibinfo {year}
  {1987})\BibitemShut {NoStop}%
\bibitem [{\citenamefont {McCutchan}\ \emph {et~al.}(2006)\citenamefont
  {McCutchan}, \citenamefont {Bonatsos},\ and\ \citenamefont
  {Zamfir}}]{mccutchan06}%
  \BibitemOpen
  \bibfield  {author} {\bibinfo {author} {\bibfnamefont {E.~A.}\ \bibnamefont
  {McCutchan}}, \bibinfo {author} {\bibfnamefont {D.}~\bibnamefont
  {Bonatsos}},\ and\ \bibinfo {author} {\bibfnamefont {N.~V.}\ \bibnamefont
  {Zamfir}},\ }\bibfield  {title} {\bibinfo {title} {{Connecting the
  X(5)-${\ensuremath{\beta}}^{2}$, X(5)-${\ensuremath{\beta}}^{4}$, and X(3)
  models to the shape/phase-transition region of the interacting boson
  model}},\ }\href {https://doi.org/10.1103/PhysRevC.74.034306} {\bibfield
  {journal} {\bibinfo  {journal} {Phys. Rev. C}\ }\textbf {\bibinfo {volume}
  {74}},\ \bibinfo {pages} {034306} (\bibinfo {year} {2006})}\BibitemShut
  {NoStop}%
\bibitem [{\citenamefont {Casten}\ \emph {et~al.}(1984)\citenamefont {Casten},
  \citenamefont {Aprahamian},\ and\ \citenamefont {Warner}}]{casten84}%
  \BibitemOpen
  \bibfield  {author} {\bibinfo {author} {\bibfnamefont {R.~F.}\ \bibnamefont
  {Casten}}, \bibinfo {author} {\bibfnamefont {A.}~\bibnamefont {Aprahamian}},\
  and\ \bibinfo {author} {\bibfnamefont {D.~D.}\ \bibnamefont {Warner}},\
  }\bibfield  {title} {\bibinfo {title} {{Axial asymmetry and the determination
  of effective $\ensuremath{\gamma}$ values in the interacting boson
  approximation}},\ }\href {https://doi.org/10.1103/PhysRevC.29.356} {\bibfield
   {journal} {\bibinfo  {journal} {Phys. Rev. C}\ }\textbf {\bibinfo {volume}
  {29}},\ \bibinfo {pages} {356} (\bibinfo {year} {1984})}\BibitemShut
  {NoStop}%
\bibitem [{\citenamefont {Zhang}\ \emph {et~al.}(2014)\citenamefont {Zhang},
  \citenamefont {Pan}, \citenamefont {Dai},\ and\ \citenamefont
  {Draayer}}]{zhang14}%
  \BibitemOpen
  \bibfield  {author} {\bibinfo {author} {\bibfnamefont {Y.}~\bibnamefont
  {Zhang}}, \bibinfo {author} {\bibfnamefont {F.}~\bibnamefont {Pan}}, \bibinfo
  {author} {\bibfnamefont {L.-R.}\ \bibnamefont {Dai}},\ and\ \bibinfo {author}
  {\bibfnamefont {J.~P.}\ \bibnamefont {Draayer}},\ }\bibfield  {title}
  {\bibinfo {title} {{Triaxial rotor in the SU(3) limit of the interacting
  boson model}},\ }\href {https://doi.org/10.1103/PhysRevC.90.044310}
  {\bibfield  {journal} {\bibinfo  {journal} {Phys. Rev. C}\ }\textbf {\bibinfo
  {volume} {90}},\ \bibinfo {pages} {044310} (\bibinfo {year}
  {2014})}\BibitemShut {NoStop}%
\bibitem [{\citenamefont {Zhang}\ \emph {et~al.}(2022)\citenamefont {Zhang},
  \citenamefont {He}, \citenamefont {Karlsson}, \citenamefont {Qi},
  \citenamefont {Pan},\ and\ \citenamefont {Draayer}}]{zhang22}%
  \BibitemOpen
  \bibfield  {author} {\bibinfo {author} {\bibfnamefont {Y.}~\bibnamefont
  {Zhang}}, \bibinfo {author} {\bibfnamefont {Y.-W.}\ \bibnamefont {He}},
  \bibinfo {author} {\bibfnamefont {D.}~\bibnamefont {Karlsson}}, \bibinfo
  {author} {\bibfnamefont {C.}~\bibnamefont {Qi}}, \bibinfo {author}
  {\bibfnamefont {F.}~\bibnamefont {Pan}},\ and\ \bibinfo {author}
  {\bibfnamefont {J.}~\bibnamefont {Draayer}},\ }\bibfield  {title} {\bibinfo
  {title} {{A theoretical interpretation of the anomalous reduced E2 transition
  probabilities along the yrast line of neutron-deficient nuclei}},\ }\href
  {https://doi.org/https://doi.org/10.1016/j.physletb.2022.137443} {\bibfield
  {journal} {\bibinfo  {journal} {Physics Letters B}\ }\textbf {\bibinfo
  {volume} {834}},\ \bibinfo {pages} {137443} (\bibinfo {year}
  {2022})}\BibitemShut {NoStop}%
\bibitem [{\citenamefont {Leschber}\ and\ \citenamefont
  {Draayer}(1987)}]{leschber87}%
  \BibitemOpen
  \bibfield  {author} {\bibinfo {author} {\bibfnamefont {Y.}~\bibnamefont
  {Leschber}}\ and\ \bibinfo {author} {\bibfnamefont {J.}~\bibnamefont
  {Draayer}},\ }\bibfield  {title} {\bibinfo {title} {{Algebraic realization of
  rotational dynamics}},\ }\href
  {https://doi.org/https://doi.org/10.1016/0370-2693(87)90829-X} {\bibfield
  {journal} {\bibinfo  {journal} {Physics Letters B}\ }\textbf {\bibinfo
  {volume} {190}},\ \bibinfo {pages} {1} (\bibinfo {year} {1987})}\BibitemShut
  {NoStop}%
\bibitem [{\citenamefont {Naqvi}\ \emph {et~al.}(1995)\citenamefont {Naqvi},
  \citenamefont {Troltenier}, \citenamefont {Draayer},\ and\ \citenamefont
  {Faessler}}]{naqvi95}%
  \BibitemOpen
  \bibfield  {author} {\bibinfo {author} {\bibfnamefont {C.}~\bibnamefont
  {Naqvi}, \bibfnamefont {H.~A.~Bahri}}, \bibinfo {author} {\bibfnamefont
  {D.}~\bibnamefont {Troltenier}}, \bibinfo {author} {\bibfnamefont {J.~P.}\
  \bibnamefont {Draayer}},\ and\ \bibinfo {author} {\bibfnamefont
  {A.}~\bibnamefont {Faessler}},\ }\bibfield  {title} {\bibinfo {title}
  {{Algebraic realization of the quantum rotor -- odd--A nuclei}},\ }\href
  {https://doi.org/https://doi.org/10.1007/BF01290907} {\bibfield  {journal}
  {\bibinfo  {journal} {Z. Phys. A}\ }\textbf {\bibinfo {volume} {351}},\
  \bibinfo {pages} {259} (\bibinfo {year} {1995})}\BibitemShut {NoStop}%
\bibitem [{\citenamefont {Smirnov}\ \emph {et~al.}(2000)\citenamefont
  {Smirnov}, \citenamefont {Smirnova},\ and\ \citenamefont
  {Van~Isacker}}]{smirnov00}%
  \BibitemOpen
  \bibfield  {author} {\bibinfo {author} {\bibfnamefont {Y.~F.}\ \bibnamefont
  {Smirnov}}, \bibinfo {author} {\bibfnamefont {N.~A.}\ \bibnamefont
  {Smirnova}},\ and\ \bibinfo {author} {\bibfnamefont {P.}~\bibnamefont
  {Van~Isacker}},\ }\bibfield  {title} {\bibinfo {title} {{SU(3) realization of
  the rigid asymmetric rotor within the interacting boson model}},\ }\href
  {https://doi.org/10.1103/PhysRevC.61.041302} {\bibfield  {journal} {\bibinfo
  {journal} {Phys. Rev. C}\ }\textbf {\bibinfo {volume} {61}},\ \bibinfo
  {pages} {041302(R)} (\bibinfo {year} {2000})}\BibitemShut {NoStop}%
\bibitem [{\citenamefont {Martinou}\ \emph
  {et~al.}(2021{\natexlab{b}})\citenamefont {Martinou}, \citenamefont
  {Bonatsos}, \citenamefont {Mertzimekis}, \citenamefont {Karakatsanis},
  \citenamefont {Assimakis}, \citenamefont {Peroulis}, \citenamefont
  {Sarantopoulou},\ and\ \citenamefont {Minkov}}]{martinou21a}%
  \BibitemOpen
  \bibfield  {author} {\bibinfo {author} {\bibfnamefont {A.}~\bibnamefont
  {Martinou}}, \bibinfo {author} {\bibfnamefont {D.}~\bibnamefont {Bonatsos}},
  \bibinfo {author} {\bibfnamefont {T.~J.}\ \bibnamefont {Mertzimekis}},
  \bibinfo {author} {\bibfnamefont {K.~E.}\ \bibnamefont {Karakatsanis}},
  \bibinfo {author} {\bibfnamefont {I.~E.}\ \bibnamefont {Assimakis}}, \bibinfo
  {author} {\bibfnamefont {S.~K.}\ \bibnamefont {Peroulis}}, \bibinfo {author}
  {\bibfnamefont {S.}~\bibnamefont {Sarantopoulou}},\ and\ \bibinfo {author}
  {\bibfnamefont {N.}~\bibnamefont {Minkov}},\ }\bibfield  {title} {\bibinfo
  {title} {{The islands of shape coexistence within the Elliott and the
  proxy-SU(3) Models}},\ }\href
  {https://doi.org/10.1140/epja/s10050-021-00396-w} {\bibfield  {journal}
  {\bibinfo  {journal} {Eur. Phys. J. A}\ }\textbf {\bibinfo {volume} {57}},\
  \bibinfo {pages} {84} (\bibinfo {year} {2021}{\natexlab{b}})}\BibitemShut
  {NoStop}%
\bibitem [{\citenamefont {Martinou}\ \emph {et~al.}(2017)\citenamefont
  {Martinou}, \citenamefont {Bonatsos}, \citenamefont {Assimakis},
  \citenamefont {Minkov}, \citenamefont {Sarantopoulou}, \citenamefont
  {Cakirli}, \citenamefont {Casten},\ and\ \citenamefont {Blaum}}]{martinou17}%
  \BibitemOpen
  \bibfield  {author} {\bibinfo {author} {\bibfnamefont {A.}~\bibnamefont
  {Martinou}}, \bibinfo {author} {\bibfnamefont {D.}~\bibnamefont {Bonatsos}},
  \bibinfo {author} {\bibfnamefont {I.~E.}\ \bibnamefont {Assimakis}}, \bibinfo
  {author} {\bibfnamefont {N.}~\bibnamefont {Minkov}}, \bibinfo {author}
  {\bibfnamefont {S.}~\bibnamefont {Sarantopoulou}}, \bibinfo {author}
  {\bibfnamefont {R.~B.}\ \bibnamefont {Cakirli}}, \bibinfo {author}
  {\bibfnamefont {R.~F.}\ \bibnamefont {Casten}},\ and\ \bibinfo {author}
  {\bibfnamefont {K.}~\bibnamefont {Blaum}},\ }\bibfield  {title} {\bibinfo
  {title} {{Parameter Free Predictions within the Proxy-SU(3) Model}},\
  }\href@noop {} {\bibfield  {journal} {\bibinfo  {journal} {Bulg. J. Phys.}\
  }\textbf {\bibinfo {volume} {44}},\ \bibinfo {pages} {407} (\bibinfo {year}
  {2017})}\BibitemShut {NoStop}%
\bibitem [{\citenamefont {Elliott}\ and\ \citenamefont
  {Wilsdon}(1968)}]{elliott68}%
  \BibitemOpen
  \bibfield  {author} {\bibinfo {author} {\bibfnamefont {J.~P.}\ \bibnamefont
  {Elliott}}\ and\ \bibinfo {author} {\bibfnamefont {C.~E.}\ \bibnamefont
  {Wilsdon}},\ }\bibfield  {title} {\bibinfo {title} {{Collective motion in the
  nuclear shell model IV. Odd-mass nuclei in the \textit{sd} shell}},\ }\href
  {https://doi.org/10.1098/rspa.1968.0033} {\bibfield  {journal} {\bibinfo
  {journal} {Proceedings of the Royal Society of London. Series A. Mathematical
  and Physical Sciences}\ }\textbf {\bibinfo {volume} {302}},\ \bibinfo {pages}
  {509} (\bibinfo {year} {1968})}\BibitemShut {NoStop}%
\bibitem [{\citenamefont {Chabanat}\ \emph {et~al.}(1998)\citenamefont
  {Chabanat}, \citenamefont {Bonche}, \citenamefont {Haensel}, \citenamefont
  {Meyer},\ and\ \citenamefont {Schaeffer}}]{chabanat98}%
  \BibitemOpen
  \bibfield  {author} {\bibinfo {author} {\bibfnamefont {E.}~\bibnamefont
  {Chabanat}}, \bibinfo {author} {\bibfnamefont {P.}~\bibnamefont {Bonche}},
  \bibinfo {author} {\bibfnamefont {P.}~\bibnamefont {Haensel}}, \bibinfo
  {author} {\bibfnamefont {J.}~\bibnamefont {Meyer}},\ and\ \bibinfo {author}
  {\bibfnamefont {R.}~\bibnamefont {Schaeffer}},\ }\bibfield  {title} {\bibinfo
  {title} {{A Skyrme parametrization from subnuclear to neutron star densities
  Part II. Nuclei far from stabilities}},\ }\href
  {https://doi.org/https://doi.org/10.1016/S0375-9474(98)00180-8} {\bibfield
  {journal} {\bibinfo  {journal} {Nuclear Physics A}\ }\textbf {\bibinfo
  {volume} {635}},\ \bibinfo {pages} {231} (\bibinfo {year}
  {1998})}\BibitemShut {NoStop}%
\bibitem [{\citenamefont {Bartel}\ \emph {et~al.}(1982)\citenamefont {Bartel},
  \citenamefont {Quentin}, \citenamefont {Brack}, \citenamefont {Guet},\ and\
  \citenamefont {Håkansson}}]{bartel82}%
  \BibitemOpen
  \bibfield  {author} {\bibinfo {author} {\bibfnamefont {J.}~\bibnamefont
  {Bartel}}, \bibinfo {author} {\bibfnamefont {P.}~\bibnamefont {Quentin}},
  \bibinfo {author} {\bibfnamefont {M.}~\bibnamefont {Brack}}, \bibinfo
  {author} {\bibfnamefont {C.}~\bibnamefont {Guet}},\ and\ \bibinfo {author}
  {\bibfnamefont {H.-B.}\ \bibnamefont {Håkansson}},\ }\bibfield  {title}
  {\bibinfo {title} {{Towards a better parametrisation of Skyrme-like effective
  forces: A critical study of the SkM force}},\ }\href
  {https://doi.org/https://doi.org/10.1016/0375-9474(82)90403-1} {\bibfield
  {journal} {\bibinfo  {journal} {Nuclear Physics A}\ }\textbf {\bibinfo
  {volume} {386}},\ \bibinfo {pages} {79} (\bibinfo {year} {1982})}\BibitemShut
  {NoStop}%
\bibitem [{\citenamefont {Berger}\ \emph {et~al.}(1984)\citenamefont {Berger},
  \citenamefont {Girod},\ and\ \citenamefont {Gogny}}]{berger84}%
  \BibitemOpen
  \bibfield  {author} {\bibinfo {author} {\bibfnamefont {J.}~\bibnamefont
  {Berger}}, \bibinfo {author} {\bibfnamefont {M.}~\bibnamefont {Girod}},\ and\
  \bibinfo {author} {\bibfnamefont {D.}~\bibnamefont {Gogny}},\ }\bibfield
  {title} {\bibinfo {title} {{Microscopic analysis of collective dynamics in
  low energy fission}},\ }\href
  {https://doi.org/https://doi.org/10.1016/0375-9474(84)90240-9} {\bibfield
  {journal} {\bibinfo  {journal} {Nuclear Physics A}\ }\textbf {\bibinfo
  {volume} {428}},\ \bibinfo {pages} {23} (\bibinfo {year} {1984})}\BibitemShut
  {NoStop}%
\bibitem [{\citenamefont {Goriely}\ \emph {et~al.}(2009)\citenamefont
  {Goriely}, \citenamefont {Hilaire}, \citenamefont {Girod},\ and\
  \citenamefont {P\'eru}}]{goriely09}%
  \BibitemOpen
  \bibfield  {author} {\bibinfo {author} {\bibfnamefont {S.}~\bibnamefont
  {Goriely}}, \bibinfo {author} {\bibfnamefont {S.}~\bibnamefont {Hilaire}},
  \bibinfo {author} {\bibfnamefont {M.}~\bibnamefont {Girod}},\ and\ \bibinfo
  {author} {\bibfnamefont {S.}~\bibnamefont {P\'eru}},\ }\bibfield  {title}
  {\bibinfo {title} {{First Gogny-Hartree-Fock-Bogoliubov Nuclear Mass
  Model}},\ }\href {https://doi.org/10.1103/PhysRevLett.102.242501} {\bibfield
  {journal} {\bibinfo  {journal} {Phys. Rev. Lett.}\ }\textbf {\bibinfo
  {volume} {102}},\ \bibinfo {pages} {242501} (\bibinfo {year}
  {2009})}\BibitemShut {NoStop}%
\bibitem [{\citenamefont {Vasileiou}\ \emph {et~al.}(2023)\citenamefont
  {Vasileiou}, \citenamefont {Mertzimekis}, \citenamefont {Mavrommatis},\ and\
  \citenamefont {Zyriliou}}]{vasileiou23}%
  \BibitemOpen
  \bibfield  {author} {\bibinfo {author} {\bibfnamefont {P.}~\bibnamefont
  {Vasileiou}}, \bibinfo {author} {\bibfnamefont {T.~J.}\ \bibnamefont
  {Mertzimekis}}, \bibinfo {author} {\bibfnamefont {E.}~\bibnamefont
  {Mavrommatis}},\ and\ \bibinfo {author} {\bibfnamefont {A.}~\bibnamefont
  {Zyriliou}},\ }\bibfield  {title} {\bibinfo {title} {{Nuclear Structure
  Investigations of Even--Even Hf Isotopes}},\ }\href
  {https://doi.org/10.3390/sym15010196} {\bibfield  {journal} {\bibinfo
  {journal} {Symmetry}\ }\textbf {\bibinfo {volume} {15}},\ \bibinfo {pages}
  {196} (\bibinfo {year} {2023})}\BibitemShut {NoStop}%
\bibitem [{\citenamefont {Pritychenko}\ \emph {et~al.}(2016)\citenamefont
  {Pritychenko}, \citenamefont {Birch}, \citenamefont {Singh},\ and\
  \citenamefont {Horoi}}]{pritychenko16}%
  \BibitemOpen
  \bibfield  {author} {\bibinfo {author} {\bibfnamefont {B.}~\bibnamefont
  {Pritychenko}}, \bibinfo {author} {\bibfnamefont {M.}~\bibnamefont {Birch}},
  \bibinfo {author} {\bibfnamefont {B.}~\bibnamefont {Singh}},\ and\ \bibinfo
  {author} {\bibfnamefont {M.}~\bibnamefont {Horoi}},\ }\bibfield  {title}
  {\bibinfo {title} {{Tables of E2 transition probabilities from the first
  $2^+$ states in even–even nuclei}},\ }\href
  {https://doi.org/https://doi.org/10.1016/j.adt.2015.10.001} {\bibfield
  {journal} {\bibinfo  {journal} {Atomic Data and Nuclear Data Tables}\
  }\textbf {\bibinfo {volume} {107}},\ \bibinfo {pages} {1} (\bibinfo {year}
  {2016})}\BibitemShut {NoStop}%
\bibitem [{nnd()}]{nndc}%
  \BibitemOpen
  \href@noop {} {\bibinfo {title} {{National Nuclear Data Center}}},\ \bibinfo
  {howpublished} {{https://www.nndc.bnl.gov/nudat3}}\BibitemShut {NoStop}%
\bibitem [{\citenamefont {Lotina}\ and\ \citenamefont
  {Nomura}(2024)}]{lotina24a}%
  \BibitemOpen
  \bibfield  {author} {\bibinfo {author} {\bibfnamefont {L.}~\bibnamefont
  {Lotina}}\ and\ \bibinfo {author} {\bibfnamefont {K.}~\bibnamefont
  {Nomura}},\ }\bibfield  {title} {\bibinfo {title} {{Impacts of hexadecapole
  deformations on the collective energy spectra of axially deformed nuclei}},\
  }\href {https://doi.org/10.1103/PhysRevC.109.034304} {\bibfield  {journal}
  {\bibinfo  {journal} {Phys. Rev. C}\ }\textbf {\bibinfo {volume} {109}},\
  \bibinfo {pages} {034304} (\bibinfo {year} {2024})}\BibitemShut {NoStop}%
\bibitem [{\citenamefont {Nomura}\ \emph {et~al.}(2012)\citenamefont {Nomura},
  \citenamefont {Shimizu}, \citenamefont {Vretenar}, \citenamefont {Nik\ifmmode
  \check{s}\else \v{s}\fi{}i\ifmmode~\acute{c}\else \'{c}\fi{}},\ and\
  \citenamefont {Otsuka}}]{nomura12}%
  \BibitemOpen
  \bibfield  {author} {\bibinfo {author} {\bibfnamefont {K.}~\bibnamefont
  {Nomura}}, \bibinfo {author} {\bibfnamefont {N.}~\bibnamefont {Shimizu}},
  \bibinfo {author} {\bibfnamefont {D.}~\bibnamefont {Vretenar}}, \bibinfo
  {author} {\bibfnamefont {T.}~\bibnamefont {Nik\ifmmode \check{s}\else
  \v{s}\fi{}i\ifmmode~\acute{c}\else \'{c}\fi{}}},\ and\ \bibinfo {author}
  {\bibfnamefont {T.}~\bibnamefont {Otsuka}},\ }\bibfield  {title} {\bibinfo
  {title} {{Robust Regularity in $\ensuremath{\gamma}$-Soft Nuclei and Its
  Microscopic Realization}},\ }\href
  {https://doi.org/10.1103/PhysRevLett.108.132501} {\bibfield  {journal}
  {\bibinfo  {journal} {Phys. Rev. Lett.}\ }\textbf {\bibinfo {volume} {108}},\
  \bibinfo {pages} {132501} (\bibinfo {year} {2012})}\BibitemShut {NoStop}%
\bibitem [{\citenamefont {Jolos}\ \emph
  {et~al.}(2005{\natexlab{a}})\citenamefont {Jolos}, \citenamefont {von
  Brentano},\ and\ \citenamefont {Pietralla}}]{jolos05a}%
  \BibitemOpen
  \bibfield  {author} {\bibinfo {author} {\bibfnamefont {R.~V.}\ \bibnamefont
  {Jolos}}, \bibinfo {author} {\bibfnamefont {P.}~\bibnamefont {von
  Brentano}},\ and\ \bibinfo {author} {\bibfnamefont {N.}~\bibnamefont
  {Pietralla}},\ }\bibfield  {title} {\bibinfo {title} {{Generalized Grodzins
  relation}},\ }\href {https://doi.org/10.1103/PhysRevC.71.044305} {\bibfield
  {journal} {\bibinfo  {journal} {Phys. Rev. C}\ }\textbf {\bibinfo {volume}
  {71}},\ \bibinfo {pages} {044305} (\bibinfo {year}
  {2005}{\natexlab{a}})}\BibitemShut {NoStop}%
\bibitem [{\citenamefont {Jolos}\ \emph
  {et~al.}(2005{\natexlab{b}})\citenamefont {Jolos}, \citenamefont {von
  Brentano}, \citenamefont {Dewald},\ and\ \citenamefont
  {Pietralla}}]{jolos05b}%
  \BibitemOpen
  \bibfield  {author} {\bibinfo {author} {\bibfnamefont {R.~V.}\ \bibnamefont
  {Jolos}}, \bibinfo {author} {\bibfnamefont {P.}~\bibnamefont {von Brentano}},
  \bibinfo {author} {\bibfnamefont {A.}~\bibnamefont {Dewald}},\ and\ \bibinfo
  {author} {\bibfnamefont {N.}~\bibnamefont {Pietralla}},\ }\bibfield  {title}
  {\bibinfo {title} {{Spin dependence of intrinsic and transition quadrupole
  moments}},\ }\href {https://doi.org/10.1103/PhysRevC.72.024310} {\bibfield
  {journal} {\bibinfo  {journal} {Phys. Rev. C}\ }\textbf {\bibinfo {volume}
  {72}},\ \bibinfo {pages} {024310} (\bibinfo {year}
  {2005}{\natexlab{b}})}\BibitemShut {NoStop}%
\bibitem [{\citenamefont {Jolos}\ and\ \citenamefont {von
  Brentano}(2006)}]{jolos06}%
  \BibitemOpen
  \bibfield  {author} {\bibinfo {author} {\bibfnamefont {R.~V.}\ \bibnamefont
  {Jolos}}\ and\ \bibinfo {author} {\bibfnamefont {P.}~\bibnamefont {von
  Brentano}},\ }\bibfield  {title} {\bibinfo {title} {{Mass coefficient and
  Grodzins relation for the ground-state band and \ensuremath{\gamma} band}},\
  }\href {https://doi.org/10.1103/PhysRevC.74.064307} {\bibfield  {journal}
  {\bibinfo  {journal} {Phys. Rev. C}\ }\textbf {\bibinfo {volume} {74}},\
  \bibinfo {pages} {064307} (\bibinfo {year} {2006})}\BibitemShut {NoStop}%
\bibitem [{\citenamefont {Jolos}\ and\ \citenamefont {von
  Brentano}(2007)}]{jolos07}%
  \BibitemOpen
  \bibfield  {author} {\bibinfo {author} {\bibfnamefont {R.~V.}\ \bibnamefont
  {Jolos}}\ and\ \bibinfo {author} {\bibfnamefont {P.}~\bibnamefont {von
  Brentano}},\ }\bibfield  {title} {\bibinfo {title} {{Bohr Hamiltonian with
  different mass coefficients for the ground- and \ensuremath{\gamma} bands
  from experimental data}},\ }\href
  {https://doi.org/10.1103/PhysRevC.76.024309} {\bibfield  {journal} {\bibinfo
  {journal} {Phys. Rev. C}\ }\textbf {\bibinfo {volume} {76}},\ \bibinfo
  {pages} {024309} (\bibinfo {year} {2007})}\BibitemShut {NoStop}%
\bibitem [{\citenamefont {Jolos}\ and\ \citenamefont
  {Kolganova}(2021)}]{jolos21}%
  \BibitemOpen
  \bibfield  {author} {\bibinfo {author} {\bibfnamefont {R.}~\bibnamefont
  {Jolos}}\ and\ \bibinfo {author} {\bibfnamefont {E.}~\bibnamefont
  {Kolganova}},\ }\bibfield  {title} {\bibinfo {title} {{Derivation of the
  Grodzins relation in collective nuclear model}},\ }\href
  {https://doi.org/https://doi.org/10.1016/j.physletb.2021.136581} {\bibfield
  {journal} {\bibinfo  {journal} {Physics Letters B}\ }\textbf {\bibinfo
  {volume} {820}},\ \bibinfo {pages} {136581} (\bibinfo {year}
  {2021})}\BibitemShut {NoStop}%
\bibitem [{\citenamefont {Rudigier}\ \emph {et~al.}(2010)\citenamefont
  {Rudigier}, \citenamefont {Régis}, \citenamefont {Jolie}, \citenamefont
  {Zell},\ and\ \citenamefont {Fransen}}]{rudigier10}%
  \BibitemOpen
  \bibfield  {author} {\bibinfo {author} {\bibfnamefont {M.}~\bibnamefont
  {Rudigier}}, \bibinfo {author} {\bibfnamefont {J.-M.}\ \bibnamefont
  {Régis}}, \bibinfo {author} {\bibfnamefont {J.}~\bibnamefont {Jolie}},
  \bibinfo {author} {\bibfnamefont {K.}~\bibnamefont {Zell}},\ and\ \bibinfo
  {author} {\bibfnamefont {C.}~\bibnamefont {Fransen}},\ }\bibfield  {title}
  {\bibinfo {title} {{Lifetime of the first excited $2^+$ state in $^{172}$W
  and $^{178}$W}},\ }\href
  {https://doi.org/https://doi.org/10.1016/j.nuclphysa.2010.07.003} {\bibfield
  {journal} {\bibinfo  {journal} {Nuclear Physics A}\ }\textbf {\bibinfo
  {volume} {847}},\ \bibinfo {pages} {89} (\bibinfo {year} {2010})}\BibitemShut
  {NoStop}%
\bibitem [{\citenamefont {McCutchan}\ and\ \citenamefont
  {Zamfir}(2005)}]{mccutchan05a}%
  \BibitemOpen
  \bibfield  {author} {\bibinfo {author} {\bibfnamefont {E.~A.}\ \bibnamefont
  {McCutchan}}\ and\ \bibinfo {author} {\bibfnamefont {N.~V.}\ \bibnamefont
  {Zamfir}},\ }\bibfield  {title} {\bibinfo {title} {{Simple description of
  light W, Os, and Pt nuclei in the interacting boson model}},\ }\href
  {https://doi.org/10.1103/PhysRevC.71.054306} {\bibfield  {journal} {\bibinfo
  {journal} {Phys. Rev. C}\ }\textbf {\bibinfo {volume} {71}},\ \bibinfo
  {pages} {054306} (\bibinfo {year} {2005})}\BibitemShut {NoStop}%
\bibitem [{\citenamefont {McCutchan}\ \emph {et~al.}(2005)\citenamefont
  {McCutchan}, \citenamefont {Casten},\ and\ \citenamefont
  {Zamfir}}]{mccutchan05b}%
  \BibitemOpen
  \bibfield  {author} {\bibinfo {author} {\bibfnamefont {E.~A.}\ \bibnamefont
  {McCutchan}}, \bibinfo {author} {\bibfnamefont {R.~F.}\ \bibnamefont
  {Casten}},\ and\ \bibinfo {author} {\bibfnamefont {N.~V.}\ \bibnamefont
  {Zamfir}},\ }\bibfield  {title} {\bibinfo {title} {{Simple interpretation of
  shape evolution in Pt isotopes without intruder states}},\ }\href
  {https://doi.org/10.1103/PhysRevC.71.061301} {\bibfield  {journal} {\bibinfo
  {journal} {Phys. Rev. C}\ }\textbf {\bibinfo {volume} {71}},\ \bibinfo
  {pages} {061301(R)} (\bibinfo {year} {2005})}\BibitemShut {NoStop}%
\bibitem [{\citenamefont {Harter}\ \emph {et~al.}(2022)\citenamefont {Harter},
  \citenamefont {Knafla}, \citenamefont {Frie\ss{}ner}, \citenamefont
  {H\"afner}, \citenamefont {Jolie}, \citenamefont {Blazhev}, \citenamefont
  {Dewald}, \citenamefont {Dunkel}, \citenamefont {Esmaylzadeh}, \citenamefont
  {Fransen}, \citenamefont {Karayonchev}, \citenamefont {Lawless},
  \citenamefont {Ley}, \citenamefont {R\'egis},\ and\ \citenamefont
  {Zell}}]{harter22}%
  \BibitemOpen
  \bibfield  {author} {\bibinfo {author} {\bibfnamefont {A.}~\bibnamefont
  {Harter}}, \bibinfo {author} {\bibfnamefont {L.}~\bibnamefont {Knafla}},
  \bibinfo {author} {\bibfnamefont {G.}~\bibnamefont {Frie\ss{}ner}}, \bibinfo
  {author} {\bibfnamefont {G.}~\bibnamefont {H\"afner}}, \bibinfo {author}
  {\bibfnamefont {J.}~\bibnamefont {Jolie}}, \bibinfo {author} {\bibfnamefont
  {A.}~\bibnamefont {Blazhev}}, \bibinfo {author} {\bibfnamefont
  {A.}~\bibnamefont {Dewald}}, \bibinfo {author} {\bibfnamefont
  {F.}~\bibnamefont {Dunkel}}, \bibinfo {author} {\bibfnamefont
  {A.}~\bibnamefont {Esmaylzadeh}}, \bibinfo {author} {\bibfnamefont
  {C.}~\bibnamefont {Fransen}}, \bibinfo {author} {\bibfnamefont
  {V.}~\bibnamefont {Karayonchev}}, \bibinfo {author} {\bibfnamefont
  {K.}~\bibnamefont {Lawless}}, \bibinfo {author} {\bibfnamefont
  {M.}~\bibnamefont {Ley}}, \bibinfo {author} {\bibfnamefont {J.-M.}\
  \bibnamefont {R\'egis}},\ and\ \bibinfo {author} {\bibfnamefont {K.~O.}\
  \bibnamefont {Zell}},\ }\bibfield  {title} {\bibinfo {title} {{Lifetime
  measurements in the tungsten isotopes $^{176,178,180}\mathrm{W}$}},\ }\href
  {https://doi.org/10.1103/PhysRevC.106.024326} {\bibfield  {journal} {\bibinfo
   {journal} {Phys. Rev. C}\ }\textbf {\bibinfo {volume} {106}},\ \bibinfo
  {pages} {024326} (\bibinfo {year} {2022})}\BibitemShut {NoStop}%
\bibitem [{\citenamefont {Wiederhold}\ \emph {et~al.}(2019)\citenamefont
  {Wiederhold}, \citenamefont {Werner}, \citenamefont {Kern}, \citenamefont
  {Pietralla}, \citenamefont {Bucurescu}, \citenamefont {Carroll},
  \citenamefont {Cooper}, \citenamefont {Daniel}, \citenamefont {Filipescu},
  \citenamefont {Florea}, \citenamefont {Gerst}, \citenamefont {Ghita},
  \citenamefont {Gurgi}, \citenamefont {Jolie}, \citenamefont {Ilieva},
  \citenamefont {Lica}, \citenamefont {Marginean}, \citenamefont {Marginean},
  \citenamefont {Mihai}, \citenamefont {Mitu}, \citenamefont {Naqvi},
  \citenamefont {Nita}, \citenamefont {Rudigier}, \citenamefont {Stegemann},
  \citenamefont {Pascu},\ and\ \citenamefont {Regan}}]{wiederhold19}%
  \BibitemOpen
  \bibfield  {author} {\bibinfo {author} {\bibfnamefont {J.}~\bibnamefont
  {Wiederhold}}, \bibinfo {author} {\bibfnamefont {V.}~\bibnamefont {Werner}},
  \bibinfo {author} {\bibfnamefont {R.}~\bibnamefont {Kern}}, \bibinfo {author}
  {\bibfnamefont {N.}~\bibnamefont {Pietralla}}, \bibinfo {author}
  {\bibfnamefont {D.}~\bibnamefont {Bucurescu}}, \bibinfo {author}
  {\bibfnamefont {R.}~\bibnamefont {Carroll}}, \bibinfo {author} {\bibfnamefont
  {N.}~\bibnamefont {Cooper}}, \bibinfo {author} {\bibfnamefont
  {T.}~\bibnamefont {Daniel}}, \bibinfo {author} {\bibfnamefont
  {D.}~\bibnamefont {Filipescu}}, \bibinfo {author} {\bibfnamefont
  {N.}~\bibnamefont {Florea}}, \bibinfo {author} {\bibfnamefont {R.-B.}\
  \bibnamefont {Gerst}}, \bibinfo {author} {\bibfnamefont {D.}~\bibnamefont
  {Ghita}}, \bibinfo {author} {\bibfnamefont {L.}~\bibnamefont {Gurgi}},
  \bibinfo {author} {\bibfnamefont {J.}~\bibnamefont {Jolie}}, \bibinfo
  {author} {\bibfnamefont {R.~S.}\ \bibnamefont {Ilieva}}, \bibinfo {author}
  {\bibfnamefont {R.}~\bibnamefont {Lica}}, \bibinfo {author} {\bibfnamefont
  {N.}~\bibnamefont {Marginean}}, \bibinfo {author} {\bibfnamefont
  {R.}~\bibnamefont {Marginean}}, \bibinfo {author} {\bibfnamefont
  {C.}~\bibnamefont {Mihai}}, \bibinfo {author} {\bibfnamefont {I.~O.}\
  \bibnamefont {Mitu}}, \bibinfo {author} {\bibfnamefont {F.}~\bibnamefont
  {Naqvi}}, \bibinfo {author} {\bibfnamefont {C.}~\bibnamefont {Nita}},
  \bibinfo {author} {\bibfnamefont {M.}~\bibnamefont {Rudigier}}, \bibinfo
  {author} {\bibfnamefont {S.}~\bibnamefont {Stegemann}}, \bibinfo {author}
  {\bibfnamefont {S.}~\bibnamefont {Pascu}},\ and\ \bibinfo {author}
  {\bibfnamefont {P.~H.}\ \bibnamefont {Regan}},\ }\bibfield  {title} {\bibinfo
  {title} {{Evolution of $E2$ strength in the rare-earth isotopes
  $^{174,176,178,180}\mathrm{Hf}$}},\ }\href
  {https://doi.org/10.1103/PhysRevC.99.024316} {\bibfield  {journal} {\bibinfo
  {journal} {Phys. Rev. C}\ }\textbf {\bibinfo {volume} {99}},\ \bibinfo
  {pages} {024316} (\bibinfo {year} {2019})}\BibitemShut {NoStop}%
\bibitem [{\citenamefont {Zamfir}\ \emph {et~al.}(1995)\citenamefont {Zamfir},
  \citenamefont {Hering}, \citenamefont {Casten},\ and\ \citenamefont
  {Paul}}]{zamfir95}%
  \BibitemOpen
  \bibfield  {author} {\bibinfo {author} {\bibfnamefont {N.}~\bibnamefont
  {Zamfir}}, \bibinfo {author} {\bibfnamefont {G.}~\bibnamefont {Hering}},
  \bibinfo {author} {\bibfnamefont {R.}~\bibnamefont {Casten}},\ and\ \bibinfo
  {author} {\bibfnamefont {P.}~\bibnamefont {Paul}},\ }\bibfield  {title}
  {\bibinfo {title} {{Hexadecapole deformations in actinide and trans-actinide
  nuclei}},\ }\href
  {https://doi.org/https://doi.org/10.1016/0370-2693(95)00978-T} {\bibfield
  {journal} {\bibinfo  {journal} {Physics Letters B}\ }\textbf {\bibinfo
  {volume} {357}},\ \bibinfo {pages} {515} (\bibinfo {year}
  {1995})}\BibitemShut {NoStop}%
\bibitem [{\citenamefont {Jolie}\ and\ \citenamefont
  {Linnemann}(2003)}]{jolie03}%
  \BibitemOpen
  \bibfield  {author} {\bibinfo {author} {\bibfnamefont {J.}~\bibnamefont
  {Jolie}}\ and\ \bibinfo {author} {\bibfnamefont {A.}~\bibnamefont
  {Linnemann}},\ }\bibfield  {title} {\bibinfo {title} {{Prolate-oblate phase
  transition in the Hf-Hg mass region}},\ }\href
  {https://doi.org/10.1103/PhysRevC.68.031301} {\bibfield  {journal} {\bibinfo
  {journal} {Phys. Rev. C}\ }\textbf {\bibinfo {volume} {68}},\ \bibinfo
  {pages} {031301(R)} (\bibinfo {year} {2003})}\BibitemShut {NoStop}%
\bibitem [{\citenamefont {Bonatsos}\ \emph {et~al.}(2004)\citenamefont
  {Bonatsos}, \citenamefont {Lenis}, \citenamefont {Petrellis},\ and\
  \citenamefont {Terziev}}]{bonatsos04}%
  \BibitemOpen
  \bibfield  {author} {\bibinfo {author} {\bibfnamefont {D.}~\bibnamefont
  {Bonatsos}}, \bibinfo {author} {\bibfnamefont {D.}~\bibnamefont {Lenis}},
  \bibinfo {author} {\bibfnamefont {D.}~\bibnamefont {Petrellis}},\ and\
  \bibinfo {author} {\bibfnamefont {P.~A.}\ \bibnamefont {Terziev}},\
  }\bibfield  {title} {\bibinfo {title} {{Z(5): critical point symmetry for the
  prolate to oblate nuclear shape phase transition}},\ }\href
  {https://doi.org/https://doi.org/10.1016/j.physletb.2004.03.029} {\bibfield
  {journal} {\bibinfo  {journal} {Physics Letters B}\ }\textbf {\bibinfo
  {volume} {588}},\ \bibinfo {pages} {172} (\bibinfo {year}
  {2004})}\BibitemShut {NoStop}%
\bibitem [{\citenamefont {Zhang}\ \emph {et~al.}(2012)\citenamefont {Zhang},
  \citenamefont {Pan}, \citenamefont {Liu}, \citenamefont {Luo},\ and\
  \citenamefont {Draayer}}]{zhang12}%
  \BibitemOpen
  \bibfield  {author} {\bibinfo {author} {\bibfnamefont {Y.}~\bibnamefont
  {Zhang}}, \bibinfo {author} {\bibfnamefont {F.}~\bibnamefont {Pan}}, \bibinfo
  {author} {\bibfnamefont {Y.-X.}\ \bibnamefont {Liu}}, \bibinfo {author}
  {\bibfnamefont {Y.-A.}\ \bibnamefont {Luo}},\ and\ \bibinfo {author}
  {\bibfnamefont {J.~P.}\ \bibnamefont {Draayer}},\ }\bibfield  {title}
  {\bibinfo {title} {{Analytically solvable prolate-oblate shape phase
  transitional description within the SU(3) limit of the interacting boson
  model}},\ }\href {https://doi.org/10.1103/PhysRevC.85.064312} {\bibfield
  {journal} {\bibinfo  {journal} {Phys. Rev. C}\ }\textbf {\bibinfo {volume}
  {85}},\ \bibinfo {pages} {064312} (\bibinfo {year} {2012})}\BibitemShut
  {NoStop}%
\bibitem [{\citenamefont {Wang}\ \emph {et~al.}(2023)\citenamefont {Wang},
  \citenamefont {He}, \citenamefont {Li},\ and\ \citenamefont {Zhou}}]{wang23}%
  \BibitemOpen
  \bibfield  {author} {\bibinfo {author} {\bibfnamefont {T.}~\bibnamefont
  {Wang}}, \bibinfo {author} {\bibfnamefont {B.}~\bibnamefont {He}}, \bibinfo
  {author} {\bibfnamefont {D.}~\bibnamefont {Li}},\ and\ \bibinfo {author}
  {\bibfnamefont {C.}~\bibnamefont {Zhou}},\ }\bibfield  {title} {\bibinfo
  {title} {{Prolate-oblate asymmetric shape phase transition in the interacting
  boson model with SU(3) higher-order interactions}},\ }\href
  {https://doi.org/10.1103/PhysRevC.107.064322} {\bibfield  {journal} {\bibinfo
   {journal} {Phys. Rev. C}\ }\textbf {\bibinfo {volume} {107}},\ \bibinfo
  {pages} {064322} (\bibinfo {year} {2023})}\BibitemShut {NoStop}%
\bibitem [{\citenamefont {Maruhn}\ \emph {et~al.}(2014)\citenamefont {Maruhn},
  \citenamefont {Reinhard}, \citenamefont {Stevenson},\ and\ \citenamefont
  {Umar}}]{maruhn14}%
  \BibitemOpen
  \bibfield  {author} {\bibinfo {author} {\bibfnamefont {J.~A.}\ \bibnamefont
  {Maruhn}}, \bibinfo {author} {\bibfnamefont {P.~G.}\ \bibnamefont
  {Reinhard}}, \bibinfo {author} {\bibfnamefont {P.~D.}\ \bibnamefont
  {Stevenson}},\ and\ \bibinfo {author} {\bibfnamefont {A.~S.}\ \bibnamefont
  {Umar}},\ }\bibfield  {title} {\bibinfo {title} {{The TDHF code Sky3D}},\
  }\href {https://doi.org/https://doi.org/10.1016/j.cpc.2014.04.008} {\bibfield
   {journal} {\bibinfo  {journal} {Computer Physics Communications}\ }\textbf
  {\bibinfo {volume} {185}},\ \bibinfo {pages} {2195} (\bibinfo {year}
  {2014})}\BibitemShut {NoStop}%
\bibitem [{\citenamefont {Schuetrumpf}\ \emph {et~al.}(2018)\citenamefont
  {Schuetrumpf}, \citenamefont {Reinhard}, \citenamefont {Stevenson},
  \citenamefont {Umar},\ and\ \citenamefont {Maruhn}}]{schuetrumpf18}%
  \BibitemOpen
  \bibfield  {author} {\bibinfo {author} {\bibfnamefont {B.}~\bibnamefont
  {Schuetrumpf}}, \bibinfo {author} {\bibfnamefont {P.~G.}\ \bibnamefont
  {Reinhard}}, \bibinfo {author} {\bibfnamefont {P.~D.}\ \bibnamefont
  {Stevenson}}, \bibinfo {author} {\bibfnamefont {A.~S.}\ \bibnamefont
  {Umar}},\ and\ \bibinfo {author} {\bibfnamefont {J.~A.}\ \bibnamefont
  {Maruhn}},\ }\bibfield  {title} {\bibinfo {title} {{The TDHF code Sky3D
  version 1.1}},\ }\href
  {https://doi.org/https://doi.org/10.1016/j.cpc.2018.03.012} {\bibfield
  {journal} {\bibinfo  {journal} {Computer Physics Communications}\ }\textbf
  {\bibinfo {volume} {229}},\ \bibinfo {pages} {211} (\bibinfo {year}
  {2018})}\BibitemShut {NoStop}%
\bibitem [{\citenamefont {Alexa}\ \emph {et~al.}(2022)\citenamefont {Alexa},
  \citenamefont {Abolghasem}, \citenamefont {Thiamova}, \citenamefont
  {Bonatsos}, \citenamefont {Rodr\'{\i}guez},\ and\ \citenamefont
  {Reinhard}}]{alexa22}%
  \BibitemOpen
  \bibfield  {author} {\bibinfo {author} {\bibfnamefont {P.}~\bibnamefont
  {Alexa}}, \bibinfo {author} {\bibfnamefont {M.}~\bibnamefont {Abolghasem}},
  \bibinfo {author} {\bibfnamefont {G.}~\bibnamefont {Thiamova}}, \bibinfo
  {author} {\bibfnamefont {D.}~\bibnamefont {Bonatsos}}, \bibinfo {author}
  {\bibfnamefont {T.~R.}\ \bibnamefont {Rodr\'{\i}guez}},\ and\ \bibinfo
  {author} {\bibfnamefont {P.-G.}\ \bibnamefont {Reinhard}},\ }\bibfield
  {title} {\bibinfo {title} {{Macroscopic and microscopic description of phase
  transition in cerium isotopes}},\ }\href
  {https://doi.org/10.1103/PhysRevC.106.054304} {\bibfield  {journal} {\bibinfo
   {journal} {Phys. Rev. C}\ }\textbf {\bibinfo {volume} {106}},\ \bibinfo
  {pages} {054304} (\bibinfo {year} {2022})}\BibitemShut {NoStop}%
\bibitem [{ame(2006{\natexlab{a}})}]{amedee1}%
  \BibitemOpen
  \href@noop {} {\bibinfo {title} {{Hartree–Fock–Bogoliubov Results based
  on the Gogny Force}}},\ \bibinfo {howpublished}
  {{http://www-phynu.cea.fr/science\_en\_ligne/carte\_potentiels\_microscopiques/carte\_potentiel\_nucleaire\\
  \_eng.htm}} (\bibinfo {year} {2006}{\natexlab{a}})\BibitemShut {NoStop}%
\bibitem [{ame(2006{\natexlab{b}})}]{amedee2}%
  \BibitemOpen
  \href@noop {} {\bibinfo {title} {{HFB+5DCH Results}}},\ \bibinfo
  {howpublished}
  {{https://www-phynu.cea.fr/science\_en\_ligne/carte\_potentiels\_microscopiques/tables/HFB-5DCH-table\_eng.htm}}
  (\bibinfo {year} {2006}{\natexlab{b}})\BibitemShut {NoStop}%
\end{thebibliography}%

\end{document}